\documentclass[a4paper,11pt]{article}

\pdfoutput=1

\usepackage{jheppub}

\usepackage{graphicx}    
\usepackage{dcolumn}     
\usepackage{bm}          
\usepackage{amssymb}     
\usepackage{subfigure}
\usepackage{verbatim}
\usepackage{environ}
\usepackage{pythonhighlight}
\usepackage{fancyvrb}

\usepackage[english]{babel}
\usepackage{xcolor}
\usepackage{graphicx}
\usepackage{amsmath}
\usepackage{dsfont}
\usepackage[makeroom]{cancel}
\usepackage{url}
\usepackage{hyperref}
\usepackage{slashed}

\usepackage{mathabx}

\newcommand{\beq}{\begin{eqnarray}}
\newcommand{\eeq}{\end{eqnarray}}

\newcommand{\as}{\alpha_{\mbox{\tiny{S}}}}
\newcommand{\eps}{\epsilon}

\newcommand{\sig}{\sigma}
\newcommand{\Gam}{\Gamma}
\newcommand{\gam}{\gamma}

\newcommand{\lam}{\lambda}

\newcommand{\npo}{{n+1}}

\newcommand{\mc}{\mathcal}

\newcommand{\lra}{\leftrightarrow}

\newcommand{\LO}{{\mbox{\tiny{LO}}}}
\newcommand{\NLO}{{\mbox{\tiny{NLO}}}}

\newcommand{\nnb}{\nonumber}

\newcommand{\Si}{{\bf S}_i}
\newcommand{\Sj}{{\bf S}_j}

\newcommand{\Cj}{{\bf C}_{ij}}

\newcommand{\q}{{q}}

\def\eq#1{Eq.~(\ref{#1})}

\newcommand{\bL}[2]{{\bf L}_{#1}^{#2}}
\newcommand{\bS}[1]{{\bf S}_{#1}}
\newcommand{\bC}[1]{{\bf C}_{#1}}

\newcommand{\bbL}[2]{\overline{\bf L}^{#1}_{#2}}
\newcommand{\bbS}[1]{\overline{\bf S}_{#1}}
\newcommand{\bbC}[1]{\overline{\bf C}_{#1}}

\newcommand{\bbHC}[1]{\overline{\bf HC}_{#1}}

\newcommand{\bSi}{\bbS{i}}
\newcommand{\bSj}{\bbS{j}}
\newcommand{\bCj}{\bbC{ij}}

\newcommand{\W}[1]{\mc{W}_{#1}}

\newcommand{\Z}[1]{\mc{Z}_{#1}}

\newcommand{\sk}[2]{\bar s_{#1}^{(#2)}}

\newcommand{\Norm}{\mc{N}_1}
\newcommand{\hc}{{\rm \, hc}}

\newcommand{\Bn}{B}
\newcommand{\Vl}{V}
\newcommand{\Rl}{R}

\newcommand{\varsi}{\varsigma}

\newcommand{\euler}{\gamma_{\!_{_E}}}

\newcommand{\zg}{({\rm 0g})}
\newcommand{\og}{({\rm 1g})}
\newcommand{\tg}{({\rm 2g})}
\newcommand{\madnk}{{\sc MadNkLO}}
\newcommand{\aMC}{{\sc MadGraph5\_aMC@NLO}}

\newcommand{\final}[1]{\theta_{{#1}\in{\rm F}}}
\newcommand{\initial}[1]{\theta_{{#1}\in{\rm I}}}

\preprint{TIF-UNIMI-2022-14}
\title{
Local analytic sector subtraction for initial- and final-state radiation at NLO in massless QCD
}

\author[a]{Gloria Bertolotti,}
\author[a]{Paolo Torrielli,}
\author[a]{Sandro Uccirati,}
\author[b]{and Marco Zaro}

\affiliation[a]{Dipartimento di Fisica, Universit\`a di Torino, and INFN, Sezione di Torino, \\ Via P. Giuria 1, I-10125 Torino, Italy}
\affiliation[b]{TIF Lab, Universit\`a di Milano, and INFN, Sezione di Milano, \\ Via Celoria 16, I-20133 Milano, Italy}

\emailAdd{gloria.bertolotti@unito.it, paolo.torrielli@unito.it, sandro.uccirati@unito.it, marco.zaro@mi.infn.it}

\abstract{
Within the framework of local analytic sector subtraction, we present the subtraction of next-to-leading-order QCD singularities for processes featuring massless coloured particles in the initial as well as in the final state. The features of the method are explained in detail, including the introduction of an optimisation procedure aiming at improving numerical stability at the cost of no extra analytic complexity. A numerical validation is provided for a variety of processes relevant to lepton as well as hadron colliders.
This work constitutes a relevant step in view of the application of our subtraction method to processes involving initial-state radiation at next-to-next-to-leading order in QCD.
}

\begin{document}
\maketitle


\section{Introduction}

In quest of a deeper understanding of fundamental interactions, and of the identification of potential new-physics effects in experimental measurements, the availability of highly accurate theoretical calculations is more and more important for a large variety of scattering processes and relevant collider observables.
In turn, this availability stems from the existence of frameworks capable of making explicit the cancellation of infrared and collinear (IRC) singularities arising in gauge theories beyond the Born approximation.

General frameworks to solve this singularity problem at next-to-leading order (NLO) in perturbation theory were developed in the `90s \cite{Frixione:1995ms,Frixione:1997np,Catani:1996vz,Nagy:2003qn,Bevilacqua:2013iha}, employing \emph{infrared subtraction}, which eventually resulted in an accuracy revolution instrumental to the success of the physics programme of the Large Hadron Collider (LHC) and other colliders.
In a subtraction method, the universal long-distance behaviour of scattering amplitudes allows to design functions (the counterterms) which approximate radiative matrix elements squared in all of their IRC-singular limits, so that the difference between complete and approximate matrix elements is regular locally in phase space. One then adds back the counterterms, analytically integrated over the radiative phase space, to the virtual-correction matrix elements. The KLN theorem \cite{Kinoshita:1962ur,Lee:1964is} ensures this sum to be finite for IRC-safe observables, hence subtracted real and virtual contributions separately lend themselves to an efficient numerical evaluation.

At variance with NLO, at next-to-NLO (NNLO) the infrared-subtraction problem has proved extremely challenging due to a steep increase in technical complexity. Although several methods  \cite{GehrmannDeRidder:2005cm,Somogyi:2005xz,Czakon:2010td,Binoth:2000ps,Anastasiou:2003gr,Caola:2017dug,Catani:2007vq,Boughezal:2015dva,Cacciari:2015jma,Sborlini:2016hat,Herzog:2018ily,Capatti:2019ypt}, both within and beyond subtraction, have been proposed that address classes of processes of high phenomenological interest, and essentially the NNLO problem is solved for the most important $2\to2$ reactions, a general solution is still elusive.

In \cite{Magnea:2018hab,Magnea:2018ebr} the ingredients were defined  of a new method, local analytic sector subtraction, aiming at a solution of the NNLO QCD subtraction problem for generic processes.
Such a method is conceived to minimise the complexity in the integration of subtraction counterterms by systematically exploiting all available freedom in their definition and parametrisation, resulting in their analytic integrability \cite{Magnea:2020trj} by means of standard (as opposed to integration-by-parts reduction) techniques in the massless case.
The framework has been so far deduced and characterised in the case of reactions not featuring QCD partons in the initial-state, i.e.~lepton-lepton collisions.

While the general proof of the method achieving its goals for such a class of NNLO processes will be given elsewhere \cite{Bertolotti:NNLO}, in this article we concentrate on extending local analytic sector subtraction at NLO to processes featuring initial-state QCD partons, thereby encompassing all possible collider types. This extension thus represents a fundamental step towards achieving a fully general local analytic sector subtraction procedure at NNLO.

On top of defining and integrating all necessary counterterms for NLO subtraction, we also propose a novel systematic optimisation of the subtraction procedure that aims at numerically improving the quality of the singularity-cancellation mechanism. While optimisation recipes are common at NLO, see for instance \cite{Frixione:1995ms,Nagy:2003tz}, they typically entail an increase in the complexity of the involved analytic integrations. In our proposal, which is applicable to any subtraction method, optimisation essentially comes without additional analytic complexity, a feature that will prove crucial when exporting the method to NNLO level.

The structure of the paper is as follows. In Section \ref{sec:NLOsubt_gen} we describe in full detail our NLO subtraction procedure, and in particular introduce the above-mentioned optimisation prescription in Section \ref{sec:localctdamping}; Section \ref{sec:Integration} deals with the analytic integration of the subtraction counterterms, enabling to show in Section \ref{sec:NLOsubtFormula} the cancellation of IRC poles for general collider processes at NLO in massless QCD; Section
\ref{sec:Numerics} documents the implementation of our method in an automated software framework, and the related validation at the level of both IRC limits and physical cross sections, for a variety of NLO processes; finally in Section \ref{sec:conclusion} we draw our conclusions.
Five technical Appendices report relevant formulae and proofs ensuring a local cancellation of IRC singularities, as well as details of the numerical implementation.


\section{NLO subtraction in presence of initial-state partons}
\label{sec:NLOsubt_gen}

\subsection{Generalities of the subtraction procedure}
We start by considering the expression of the differential cross section for a hadron-initiated scattering process,
\beq
d \sig_{AB}(p_A, p_B)
& = &
\sum_{a,b} \,
\int_0^1 d\eta_a \, 
f_{a / A}(\eta_a,\mu^2_F) 
\int_0^1 d\eta_b \,
f_{b / B}(\eta_b,\mu^2_F) \, 
d \hat{\sig}_{ab} (k_a, k_b, \mu^2_F)
\, ,
\label{eq:hxsec}
\eeq
where $a$ and $b$ represent the flavours of the incoming partons carrying the longitudinal momentum fractions $\eta_a, \eta_b$ of the respective incoming hadrons $A$ and $B$, with $k_a = \eta_a \, p_A$ and $k_b = \eta_b \, p_B$. Upon neglecting non-perturbative corrections $\mc O((\Lambda_{\rm QCD}/Q)^p)$, the cross section is factorised into a long-distance contribution, encoded by the parton density functions (PDFs) $f_{i/I}$, times the short-distance partonic cross section $d \hat{\sig}_{ab} (k_a,k_b)$. The boundary between the long- and the short-distance regimes is set by the factorisation scale $\mu_F$, leftover of the PDF-renormalisation procedure that allows to reabsorb initial-state collinear singularities, and whose dependence in the partonic cross section is compensated by that in the PDFs order by order in perturbation theory.

We are interested in the NLO prediction for the partonic cross section, differential with respect to a generic IRC-safe observable $X$. 
Henceforth, we will refer to $\hat{\sigma}_{ab}$ as $\sigma$ and we will focus on reactions that at Born level feature $n$ massless coloured partons (as well as an arbitrary number of massless or massive colourless particles), of which up to two in the initial state. 
Scattering amplitudes for such processes can be expanded in perturbation theory as
\beq
{\cal A}_n 
\, = \, 
{\cal A}_n^{(0)} 
\, + \, 
{\cal A}_n^{(1)} 
\, + \,
{\cal A}_n^{(2)} 
\, + \, 
\ldots 
\,\, ,
\label{pertexpA}
\eeq
where the superscripts denote the loop order.
The expressions of the Born, real emission, and ($\overline{\rm MS}$-renormalised) virtual contributions,
\beq
&& 
\hspace{2mm} 
\Bn 
\, = \, 
\left| {\cal A}_n^{(0)} \right|^2 
\, , 
\qquad 
\Rl 
\, = \, 
\left| {\cal A}_\npo^{(0)} \right|^2 
\, ,  
\qquad
\Vl 
\, = \, 
2 \, {\bf Re}
\left[ 
{\cal A}_n^{(0) *} \, {\cal A}_n^{(1)} 
\right] 
\, ,
\label{pertsigNLOparts}
\eeq
allow one to write the LO and NLO coefficients of the differential partonic cross section as
\beq
\frac{d \sig_\LO}{dX} 
& = & 
\int d \Phi_n \, \Bn \, \delta_{X_n} 
\, , 
\label{eq:LO_struct} 
\\
\frac{d \sig_\NLO - d \sig_\LO}{d X} 
& = & 
\int d \Phi_n \, \Vl \, \delta_{X_n} 
+ 
\int d \Phi_\npo \, \Rl \, \delta_{X_{n+1}} 
+ 
\int d \Phi_n^{x\hat x} \, C(x,\hat x) \, \delta_{X_n}
\, ,
\label{eq:NLO_struct}
\eeq
where $\delta_{X_i} \equiv \delta(X-X_i)$, $X_i$ standing for the observable computed with $i$-body kinematics, and $ d \Phi_j = d \Phi_j (k_a,k_b)$ is the $j$-body phase space, including suitable polarisation sums/averages and flux factors; the convolution phase space $d \Phi_n^{x\hat x}$, defined by
\beq
\int 
d \Phi_n^{x\hat x}
\, \equiv \,
\int_0^1 \! \frac{d x}{x} 
\int_0^1 \!\frac{d \hat x}{\hat x} 
\int d \Phi_n(x k_a, \hat x k_b)
\, ,
\eeq
shows a dependence on rescaled initial-state partonic momenta $x k_a$ and $\hat x k_b$, with $0 \leq x, \hat x \leq 1$. The PDF collinear counterterm $C(x,\hat x)$, encoding the full $\mu_F$ dependence of the partonic cross section, is defined in $\overline{\rm MS}$ as
\beq
C(x,\hat x)
& = &
\frac{\as}{2 \pi} \, \frac{1}{\eps} \,
\frac{(e^{\gamma_E})^{\eps}}{\Gamma(1-\eps)}
\left(\frac{\mu^2}{\mu_F^2}\right)^{\!\!\eps} 
\Big[ 
\bar P_a(x)\,\delta(1-\hat x) 
+
\bar P_b(\hat x)\,\delta(1-x)
\Big] 
\, 
\Bn(x k_a, \hat x k_b)  
\, ,
\label{eq:PDFct}
\eeq
where $\bar P_i(x)$ represent the lowest-order four-dimensional regularised Altarelli-Parisi splitting kernels (see Appendix \ref{app:AP} for their explicit expressions).

While the finiteness of the NLO correction in Eq.~(\ref{eq:NLO_struct}) is ensured by the  KLN theorem \cite{Kinoshita:1962ur,Lee:1964is} supplemented with PDF renormalisation, as well as by the  IRC-safety of $X$, the $n$-body and $(n+1)$-body contributions are separately divergent. In dimensional regularisation, where amplitudes are evaluated in $d=4-2\eps$ space-time dimensions, such divergences arise at NLO as double and single $1/\eps$ poles in the expression of $V$; correspondingly, the real contribution $R$, which is finite for $\eps\to0$, features IRC phase-space singularities which translate into double and single $1/\eps$ poles upon integration over the radiative phase space. 

The procedure of infrared subtraction allows to achieve the cancellation of such poles by adding and subtracting to Eq.~(\ref{eq:NLO_struct}) a counterterm cross section
\beq
\left. \frac{d \sig_\NLO}{d X} \right|_{\rm ct} 
& \equiv & 
\int d \Phi_{n+1} \,  K \, \delta_{X_n} 
\nnb\\
& \equiv & 
\int d {\Phi}_n \, I  \, \delta_{X_n} 
+ 
\int d \Phi_n^{x \hat x} \, J(x,\hat x) \, \delta_{X_n} 
\, .
\label{Ct}
\eeq
The counterterm $K$ is designed so as to reproduce point by point all phase-space singularities of the real contribution and, at the same time, to lend itself to a sufficiently simple analytical integration over the radiative phase space.
The outcome of this integration can be recast into the sum of an $(x, \hat x)$-independent contribution $I$ and an $(x, \hat x)$-dependent contribution $J$, which display the same $1/\eps$ pole content (with opposite signs) as $V$ and $C(x, \hat x)$, respectively.
At this point, the NLO correction to the partonic cross section can be rewritten as 
\beq
\frac{d \sig_\NLO - d \sig_\LO }{dX} 
& = & 
\int d \Phi_n \, \Big( \Vl +  I  \Big) \, \delta_{X_n} 
\nnb\\
&&
\hspace{-4mm}
+ \, 
\int d \Phi_n^{x \hat x} \, 
\Big( C(x,\hat x) + J(x,\hat x) \Big) \, \delta_{X_n} 
\nnb\\
&&
\hspace{-4mm}
+ \, 
\int d \Phi_\npo 
\Big(  
\Rl  \, \delta_{X_{n+1}} 
- 
K \, \delta_{X_n} 
\Big) 
\, ,
\label{sigNLOsub}
\eeq
where each line is separately finite in $d = 4$ dimensions, and free from phase-space divergences, thus suitable for numerical integration.

We stress that in case of lepton-hadron collisions, the above discussion carries over identically, up to the formal replacements
\beq
\int 
d \Phi_n^{x\hat x}
& \to &
\int 
d \Phi_n^{x}
\, \equiv \,
\int_0^1 \! \frac{d x}{x} 
\int d \Phi_n(x k_a)
\, ,
\nnb\\
C(x,\hat x)
& \to &
C(x)
\, \equiv \,
\frac{\as}{2 \pi} \, \frac{1}{\eps} \,
\frac{(e^{\gamma_E})^{\eps}}{\Gamma(1-\eps)}
\left(\frac{\mu^2}{\mu_F^2}\right)^{\!\!\eps} 
\bar P_a(x)
\, 
\Bn(x k_a)  
\, ,
\label{eq:PDFct2}
\eeq
which in turn require defining a single-argument counterterm $J(x)$ instead of $J(x,\hat x)$. For lepton-lepton collisions, as well, one just sets the second line of 
\eq{sigNLOsub} to zero.


\subsection{Sector functions}
\label{NLOsecfun}
The specification of the subtraction counterterm $K$ completely defines a subtraction scheme. Local analytic sector subtraction is based on the well known idea \cite{Frixione:1995ms,Frixione:1997np} of dividing the radiative phase space into regions, each of which associated with the IRC singularities stemming from an identified set of partons (two at NLO). This can be achieved through the introduction of a unitary phase-space partition
\beq
\label{eq:sf_rel1}
\sum_i  \sum_{j \neq i} \mc W_{i j} 
& = & 
1 
\, , 
\eeq
by means of kinematic \emph{sector functions} $\mc W_{ij}$ with the following properties: 
\beq
\label{eq:sf_rel2}
\Si  \, \mc W_{ab} 
& = & 
0
\, ,
\quad \quad \quad
\forall \, i \neq a
\, ,
\\
\label{eq:sf_rel3}
\Cj  \, \mc W_{ab} 
& = &
0  
\, ,
\quad \quad \quad
\forall \, ab \not\in \{ij, \, ji\}
\, ,
\eeq
\beq
\label{eq:sf_rel4}
&&
\Si \,
\sum_{\substack{ k \neq i}}  \mc W_{ik} 
\, = \, 
\final{i}
\, ,
\\
\label{eq:sf_rel5}
&&
\Cj  \,
\big(
\mc W_{ij} 
+
\mc W_{ji} 
\big)
\, = \, 
1-\initial{i} \, \initial{j}
\, ,
\\
\label{eq:sf_rel6}
&&
\Si \, \Cj  \,
\mc W_{ij} 
\, = \, 
\final{i}
\, .
\eeq
$\Si$ and $\Cj$ are projection operators that select the leading behaviour of functions in the limit in which parton $i$ becomes soft (i.e.~its energy vanishes), and partons $i$ and $j$ become collinear (i.e.~their relative angle vanishes), respectively; 
the symbol $\theta_{\cal C}$ is $1$ or $0$ if condition $\cal C$ is or is not fulfilled, so that $\final{a}$ ($\initial{a}$) enforces parton $a$ to belong to the final (initial) state. 
\eq{eq:sf_rel2} and \eq{eq:sf_rel3} identify the singular pair in a given sector; the properties in Eqs.~(\ref{eq:sf_rel4} - \ref{eq:sf_rel6}), which we dub \emph{sum rules}, express that the sum over all sectors that share a given soft or collinear singularity reduces to unity in that singular limit (in all physically meaningful cases): this allows to eliminate sector functions upon suitable combination of particle labels, which will prove crucial in view of analytic counterterm integration, as detailed below.

The actual definition of sector functions $\mc W_{ij}$ is largely arbitrary, provided it satisfies the defining relations (\ref{eq:sf_rel1} - \ref{eq:sf_rel6}). In terms of the partonic centre-of-mass (CM) four-momentum $\q^\mu = (\sqrt s,\vec 0\,)$ and of parton momenta $k_i^\mu$, we start defining dot products
\beq
s_{\q i} 
& = & 
2 \, \q \cdot k_i 
\, ,
\qquad 
s_{ij} 
\, = \, 
(k_i + k_j)^2 
\, = \, 
2 \, k_i \cdot k_j 
\, , 
\label{eq:sEwdef}
\eeq
and dimensionless invariants associated with the energy of parton $i$ and the angle $\theta_{ij}$ between $i$ and $j$ in the CM frame, namely
\beq
e_i 
& = & 
\frac{s_{\q i}}{s} 
\, , 
\qquad \quad 
\hspace{1pt}  
w_{ij} 
\, = \, 
\frac{s \, s_{ij}}{s_{\q i} \, s_{\q j}} 
\, = \, 
\frac{1-\cos \theta_{ij}}{2}
\, .
\label{eq:Ewdef}
\eeq 
Our choice of NLO sector functions is then
\beq
\label{eq:sfunNLO}
\mc W_{ij} 
\, = \, 
\frac{\sig_{ij}}{\sig}  
\, , 
\qquad \qquad
\sig_{ij} 
\, = \, 
\frac{\final{i}}{e_i \, w_{ij}} 
\, = \, 
\final{i} \,
\frac{s_{qj}}{s_{ij}} 
\, ,
\qquad \qquad
\sig 
\, = \, 
\sum_k \sum_{l \neq k} \, \sig_{kl} 
\, ,
\eeq
where the sums run over all massless initial- and final-state QCD particles. In particular, the action of the soft and collinear projection operators on sector functions is 
\beq
\mc W_{\textrm{s}, ij}
& \equiv &
\Si \, \mc W_{ij} 
\, = \,
\final{i}
\,
\frac{1/w_{ij}}{\sum\limits_{l \neq i} 1 / w_{il}}
\, ,
\\
\mc W_{\textrm{c}, ij}
& \equiv &
\Cj \, \mc W_{ij} 
\, = \,
\final{i}
\Big(
\final{j}
\frac{e_j}{e_i+e_j}
+
\initial{j}
\Big)
\, ,
\\
\mc W_{\textrm{sc}, ij}
& \equiv &
\Si \, \Cj \, \mc W_{ij}  
\, = \,
\final{i}
\, .
\label{eq:Wij_lim2}
\eeq

\subsection{Definition of candidate local counterterms}
\label{sec:localct}
After partitioning the radiative phase space, a \emph{candidate} local counterterm $\hat K$ is obtained considering one partition $\W{ij}$ at a time, and collecting the action of all singular projectors $\bS{i}$ and $\bC{ij}$ on $R \, \W{ij}$:
\beq
\label{eq:NLOK}
\hat K_{ij}
& \equiv &
\Big[
\Si 
+
\Cj 
-
\Si \, \Cj 
\Big]
\, R \,  \mc W_{ij} 
\, ,
\qquad \qquad
\hat K
\, \equiv \,
\sum_i  \sum_{j \neq i} 
\hat K_{i j}
\, ,
\\
R - \hat K
& \equiv &
\sum_i  \sum_{j \neq i} 
\Big[
R \, \mc W_{i j} 
- 
\hat K_{i j} 
\Big]
\, = \,
\sum_i  \sum_{j \neq i} 
\big( 1 - \bS{i} \big)
\big( 1 - \bC{ij} \big)
R \, \mc W_{i j}
\, = \,
\rm{finite} 
\, ,
\eeq
where the $- \, \bS{i}\,\bC{ij}$ term in brackets removes the double-counting of soft-collinear configurations introduced by the incoherent sum of soft and collinear limits.
The action of soft and collinear limits on $R$ gives rise to the universal singular kernels described below.

\subsubsection{Soft limit} 
The soft limit $\bS{i}$ on the real matrix element squared can be written as
\beq
\Si \, \Rl 
& = &
 - \, \Norm
\sum_{k \neq i}
\sum_{l \neq i} \,
\mc I_{k l }^{(i)} \,
\Bn_{k l} \! \left( \{ k \}_{\slashed i} \right) 
\, ,
\label{eq:SiR} 
\eeq
where the eikonal kernel
\beq
\mc I_{k l }^{(i)}
\, = \, 
\final{i} \, \delta_{f_i g} \, \frac{s_{k l }}{s_{i k} \, s_{i l}}
\label{eq:eikkern}
\eeq
is non-vanishing only if final-state parton $i$, with flavour $f_i$, is a gluon. The soft kinematics $\{k\}_\slashed{i}$ is the set of real-radiation momenta after removal of soft momentum $k_i$. 
The colour-correlated Born matrix element is defined schematically as
\beq
\Bn_{kl} \, = \, {\cal A}_n^{(0)*} \, ({\bf T}_k \cdot {\bf T}_l) \, {\cal A}_n^{(0)} \, ,
\eeq
where ${\cal A}_n$ is understood as a ket in colour space \cite{Catani:1996vz} transforming non-trivially under the action of the SU($N_c$) generators ${\bf T}_a$.
Finally, the coefficient $\Norm$ is defined as
\beq
\Norm 
\, = \, 
8 \pi \as \left( \frac{\mu^2 e^{\gam_E}}{4 \pi} \right)^{\eps} \, .
\label{norm1}
\eeq

\subsubsection{Collinear limit} 
\label{subsec:Cij}
To describe the collinear limit $\bC{ij}$ in case both $i$ and $j$ are outgoing, i.e. for the splitting $[ij]\to i + j$, we introduce a Sudakov parametrisation
\beq
k_i^\mu 
& = & 
z_i \, \bar{k}_{[ij]}^\mu 
+ 
\widetilde{k}_{\rm F} ^\mu 
- 
\frac{1}{z_i} \frac{\widetilde{k}_{\rm F} ^2}{s_{[ij]r}} k_r ^\mu 
\, ,
\nnb\\
k_j^\mu 
& = & 
z_j \, \bar{k}_{[ij]}^\mu 
- 
\widetilde{k}_{\rm F} ^\mu 
- 
\frac{1}{z_j} \frac{\widetilde{k}_{\rm F} ^2}{s_{[ij]r}} k_r ^\mu 
\, ,
\nnb\\
k_{[ij]}^\mu
& \equiv &
k_i ^{\mu} 
+ 
k_j ^{\mu} 
\, ,
\qquad \qquad
s_{[ij]r}
\, \equiv \,
s_{ir} + s_{jr}
\, , 
\nnb\\
\bar{k}_{[ij]}^\mu 
& = & 
k_{[ij]}^\mu 
- 
\frac{s_{ij}}{s_{[ij]r}} \ k_r^{\mu} \, 
\quad 
(r \neq i, \, j) 
\, ,
\label{colldir}
\eeq
where massless vector $\bar{k}_{[ij]}^\mu$ defines the collinear direction, while $k_r^{\mu}$ is a light-like reference vector chosen from the set of on-shell momenta $\{k\} = \{k_1, \cdots, k_{n+1}\}$;
$z_i$ and $\widetilde{k}_{\rm F}^{\mu}$ are the longitudinal momentum fraction and the transverse momentum of parton $i$ with respect to the collinear direction, respectively,
\beq
z_i 
\, = \, 
\frac{s_{ir}}{s_{[ij]r}}
\, ,
\qquad \qquad
\widetilde{k}_{\rm F} ^{\mu}
\, = \,
k_i ^{\mu} 
- 
z_i \, k_{[ij]}^{\mu} 
-  
(1-2z_i)
\frac{s_{ij}}{s_{[ij]r}}  k_r ^{\mu}
\, ,
\eeq
satisfying $z_i + z_j = 1$, $\widetilde{k}_{\rm F} \cdot \bar{k}_{[ij]} =  \widetilde{k}_{\rm F} \cdot k_r = 0$.
On the other hand, when a final-state parton $i$ is collinear to an incoming parton $j$, relevant to the $j\to [ij] + i$ splitting, momentum $k_i^\mu$ is parametrised in terms of its transverse momentum $\widetilde{k}_{\rm I} ^\mu$ and longitudinal momentum fraction $x_{i}$ as
\beq
k_i^\mu 
\, = \, 
x_{i} \ k_j^\mu 
+ 
\widetilde{k}_{\rm I} ^\mu 
- 
\frac{1}{x_i} \frac{\widetilde{k}_{\rm I} ^2}{s_{jr}} k_r ^\mu 
\, ,
\label{remapped_ki}
\eeq
where
\beq
x_{i} 
\, = \, 
\frac{s_{ir}}{s_{jr}} 
\, ,
\qquad \qquad
\widetilde{k}_{\rm I}^\mu 
\, = \, 
k_i ^{\mu} 
-
x_i \, k_j ^{\mu}
-
\frac{s_{ij}}{s_{jr}} \, k_r ^{\mu}
\, ,
\eeq
satisfying $x_{[ij]} + x_i  =  1$, $\widetilde{k}_{\rm I}\cdot  k_r  = 
\widetilde{k}_{\rm I}\cdot  k_j = 0$.
The collinear direction in this case is identified as
\beq
\bar{k}_{[ij]}^\mu  
\, = \, 
x_{[ij]} \, k_j^\mu 
-
\widetilde{k}_{\rm  I}^\mu 
- 
\frac{1}{x_{[ij]}} \frac{\widetilde{k}_{\rm I}^2}{s_{jr}} k_r^\mu 
\, .
\eeq
The universal (un-regularised, $d$-dimensional)  Altarelli-Parisi splitting kernels \cite{Gribov:1972ri,Altarelli:1977zs,Dokshitzer:1977sg} encoding the collinear behaviour of $R$ are matrices in spin space and can be compactly written as
\beq
P_{ab,	\star}^{\mu \nu}(\xi)
& = &
P_{ab}(\xi) \,
\big(\!-\!g^{\mu\nu} \big)
+
Q_{ab,\star}(\xi) 
\left[  
- g^{\mu\nu} 
+ 
(d - 2) \, 
\frac{\widetilde{k}_{\star}^\mu \, \widetilde{k}_{\star}^\nu}{\widetilde{k}_{\star}^2}
\right] 
\label{eq:APkernels_munu}
\, ,
\eeq
where $\xi$ is the longitudinal momentum fraction of splitting parton $a$, and the dependence on $\star = \rm{I,F}$ will be specified shortly. In a flavour-symmetric notation, the spin-averaged components $P_{ab}(\xi)$ read
\beq
\label{eq:APkernels1}
P_{ab}(\xi)
& = &
\delta_{f_a g} \delta_{f_b g} \,
2 \, C_A 
\left[ \frac{\xi}{1-\xi} + \frac{1-\xi}{\xi} + \xi  (1-\xi) \right]
+
\delta_{ \{f_a f_b\} \{q \bar q\} } \, T_R
\left[ 1 - \frac{2 \, \xi (1-\xi)}{1 - \eps} \right]
\\
& + &
\delta_{f_a \{q ,\bar q\}} \delta_{f_b g} \, C_F 
\left[ 2 \frac{\xi}{1-\xi} + (1 - \eps) (1-\xi) \right] 
+
\delta_{f_a g} \delta_{f_b \{q , \bar q\}} \, C_F
\left[ 2 \frac{1 - \xi}{\xi} + (1 - \eps) \, \xi \right]
\, ,
\nnb
\eeq
where we defined flavour delta functions as $\delta_{f_a \{q , \bar q\}} \equiv \delta_{f_a q} + \delta_{f_a \bar q}$ and $\delta_{ \{f_a f_b\} \{q \bar q\} } \equiv \delta_{f_a q} \, \delta_{f_b \bar q} + \delta_{f_a \bar q} \, \delta_{f_b q}$. By QCD helicity conservation, the collinear azimuthal kernels $Q_{ab,\star}(\xi)$ are non-vanishing only when the virtual parton involved in the splitting is a gluon: the expression for $Q_{ab,\star}(\xi)$ thus depends on whether the virtual gluon is the splitting ancestor ($\star=\rm F$),
\beq
Q_{ab, \rm F} (\xi)  
& = & 
-  \, 
\delta_{f_a g} \delta_{f_b g}\, 
2 \, C_A \, \xi (1-\xi) 
+ 
\delta_{ \{f_a f_b\} \{q \bar q\} }  \, 
T_R \,  
\frac{2 \, \xi (1-\xi)}{1 - \eps} 
\ , 
\label{eq:APkernels2a}
\eeq
or is one of the splitting siblings ($\star=\rm I$)
\beq
Q_{ab, \rm I} (\xi) 
& = & 
-  \, 
\delta_{f_a g} \delta_{f_b g} \, 
2 \, C_A \, 
\frac{1-\xi}{\xi} 
-
\delta_{f_a g} \delta_{f_b \{q , \bar q\}} \, 
2 \, C_F \,  
\frac{1- \xi}{\xi}
\, ,
\label{eq:APkernels2b}
\eeq
where the notation is reminiscent of the fact that at NLO the two cases apply to final- and initial-state splittings, respectively.

\begin{figure}[h]
\centering
\includegraphics[width=0.6\linewidth]{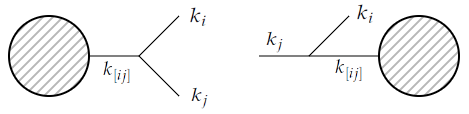}
\caption{
Final-state (left) and initial-state (right) splittings.
}
\label{collinearsplitting}
\end{figure}

In terms of such kernels, the collinear $\Cj$ limit of the real matrix element can be finally written as
\beq
\Cj \, \Rl  
& = & 
\frac{\Norm}{s_{ij}} \, 
\Bigg[
\final{i} \, \final{j} \,
P_{ij, \rm F}^{\mu \nu}(z_i) \, \Bn_{\mu \nu} \! \left( \{ k \}_{\slashed i \slashed j}, k_{[ij]} \right)
\nnb\\
&&
\qquad
+ \,
\final{i} \, \initial{j} \,
\frac{P_{[ij]i, \rm I}^{\mu \nu}(x_{[ij]})}
{x_{[ij]}}
\, \Bn_{\mu \nu} \! \left( \{ k \}_{\slashed i \slashed j}, x_{[ij]} k_{j} \right) 
\nnb\\
&&
\qquad
+ \,
\final{j} \, \initial{i} \,
\frac{P_{[ji]j, \rm I}^{\mu \nu}(x_{[ji]})}
{x_{[ji]}}
\, \Bn_{\mu \nu} \! \left( \{ k \}_{\slashed i \slashed j}, x_{[ji]} k_{i} \right) 
\Bigg]
\, ,
\label{eq:CijR1} 
\eeq
where $B_{\mu\nu}$ is the spin-correlated Born amplitude, while $( \{ k \}_{\slashed a \slashed b}, k_{c})$ is the radiative kinematics with $k_a$ and $k_b$ removed and replaced by $k_c$. The first two lines of \eq{eq:CijR1} are pictorially represented in the left and right panels of Figure \ref{collinearsplitting}, respectively, while the third line is obtained from the second upon $i \lra j$ exchange.

\subsubsection{Soft-collinear limit}
In the soft-collinear $\Si\,\Cj$ limit, final-state gluon $i$ becomes both soft and collinear to initial- or final-state parton $j$.
The corresponding kernel is
\beq
\Si \, \Cj \, \Rl  
\, = \, 
\Cj \, \Si \, \Rl 
\, = \,  
\mathcal{N}_1 \, 
2 \, C_{f_j} \, \mc I_{jr}^{(i)}  \, B(\{ k\}_{\slashed{i}})
\, ,
\label{eq:SiCijR}
\eeq
where $C_{f_j} = C_A \, \delta_{f_j g} + C_F \, \delta_{f_j \{q , \bar q\}}$ is the SU($N_c$) Casimir operator associated to flavour $f_j$, and $f_j = f_{[ij]}$ since $i$ is a gluon.

For later convenience, we define hard-collinear kernels upon subtracting from the collinear Altarelli-Parisi  kernels in  \eq{eq:APkernels1} their respective soft limits: for a final-state splitting, both collinear siblings $i$ and $j$ can give rise to a soft singularity, thus
\beq
P^{\hc}_{ij,\rm F} (z_i) 
& \equiv & 
\big(
1 -  \, \Si - \, \Sj
\big) \, 
P_{ij} (z_i) 
\nnb \\
& = &
\delta_{f_i g} \delta_{f_j g} \, 2  \, C_A \, z_i  \, z_j 
+  
\delta_{\{f_i f_j \}\{q \bar q\}} \, T_R  
\left( 1 - \frac{2 \, z_i   z_j}{1 - \epsilon} \right)
\nnb \\
&&
+ \,
\delta_{f_i \{q , \bar q\}} \delta_{f_j  g} \, C_F (1-\epsilon) \, z_j 
+ 
\delta_{f_i g} \delta_{f_j  \{q , \bar q\}} \, C_F (1-\epsilon) \, z_i 
\, ,
\eeq
while for an initial-state splitting just one of the two, $i$, can be soft, so
\beq
\label{eq:hcPI}
P_{[ij] i,\rm I}^{ \hc}(x_{[ij]})
& \equiv & 
x_{[ij]}
\big(
1 - \, \Si
\big) \,
\frac{P_{[ij] i}(x_{[ij]})}{x_{[ij]}} 
\\
& = &
\delta_{f_{[ij]} g} \delta_{f_i g} \,  2 C_A 
\biggl[ \frac{x_i}{x_{[ij]}} + x_{[ij]} x_i \biggr]  
+ 
\delta_{\{f_{[ij]} f_i \}\{q \bar q\}} \, T_R 
\biggl[1 - \frac{2 \ x_{[ij]} x_i}{1-\epsilon} \biggr] 
\nnb \\
&&
+ \, 
\delta_{f_{[ij]}\{q, \bar q\}} \delta_{f_i  g} \, C_F  (1- \epsilon) \, x_i  
+ 
\delta_{f_{[ij]} g} \delta_{f_i  \{q, \bar q\}} \, C_F 
\biggl[2\frac{x_i}{x_{[ij]}} +(1-\epsilon) \, x_{[ij]} \biggr] 
\, .
\nnb
\eeq
In analogy with \eq{eq:APkernels_munu}, we introduce
\beq
P_{ab, \star}^{\mu \nu, \hc}(\xi)
& = &
P_{ab, \star}^{\hc}(\xi) \,
\big(\!-\!g^{\mu\nu} \big)
+
Q_{ab,\star}(\xi) 
\left[  
- g^{\mu\nu} 
+ 
(d - 2) \, 
\frac{\widetilde{k}_{\star}^\mu \, \widetilde{k}_{\star}^\nu}{\widetilde{k}_{\star}^2}
\right] 
\label{eq:APkernels_munu_hc}
\, .
\eeq

\subsection{Phase-space mappings and counterterm definitions}
\label{sec:localctdef}
Although the \emph{candidate} counterterm locally reproduces all real phase-space singularities, it embodies Born matrix elements that are evaluated with kinematics that either do not satisfy $n$-body momentum conservation (in the soft case, $\{k\}_{\slashed{i}}$), or feature an off-shell leg (in the collinear case, ($\{k\}_{\slashed{a}  \slashed{b}}, k_{c}$)).
Conversely, it is desirable that the Born matrix elements appearing in counterterms have a physical (i.e.~on-shell and momentum conserving) $n$-body kinematics for all choices of the $\npo$ radiative momenta, and not only for specific singular configurations, whence a kinematic mapping of momenta is required. In turn, such a mapping operation entails a factorisation of the $(\npo)$-body phase space $d\Phi_\npo$ into a remapped $n$-body phase space $d\Phi_n$ times a single-radiative measure $d\Phi_{\rm rad}$, which allows the analytic integration of the radiative degrees of freedom at fixed underlying Born kinematics.

\begin{figure}[h]
\centering
\includegraphics[width=0.8\linewidth]{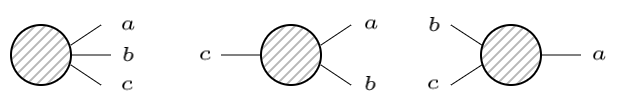}
\caption{
Final-final (left), final-initial (middle), and initial-initial (right) dipoles. 
}
\label{remapping_setting}
\end{figure}

A convenient way of achieving phase-space factorisation is through Catani-Seymour dipole mappings \cite{Catani:1996vz}, in which a triplet of massless momenta $k_a$, $k_b$, and $k_c$ (the emitted, emitter, and recoiler parton, respectively) are mapped onto a dipole of Born-level momenta $\bar k_b^{(abc)}$ and $\bar k_c^{(abc)}$, according to
\beq
&&
\final{a} \, \final{b} \, \final{c}:
\qquad
\bar{k}_{b}^{(abc)} 
+ 
\bar{k}_c ^{(abc)} 
\, = \,
k_a + k_b + k_c
\, ,
\nnb\\ [3pt]
&&
\final{a} \, \final{b} \, \initial{c}:
\qquad
\bar{k}_{b}^{(abc)} 
-
\bar{k}_c ^{(abc)} 
\, = \,
k_a + k_b - k_c
\, ,
\nnb\\ [3pt]
&&
\final{a} \, \initial{b} \, \initial{c}:
\qquad
\sum_{\substack{i \in \rm F \\ i \neq a}} \, \bar{k}_i ^{(abc)}
-
\bar{k}_{b}^{(abc)} 
-
\bar{k}_c ^{(abc)} 
\, = \,
\sum_{\substack{i \in \rm F \\ i \neq a}} \, k_i
+
k_a
-
k_b
-
k_c
\, .
\eeq
The three assignments are represented pictorially in Figure \ref{remapping_setting}; details on the mappings, parametrisations and corresponding phase-space factorisation are given in Appendix~\ref{app:mappings}.

There is ample freedom in the choice of mapping dipoles, as long as this is compatible with the locality of subtraction: in particular, the choice can be adapted to the identity of the partons involved in the different singular kernels. In the soft limit, each eikonal kernel $\mc I_{kl}^{(i)}$ leads naturally to the choice $(abc) = (ikl)$ or $(abc) = (ilk)$, while in the collinear limits the most natural mapping involves the splitting partons and the recoiler, $(abc) = (ijr)$ or $(abc) = (irj)$.
Denoting mapped limits with a bar, we thus define the soft counterterm to be
\beq
\label{eq:SiRdef}
\bSi \, \Rl 
& = &
- \, 2 \, \Norm
\sum_{k \neq i}
\sum_{\substack{l \neq i \\ l < k }}
\mc I_{kl}^{(i)} 
\bigg[
\big(
\initial{k} \, \initial{l}
+
\final{k} \, \initial{l}
+
\final{k} \, \final{l}
\big) \,
\bar{\Bn}^{(ikl)}_{kl} 
+
\initial{k} \, 
\final{l} \,
\bar{\Bn}^{(ilk)}_{kl} 
\bigg] 
\, ,
\qquad
\eeq
where $\bar{\Bn}_{\dots}^{(abc)} \equiv \Bn_{\dots}( \{\bar{k}\}^{(abc)})$.
As for collinear and soft-collinear kernels, we define
\beq
\label{eq:CijRdef}
\bCj \, \Rl
& = &
\frac{\Norm}{s_{ij}} \,
\Bigg[
\final{i} \, \final{j} \,
P_{ij, \rm F}^{\mu \nu} (z) \, \bar{\Bn}^{(ijr)}_{\mu \nu}
\nnb\\
&&
\qquad
+ \,
\final{i} \, \initial{j} \,
\frac{P_{[ij]i, \rm I}^{\mu \nu} (x)}x \,
\Big(
\final{r} \, \bar{\Bn}^{(irj)}_{\mu \nu}
+
\initial{r} \, \bar{\Bn}^{(ijr)}_{\mu \nu}
\Big)
\nnb\\
&&
\qquad
+ \,
\final{j} \, \initial{i} \,
\frac{P_{[ji]j, \rm I}^{\mu \nu} (x)}x \,
\Big(
\final{r} \, \bar{\Bn}^{(jri)}_{\mu \nu}
+
\initial{r} \, \bar{\Bn}^{(jir)}_{\mu \nu}
\Big)
\Bigg]
\, , ~
\\
\label{eq:SiCijRdef}
\bSi \, \bCj \, \Rl
& = &
\Norm \, 2 \, C_{f_j} \, \mc I_{jr}^{(i)}
\Bigg[
\final{j} \, \bar{\Bn}^{(ijr)} + \initial{j} \,
\Big(
\frac{\final{r}}{1-z} \, \bar{\Bn}^{(irj)} + \initial{r} \, (1-v) \,
\bar{\Bn}^{(ijr)}
\Big)
\Bigg] \, ,
\eeq
where the soft-collinear
contributions for $j\in\rm I$ feature kinematical factors, written in
terms of variables $z$ and $v$ defined in Appendix
\ref{app:mappings},\footnote{We stress that the definitions of $z$,
  $x$, and $v$ in the previous equations are mapping-dependent: for
  instance, one should correctly interpret the notation
  $f(x) ( \final{r} \, \bar{\Bn}^{(irj)}_{\mu \nu} + \initial{r} \,
  \bar{\Bn}^{(ijr)}_{\mu \nu})$ to mean
  $\final{r} \, f(x^{(irj)}) \, \bar{\Bn}^{(irj)}_{\mu \nu} +
  \initial{r} \, f(x^{(ijr)}) \, \bar{\Bn}^{(ijr)}_{\mu \nu}$, and
  similarly for the other terms.}  whose purpose is to reconstruct the
hard-collinear kernels of \eq{eq:hcPI}:
\beq
\label{eq:HCijRdef}
\bbHC{ij} \, \Rl
& \equiv &
(1-\bSi-\bSj) \, \bCj \, \Rl
\nnb\\
& = &
\frac{\Norm}{s_{ij}} \,
\Bigg[
\final{i} \, \final{j} \, P_{ij, \rm F}^{\mu \nu, \hc} (z) \,
\bar{\Bn}^{(ijr)}_{\mu \nu}
\nnb\\[-2mm]
&&
\hspace{6mm}
+ \,
\final{i} \, \initial{j} \, \frac{P_{[ij]i, \rm I}^{\mu \nu, \hc} (x)}x \,
\Big(
\final{r} \,
\bar{\Bn}^{(irj)}_{\mu \nu} +
\initial{r} \, \bar{\Bn}^{(ijr)}_{\mu\nu}
\Big)
\nnb\\
&&
\hspace{6mm}
+ \,
\final{j} \, \initial{i} \, \frac{P_{[ji]j, \rm I}^{\mu \nu, \hc} (x)}x \,
\Big(
\final{r} \,
\bar{\Bn}^{(jri)}_{\mu \nu} +
\initial{r} \,
\bar{\Bn}^{(jir)}_{\mu \nu}
\Big)
\Bigg]
\, ,
\eeq
where we have defined
$\bSj \, \bCj \equiv \bSj \, \bbC{ji}$.  It can be checked (see
Appendix~\ref{app:consistency}) that the definitions in
Eqs.~(\ref{eq:SiRdef} - \ref{eq:SiCijRdef}) satisfy the following set
of \emph{consistency relations}
\begin{align}
\label{limits}
\Si \, \bSi \, R  \, =& \, \, \Si \, R \, , 
& 
\Cj \, \bCj \, R \,  =& \, \, \Cj \, R \, ,
\nnb\\
\Si \, \bSi \, \bCj \ R \, =& \, \, \Si \, \bCj \, R \, ,
& 
\Cj \, \bSi \, \bCj \, R  \, =& \, \, \Cj \, \bSi \, R \, ,
\end{align}
ensuring that the application of the mappings detailed above preserves
the local cancellation of singularities.  This leads to defining the
sought local counterterm $K$ as
\beq
\label{eq:NLOKW}
K & \equiv & \sum_i \sum_{j \neq i} K_{i j} \, , \qquad K_{ij} \;
\equiv \; \Big[ \bSi + \bCj - \bSi \, \bCj \Big] \, R \, \mc W_{ij} \,
,
\eeq
where, introducing the collective notation $\bbL{}{} = \bbS{i}, \, \bbC{ij}, \, \bbS{i}\, \bbC{ij}$, we have defined $\bbL{} \, R \, \mc W_{ij} \equiv (\bbL{}\,R) \, (\bbL{}\,\mc W_{ij})$.
The
definition in \eq{eq:NLOKW} is thus complete only after specifying
the action $\bbL{}\,\mc W_{ij}$ of the barred limits on sector
functions.  The simplest choice is
$\bbL{} \, \mc W_{ab} \equiv \bL{} \, \mc W_{ab}$, resulting in
\beq
R - K & = & \sum_i \sum_{j \neq i} \Big[ R \, \mc W_{ij} - K_{ij}
\Big] \, = \, \rm{finite} \, .
\eeq
With this choice, however, the quantity $K_{ij}$ defined in
\eq{eq:NLOKW} features \emph{spurious} singularities in the collinear
$\bC{ir}$ and $\bC{jr}$ limits, which are not present in
$R \, \W{ij}$.
Such singularities, generated by the denominators of
the Altarelli-Parisi kernels, drop out only in the sum of mirror
sectors $\W{ij}+\W{ji}$, ensuring the finiteness of $R-K$.  One could
envisage removing spurious singularities at the level of the single
$\W{ij}$ partition, which entails redefining the action of the barred
limits on sector functions. For instance, it is straightforward to
check that by choosing
\beq \overline{\mc W}_{\textrm{s}, ij} & \equiv
& \bSi \, \mc W_{ij} \, \equiv \, \final{i} \,
\frac{1/w_{ij}}{\sum\limits_{l \neq i} 1 / w_{il}} \, ,
\nnb\\
\overline{\mc W}_{\textrm{c}, ij} & \equiv & \bCj \, \mc W_{ij} \,
\equiv \, \final{i} \left( \final{j} \frac{e_j \, w_{jr}}{e_i \,
    w_{ir}+e_j \, w_{jr}} + \initial{j} \right) \, ,
\nnb\\[2mm]
\overline{\mc W}_{\textrm{sc}, ij} & \equiv & \bSi \, \bCj \, \mc
W_{ij} \, \equiv \, \final{i} \, ,
\label{eq:Wij_lim}
\eeq one achieves \emph{for all $ij$ pairs} \beq R \, \mc W_{ij} -
K_{i j} \, = \, \big( 1 - \bSi \big) \big( 1 - \bCj \big) \, R \, \mc
W_{ij} \, = \, \rm{finite} \, . 
\eeq
Another possibility, which preserves the flexibility of the numerical
implementation of the method reducing the number of sectors, is to
introduce symmetrised sector functions in the first place
\beq
\mc Z_{ij}
& \equiv &
\mc W_{ij} + \mc W_{ji}
\, ,
\eeq
along with
\beq
\Z{{\rm s},\,ij}
& \equiv &
\bbS{i} \, \Z{ij}
\; \equiv \;
\bbS{i} \, \W{ij}
\, = \,
\final{i} \,
\frac{1/w_{ij}}{\sum\limits_{l \neq i} 1 / w_{il}}
\, ,
\qquad
\Z{{\rm s},\,ji}
\; \equiv \;
\bbS{j} \, \Z{ij}
\, = \,
\final{j} \,
\frac{1/w_{ij}}{\sum\limits_{l \neq j} 1 / w_{jl}}
\, ,
\nnb\\
\Z{{\rm c},\,ij}
& \equiv &
\bbC{ij} \, \Z{ij}
\; \equiv \;
\bbC{ij} \, \W{ij} + \bbC{ij} \, \W{ji}
\, = \,
1 - \initial{i} \, \initial{j}
\, ,
\nnb\\[2mm]
\Z{{\rm sc},\,ij}
& \equiv &
\bbS{i} \, \bbC{ij} \, \Z{ij}
\; \equiv \;
\bbS{i} \, \bbC{ij} \, \W{ij}
\, = \,
\final{i} \, ,
\qquad \Z{{\rm sc},\,ji}
\; \equiv \;
\bbS{j} \, \bbC{ij} \, \Z{ij}
\, = \,
\final{j}
\, .
\eeq
We can then define the symmetrised counterterms
\beq
\label{eq:NLOKZ}
K_{\{ij\}}
& \equiv &
K_{ij} + K_{ji}
\, = \,
\Big[
\bbS{i} + \bbS{j}
+ \bbC{ij} \big( 1 - \bbS{i} - \bbS{j} \big)
\Big] \, R \, \mc Z_{ij}
\nnb\\[2mm]
& = &
\big(
\bbS{i} \, R \big) \mc Z_{{\rm s},ij} + \big( \bbS{j} \, R
\big) \mc Z_{{\rm s},ji} +
\bbHC{ij} \, R
\, ,
\nnb\\[2mm]
K
& = &
\sum_i \sum_{j < i} K_{\{ij\}}
\, ,
\eeq
satisfying
\beq
R \, \mc Z_{ij} - K_{\{ij\}}
& = &
\Big[
\big( 1 - \bbC{ij} \big)
\big( 1 - \bbS{i} - \bbS{j} \big)
\Big] \, R \, \mc Z_{ij}
\, = \,
\rm{finite}
\, ,
\nnb\\[2mm]
R - K
& = &
\sum_i \sum_{j < i}
\Big[
R \, \mc Z_{ij} - K_{\{ij\}}
\Big]
\, = \,
\rm{finite}
\, .
\eeq
We stress that, in any case, the
sum rules in (\ref{eq:sf_rel4} - \ref{eq:sf_rel6}) allow to get rid of
sector functions and to write the local counterterm purely as a
collection of universal soft and collinear NLO kernels:
\beq
\label{eq:NLOK}
K
& = &
\sum_i 
\bigg[
\bbS{i} \, R
+
\sum_{j < i} \,
\bbHC{ij} \, R
\bigg]
\, .
\eeq


\subsection{Local counterterms with damping factors}
\label{sec:localctdamping}
At this stage, we have all the ingredients to build a counterterm $K$
which, along with its integration over the radiative phase space,
achieves a local subtraction. Nonetheless, it is worth investigating a
systematic optimisation of the above definitions, with a view to
improving the efficiency of the method.  Since the subtraction
procedure is necessary only in the IRC corners of the phase space, one
is allowed to tune the counterterm contribution in the non-singular regions, thereby reducing numerical instabilities. This is customarily achieved
in the literature by introducing parameters (such as the $\alpha$
parameter in CS \cite{Nagy:2003tz}, and the $\delta$ and
$\xi_{\rm cut}$ parameters in FKS \cite{Frixione:1995ms}) that set a
hard boundary to the phase space allowed for counterterms. The
enhanced numerical stability of this procedure in general comes at the
price of a more cumbersome analytic counterterm integration, which may
become untenable at NNLO.

What we propose in this article is instead to multiply the local
counterterms in Eqs.~(\ref{eq:SiRdef} - \ref{eq:SiCijRdef}) by means
of smooth \emph{damping factors} (as opposed to hard step functions)
in order to gradually turn them off away from the singular regions.
Although one has some freedom in constructing such damping factors, provided the validity of Eqs.~(\ref{limits}) is not spoiled, it is particularly convenient to define them as powers, with tunable exponents, of the kinematic invariants proper of the chosen phase-space parametrisation. By doing so, one is essentially including in a controlled way subleading power terms in the normal variables through which the IRC kernels are already written. As a result, the presence of damping factors does not affect the complexity of the analytic integrations, which is crucial for exporting this optimisation to higher perturbative orders.
The explicit dependence of (the finite part of) the integrated counterterms upon the damping parameters, namely the above-mentioned tunable exponents, must cancel against an analogous dependence in the local counterterms, which is known to offer a powerful handle to check the numerical implementation of the subtraction method. 

We start by including damping factors in the soft counterterm, \eq{eq:SiRdef}:
\beq
\label{eq:dampedS}
\bSi \, \Rl 
& = &
- \, 2 \, \Norm
\sum_{k \neq i}
\sum_{\substack{l \neq i \\ l < k }}
\mc I_{kl}^{(i)} 
\bigg\{
\final{k} \,
(1-z)^\alpha \,
\Big[
\final{l} \, 
(1-y)^\alpha 
+
\initial{l} \, 
x^\alpha
\Big]
\bar{\Bn}^{(ikl)}_{kl} 
\nnb\\[-2mm]
&&
\hspace{29mm}
+ \,
\initial{k} \, 
x^\alpha \,
\Big[
\final{l} \,
(1-z)^\alpha \,
\bar{\Bn}^{(ilk)}_{kl} 
+
\initial{l} \, 
\bar{\Bn}^{(ikl)}_{kl} 
\Big]
\bigg\} \, 
\, ,
\eeq
where $\alpha\geq 0$, and the $x$, $y$, $z$ kinematic variables are those associated to the $(ikl)$ or $(ilk)$ phase-space mappings, i.e.~they are different for each term in the eikonal double sum. In detail, they are defined as in \eq{eq:CSparam1}, \eq{eq:CSparam2}, \eq{eq:CSparam3} for $(kl)= \rm FF, FI, IF, II$, respectively. The case with no damping, \eq{eq:SiRdef}, is simply obtained setting $\alpha=0$.

As far as collinear and soft-collinear contributions are concerned, \eq{eq:CijRdef} and \eq{eq:SiCijRdef} are modified as
\beq
\label{eq:dampedC}
\bCj \, \Rl 
& = &
\frac{\Norm}{s_{ij}} \,
\Bigg\{ \;
\final{i} \, \final{j} \,
P_{ij, \rm F}^{\mu \nu} (z) \,
\Big[
\final{r} \,
(1-y)^\beta
+
\initial{r} \,
x^\beta \,
\Big] \,
\bar{\Bn}^{(ijr)}_{\mu \nu}
\nnb\\[-2mm]
&&
\hspace{7mm}
+ \,
\final{i} \, \initial{j} \,
\frac{P_{[ij]i, \rm I}^{\mu \nu} (x)}{x} \,
\Big[
\final{r} \,
(1-z)^\gamma \,
\bar{\Bn}^{(irj)}_{\mu \nu}
+
\initial{r} \,
(1-v)^\gamma \,
\bar{\Bn}^{(ijr)}_{\mu \nu}
\Big]
\nnb\\
&&
\hspace{7mm}
+ \,
\final{j} \, \initial{i} \,
\frac{P_{[ji]j, \rm I}^{\mu \nu} (x)}{x} \,
\Big[
\final{r} \,
(1-z)^\gamma \,
\bar{\Bn}^{(jri)}_{\mu \nu}
+
\initial{r} \,
(1-v)^\gamma \,
\bar{\Bn}^{(jir)}_{\mu \nu}
\Big]
\Bigg\}
\, ,
\qquad
\\
\label{eq:dampedSC}
\bSi \, \bCj \, \Rl 
& = & 
\Norm \, 2 \, C_{f_j} \, 
\mc I_{jr}^{(i)} 
\bigg\{ \;
\final{j} \,
(1-z)^\alpha \,
\Big[
\final{r} \,
(1-y)^\beta
+
\initial{r} \,
x^\beta
\Big]
\bar{\Bn}^{(ijr)}
\nnb\\[-2mm]
&&
\hspace{21mm}
+ \,
\initial{j} \,
x^\alpha 
\Big[
\final{r} \,
(1-z)^{\gamma-1} 
\bar{\Bn}^{(irj)}
+
\initial{r} \,
(1-v)^{\gamma+1} 
\bar{\Bn}^{(ijr)}
\Big]
\bigg\}
\, ,
\quad
\eeq
where $\alpha$ is the same exponent appearing in the damped soft counterterm, \eq{eq:dampedS}, while $\beta, \gamma\geq 0$ are relevant for final- and initial-state collinear emission, respectively. The kinematic variables building the damping factors depend on the mapping appearing in the relevant Born matrix element, as in the soft case. The un-damped limits are obtained upon setting $\alpha=\beta=\gamma=0$.

Following the same steps detailed in Appendix~\ref{app:consistency}, it can be checked that the damped counterterm definitions in Eqs.~(\ref{eq:dampedS} - \ref{eq:dampedSC}) correctly satisfy the consistency relations in Eqs.~(\ref{limits}). It will be moreover shown in the next Section that, as expected, the $\eps$ poles of the integrated counterterms do not feature any dependence on the arbitrary parameters $\alpha, \beta, \gamma$, which thus appear only in the finite part $\mc O(\eps^0)$.

We point out that the structure of the local counterterm $K$ and of its sector components $K_{ij}$, $K_{\{ij\}}$ given in Eqs.~(\ref{eq:NLOKW}, \ref{eq:NLOKZ}, \ref{eq:NLOK}) is not affected by the presence of damping factors and remains formally valid for arbitrary values of $\alpha, \beta, \gamma$. 
The damped $\bbHC{ij}\,R$ counterterms can still be written in terms of the hard-collinear kernels $P^{\hc}_{ij,\star}$ 
\beq
\label{eq:HCijRdamped}
\bbHC{ij} \, \Rl
& \equiv &
(1-\bSi-\bSj) \, \bCj \, \Rl
\nnb\\
& = &
\final{i} \, \final{j} \,
\bbHC{ij}^{\,\rm F} \, \Rl
+
\final{i} \,
\initial{j} \, 
\bbHC{ij}^{\,\rm I} \, \Rl
+
\final{j} \,
\initial{i} \, 
\bbHC{ji}^{\,\rm I} \, \Rl
\, ,
\nnb\\[2mm]
\bbHC{ij}^{\,\rm F} \, \Rl
& \equiv &
\Norm
\Big[
\final{r} \,
(1-y)^\beta
+
\initial{r} \,
x^\beta \,
\Big]
\nnb\\
&&
\hspace{5mm}
\Bigg[
\frac{P_{ij, \rm F}^{\mu \nu,{\rm hc}} (z)\!}{s_{ij}}
\bar{\Bn}^{(ijr)}_{\mu \nu}
+
2 \,
\Big[
C_{\!f_j} \mc I_{jr}^{(i)} 
\Big(\! 1 \!-\! (1\!-\!z)^\alpha \!\Big)
+
C_{\!f_i} \mc I_{ir}^{(j)} 
( 1 \!-\! z^\alpha )
\Big]
\bar{\Bn}^{(ijr)}
\!\Bigg]
\, ,
\nnb\\
\bbHC{ij}^{\,\rm I} \, \Rl
& \equiv &
\Norm\, 
\Bigg[
\final{r} \,
(1 \!-\! z)^\gamma 
\Bigg(\!
\frac{P_{\![ij]i, \rm I}^{\mu \nu,{\rm hc}} (x)\!}{x\,s_{ij}} 
\bar{\Bn}^{(irj)}_{\mu \nu}
+
2 C_{\!f_j} \, \mc I_{jr}^{(i)} 
\frac{1 - x^\alpha}{1-z}
\bar{\Bn}^{(irj)}
\!\Bigg)
\nnb\\
&&
\hspace{5mm}
+ \,
\initial{r} \,
(1 \!-\! v)^\gamma 
\Bigg(\!
\frac{P_{\![ij]i, \rm I}^{\mu \nu,{\rm hc}} (x)\!}{x\,s_{ij}} 
\bar{\Bn}^{(ijr)}_{\mu \nu}
+
2 C_{\!f_j} \mc I_{jr}^{(i)} 
(1 \!-\! x^\alpha)(1 \!-\! v)
\bar{\Bn}^{(ijr)}
\!\Bigg)
\!\Bigg]
\, ,
\eeq
which will be integrated in the next Section.


\section{Counterterm integration}
\label{sec:Integration}

In order to analytically integrate the counterterms, it is convenient
to start from \eq{eq:NLOK}, relying on the kernel definitions in
Eqs.~(\ref{eq:dampedS}, \ref{eq:HCijRdamped}). The counterterm
expression is summed over sectors, compatibly with the fact that its
integral must reproduce the poles of the virtual matrix element, which
is not partitioned. 
We split $K$ into soft, final-state hard-collinear and initial-state hard-collinear contributions, 
\beq
K  
\, & \equiv & \,
K_{\rm s}
\, + \,
K_{\hc,\rm F}
\, + \,
K_{\hc,\rm I}
\, ,
\label{eq:NLOKcompact}
\eeq
defined as
\beq
\label{eq:NLOKcompactS}
K_{\rm s}
& \equiv & 
\sum_{i} \,
\bSi \, R
\, ,
\\
\label{eq:NLOKcompactHCF}
K_{\hc,\rm F}
& \equiv &
\sum_{i} \,
\sum_{j < i} \, 
\final{i} \, \final{j} \,
\bbHC{ij}^{\,\rm F} \, \Rl
\, ,   
\\ 
\label{eq:NLOKcompactHCI}
K_{\hc,\rm I}
& \equiv &
\sum_{i} \,
\sum_{j < i} \,
\Big[
\final{i} \, \initial{j} \,
\bbHC{ij}^{\,\rm I} \, \Rl
+
\final{j} \, \initial{i} \,
\bbHC{ji}^{\,\rm I} \, \Rl
\Big]
\, .
\eeq
The phase-space measures used for integration are given and described in full detail in Appendix~\ref{app:mappings}, for all cases of initial- and final-state radiation.

We start with the integration of the soft counterterm $K_{\rm s}$ in \eq{eq:NLOKcompactS}, yielding
\beq
\label{eq:ints}
&&
\hspace{-5mm}
\int d\Phi_\npo \,
\bSi \, \Rl
\, = \,
\\
& = &
- \, 2 \, \Norm \,
\frac{\varsi_\npo}{\varsi_n} \,
\sum_{k \neq i}
\sum_{\substack{l \neq i \\ l < k }} \,
\mc I_{kl}^{(i)} \,
\bigg[
\final{k} \, \final{l} \,
\int d \Phi_n^{(ikl)} \!
\int d \Phi_{\rm rad}^{(ikl)} \,
(1-y)^\alpha (1-z)^\alpha \,
\bar{\Bn}^{(ikl)}_{kl}
\nnb\\[-3mm]
&&
\hspace{39mm}
+ \, 
\final{k} \, \initial{l} \,
\int \! \int d\Phi_{n}^{(ikl)}(xk_l) \, 
d \Phi_{\rm rad}^{(ikl)} \,
x^{\alpha} \,  (1-z) ^{\alpha} \,
\bar{\Bn}^{(ikl)}_{kl}
\nnb\\
&&
\hspace{39mm}
+ \, 
\final{l} \, \initial{k} \,
\int \! \int d\Phi_{n}^{(ilk)}(xk_k) \, 
d \Phi_{\rm rad}^{(ilk)} \,
x^{\alpha} \,  (1-z) ^{\alpha} \,
\bar{\Bn}^{(ilk)}_{kl}
\nnb\\
&&
\hspace{39mm}
+ \, 
\initial{k} \, \initial{l} \,
\int \! \int d\Phi_{n}^{(ikl)} (x k_k, k_l) \, 
d \Phi_{\rm rad}^{(ikl)} \, 
x^{\alpha} \,
\bar{\Bn}^{(ikl)}_{kl}
\bigg]
\nnb\\
& \equiv &
- \, 2 \, 
\frac{\varsi_\npo}{\varsi_n} 
\sum_{k \neq i}  \sum_{\substack{l \neq i \\ l < k }}
\bigg\{ \,
\final{k} \, \final{l} \,
\int d \Phi_n^{(ikl)}  \, 
I_{\rm s, \rm F \rm F}^{ikl} \,
\bar{\Bn}^{(ikl)}_{kl} 
\nnb\\[-4mm]
&&
\hspace{26mm}
+ \,
\final{k} \, \initial{l} 
\bigg[\!
\int \! d\Phi_{n}^{(ikl)} (k_l) \, 
I_{\rm s, \rm F \rm I}^{ikl}  
+ 
\int_0^1 \! \frac{dx}{x} 
\! \int \! d\Phi_{n}^{(ikl)} (x k_l) \,  
J_{\rm s, \rm F \rm I}^{ikl}(x)
\!\bigg]
\bar{\Bn}^{(ikl)}_{kl} 
\nnb\\
&&
\hspace{26mm}
+ \,
\final{l} \, \initial{k} 
\bigg[\!
\int \! d\Phi_{n}^{(ilk)} (k_k) \, 
I_{\rm s, \rm F \rm I}^{ilk}  
+ 
\int_0^1 \! \frac{dx}{x} 
\! \int \! d\Phi_{n}^{(ilk)} (x k_k) \,  
J_{\rm s, \rm F \rm I}^{ilk}(x)
\!\bigg]
\bar{\Bn}^{(ilk)}_{kl} 
\nnb\\
&&
\hspace{26mm}
+ \,
\initial{k} \, \initial{l}
\bigg[\!
\int \! d\Phi_{n}^{(ikl)\!} (k_k, k_l)  \,  
I_{\rm s, \rm I \rm I}^{ikl} 
+  
\int_0^1 \! \frac{dx}{x} 
\! \int \! d\Phi_{n}^{(ikl)\!} (x k_k, k_l)  \, 
J_{\rm s, \rm I \rm I}^{ikl}(x) 
\!\bigg]
\bar{\Bn}^{(ikl)}_{kl} 
\bigg\}
.
\nnb
\eeq
The expressions for the integrals $I_{\rm s, \star \star}^{iab}$ and $J_{\rm s, \star \star}^{iab}(x)$, are reported in Appendix \ref{app:softMI}, where the latter (former) collect $x$-(in)dependent contributions.

Moving to the hard-collinear counterterms $K_{\hc, \star}$ in Eqs.~(\ref{eq:NLOKcompactHCI} - \ref{eq:NLOKcompactHCF}), we notice that the azimuthal contribution multiplying $Q_{ab, \star}$ in the collinear kernels vanishes upon integration, hence only unpolarised Altarelli-Parisi kernels need to be integrated. 
For a final-state $j$, relevant to $K_{\hc, \rm F}$, one has
\beq
\label{eq:inthcF}
&&
\hspace{-5mm}
\int d\Phi_\npo \,
\bbHC{ij}^{\,\rm F} \, \Rl
\, = \,
\nnb\\
& = &
\Norm \, 
\frac{\varsi_\npo}{\varsi_n} \,
\bigg[
\final{r} \,  
\int d \Phi_n^{(ijr)} \!
\int d \Phi_{\rm rad}^{(ijr)} \,
(1-y)^\beta
+
\initial{r} \,  
\int \int d\Phi_n^{(ijr)} (xk_r) \,
d\Phi_{\rm rad}^{(ijr)} \,
x^\beta
\bigg]
\nnb\\
&&
\hspace{14mm}
\Bigg[
\frac{P_{ij, \rm F}^{{\rm hc}} (z)\!}{s_{ij}}
+
2 \,
\Big[
C_{\!f_j} \,\mc I_{jr}^{(i)} 
\Big( 1 \!-\! (1\!-\!z)^\alpha \Big)
+
C_{\!f_i} \,\mc I_{ir}^{(j)} 
( 1 \!-\! z^\alpha )
\Big]
\!\Bigg]
\bar{\Bn}^{(ijr)}
\nnb\\
& \equiv &
\frac{\varsi_\npo}{\varsi_n} \,
\bigg[ \,
\final{r}  
\! \int \! d\Phi_n^{(ijr)}
\Big( 
I_{\rm hc, \rm F \rm F}^{ijr} 
+ 
I_{\rm sc, \rm F \rm F}^{ijr} 
+ 
I_{\rm sc, \rm F \rm F}^{jir} 
\Big)
\nnb\\
&&
\hspace{8mm}
+ \,
\initial{r} 
\! \int \! d\Phi_n^{(ijr)} (k_r) 
\Big(
I_{\rm hc, \rm F \rm I}^{ijr} +
I_{\rm sc, \rm F \rm I}^{ijr} +
I_{\rm sc, \rm F \rm I}^{jir}
\Big)
\nnb\\
&&
\hspace{8mm}
+ \,
\initial{r} \,  
\int_0^1 \frac{dx}{x} 
\int d\Phi_n^{(ijr)} (xk_r) \, 
\Big(
J_{\rm hc, \rm F \rm I}^{ijr}(x) +
J_{\rm sc, \rm F \rm I}^{ijr}(x) +
J_{\rm sc, \rm F \rm I}^{jir}(x)
\Big)
\bigg] \,
\bar{\Bn}^{(ijr)}
\, ,
\eeq
where the contributions proportional to $\final{r}$ or $\initial{r}$ correspond to different prescriptions for the position of the recoiler particle.
Likewise, the integration of the constituents of $K_{\hc,\rm I}$ gives
\beq
\label{eq:inthcI}
\!\!
&&
\hspace{-4mm}
\int d\Phi_\npo \,
\bbHC{ij}^{\,\rm I} \, \Rl
\, = \,
\\
\!\!
& = &
\Norm 
\frac{\varsi_\npo}{\varsi_n} 
\Bigg[
\final{r} 
\! \int \!\! \int \! d\Phi_n^{(irj)} (xk_j) \,
d\Phi_{\rm rad}^{(irj)} 
(1 \!-\! z)^\gamma 
\Bigg(\!
\frac{P_{\![ij]i, \rm I}^{{\rm hc}} (x)\!}{x\,s_{ij}} 
+
2 C_{\!f_j} \mc I_{jr}^{(i)} 
\frac{1 \!-\! x^\alpha}{1 \!-\! z}
\!\Bigg) 
\bar{\Bn}^{(irj)}
\nnb\\
\!\!
&&
\hspace{13mm}
+ \,
\initial{r} 
\! \int \!\! \int \! d\Phi_n^{(ijr)\!} (xk_j,k_r) \,
d\Phi_{\rm rad}^{(ijr)} 
(1 \!-\! v)^\gamma \!
\Bigg(\!
\frac{P_{\![ij]i, \rm I}^{{\rm hc}} (x)\!}{x\,s_{ij}} 
+
2 C_{\!f_j} \mc I_{jr}^{(i)} 
(1 \!-\! x^\alpha)(1 \!-\! v)
\!\Bigg)
\bar{\Bn}^{(ijr)}
\!\Bigg]
\nnb\\
\!\!
& \equiv &
\frac{\varsi_\npo}{\varsi_n}
\bigg\{
\final{r} 
\bigg[
\int_0^1 \! \frac{dx}{x} \! \int \! d\Phi_n^{(irj)} (xk_j) 
\Big( J_{\rm hc, \rm I \rm F}^{irj}(x) + J_{\rm sc, \rm I \rm F}^{irj}(x) \Big)
+
\int \! d\Phi_n^{(irj)} (k_j) \,
I_{\rm sc, \rm I \rm F}^{irj}
\bigg]
\bar{\Bn}^{(irj)}
\nnb\\
\!\!
&&
\hspace{8mm}
+ \,
\initial{r} \,
\bigg[
\int_0^1 \! \frac{dx}{x} \! 
\int \! d\Phi_n^{(ijr)\!} (xk_j, k_r) 
\Big( J_{\rm hc, \rm I \rm I}^{ijr}(x) + J_{\rm sc, \rm I \rm I}^{ijr}(x) \Big)
+
\int \! d\Phi_n^{(ijr)\!} (k_j, k_r) \,
I_{\rm sc, \rm I \rm I}^{ijr}
\bigg]
\bar{\Bn}^{(ijr)}
\!\Bigg\}
.
\nnb
\eeq
All integrals $I^{iab}_{\rm hc/\rm sc,\star\star}$ and $J^{iab}_{\rm hc/\rm sc,\star\star}(x)$ featuring in the previous equations are collected in Appendix \ref{app:collinearMI}.

In order to obtain the integrated counterterms $I$ and $J$, two final steps are required. First, all various Born-level parametrisations are identified, as the corresponding phase spaces have identical support, which amounts to the following relabelings:
\beq
\{\bar k\}^{(abc)}
\, \to \,
\{k\}
\, ,
\qquad \qquad
d\Phi_n^{(abc)}
\, \to \,
d\Phi_n
\, ,
\qquad \qquad
\bar{\Bn}^{(abc)}_{\dots}
\, \to \,
\Bn_{\dots}
\, .
\eeq
Then, sums over $(n+1)$-body labels must be converted into Born-level sums. When a final-state gluon $i$ is removed, relevant to the soft case, one has
\beq
\frac{\varsi_\npo}{\varsi_n} \,
\sum_{i \in \rm F} \, \delta_{f_i g}
& = &
1
\, ;
\eeq
when two final-state particles $i$ and $j$ are replaced by the \textit{parent} particle $p$, the sums over $i$ and  $j$ can be recast as a sum over $p$ according to 
\beq
\frac{\varsi_\npo}{\varsi_n} \,
\sum_{i \in \rm F}
\sum_{\substack{j \in \rm F \\ j < i}} \,
\delta_{\{f_i f_j\}\{q \bar q\}}
& = &
N_f \, \sum_{p \in \rm F} \, \delta_{f_p g}
\, ,
\nnb\\
\frac{\varsi_\npo}{\varsi_n} \,
\sum_{i \in \rm F}
\sum_{\substack{j \in \rm F \\ j < i}} \, 
\left( 
\delta_{f_i\{q, \bar q\}}
\delta_{ f_j g}
+
\delta_{f_j\{q, \bar q\}}
\delta_{ f_i g}
\right)
& = &
\sum_{p \in \rm F} \,
\delta_{f_p \{q, \bar q\}}
\, ,
\nnb\\
\frac{\varsi_\npo}{\varsi_n} \,
\sum_{i \in \rm F}
\sum_{\substack{j \in \rm F \\ j < i}} \,
\delta_{f_i g} \,
\delta_{f_j g}
& = &
\frac{1}{2}
\sum_{p \in \rm F} \, \delta_{f_p g}
\, ,
\eeq
where $N_f$ is the number of light active flavours; in the case of a final-state particle $i$ and an initial-state particle $j$ replaced by the resulting initial-state particle $a$, the relevant relations are
\beq
\frac{\varsi_\npo}{\varsi_n} \,
\sum_{i \in \rm F}
\sum_{j \in \rm I} \,
\delta_{\{f_{[ij]} f_i\}\{q \bar q\}}
& = &
\sum_{a \in \rm I} \,
\delta_{f_a g}
\, ,
\nnb\\
\frac{\varsi_\npo}{\varsi_n} \,
\sum_{i \in \rm F}
\sum_{j \in \rm I}  \,
\delta_{f_{[ij]}\{q, \bar q\}} \,
\delta_{f_i g}
& = &
\sum_{a \in \rm I} \,
\delta_{f_a \{q, \bar q\}}
\, ,
\nnb\\
\frac{\varsi_\npo}{\varsi_n} \,
\sum_{i \in \rm F}
\sum_{j \in \rm I}  \,
\delta_{f_{[ij]} g} \,
\delta_{ f_i \{q, \bar q\}}
& = &
\sum_{a \in \rm I} \,
\delta_{f_a \{q, \bar q\}}
\, ,
\nnb\\
\frac{\varsi_\npo}{\varsi_n} \,
\sum_{i \in \rm F}
\sum_{j \in \rm I} \,
\delta_{f_{[ij]} g} \,
\delta_{f_i g}
& = &
\sum_{a \in \rm I} \,
\delta_{f_a g}
\, .
\eeq
After such a procedure, all above integrals are naturally written in terms of Born-level quantities. With $\star = \rm F, \rm I$, one has
\beq
I_{\rm s, \star \star}^{abc}
& \to &
I_{\rm s, \star \star} \left(s_{bc} \right)
\, ,
\qquad
J_{\rm s, \star \star}^{abc}(x) 
\, \to \,
J_{\rm s, \star \star} \left(s_{bc}, x \right)
\, ,
\nnb\\
\nnb\\
I_{\rm sc, \star \star}^{abc}
& \to &
2 \,  C_{f_b} \,
I_{\rm sc, \star \star}
\left(s_{bc} \right)
\, ,
\qquad
J_{\rm sc, \star \star}^{abc}(x) 
\,  \to \,
2 \, C_{f_b} \,
J_{\rm sc, \star \star}
\left(s_{bc}, x \right)
\, ,
\nnb\\
\nnb\\
I_{\rm hc, \rm F \star}^{abc}
& \to &
\delta_{f_b g} \, 
\bigg[
\frac{1}{2} \,
I_{\rm hc, \rm F \star}^{\tg}
\left(s_{bc} \right)
+
N_f \,  I_{\rm hc, \rm F \star}^{\zg}
\left(s_{bc} \right)
\bigg]
+ 
\delta_{f_b \{q, \bar q\}} \,
I_{\rm hc, \rm F \star}^{\og}
\left(s_{bc} \right)
\, ,
\nnb\\
J_{\rm hc, \rm F \star}^{abc}(x) 
& \to &
\delta_{f_b g} \, 
\bigg[
\frac{1}{2} \,
J_{\rm hc, \rm F \star}^{\tg}
\left(s_{bc}, x\right)
+
N_f \, J_{\rm hc, \rm F \star}^{\zg}
\left(s_{bc}, x \right)
\bigg]
+ 
\delta_{f_b \{q, \bar q\}} \,
J_{\rm hc, \rm F \star}^{\og}
\left(s_{bc}, x\right)
\, ,
\nnb\\
J_{\rm hc, \rm I \star}^{abc}(x) 
& \to &
\delta_{f_b g} \, 
\bigg[
J_{\rm hc, \rm I \star}^{\tg}
\left(s_{bc}, x \right)
+
J_{\rm hc, \rm I \star}^{\zg} \left(s_{bc}, x \right)
\bigg]
+
\delta_{f_b \{q, \bar q\}} \,
J_{\rm hc, \rm I \star}^{\og}
\left(s_{bc}, x \right)
\, ,
\eeq
where, on the right-hand sides, $b$ and $c$ are Born-level labels.
The quantities $I_{\rm s/sc/hc,\star\star}(s)$ and  $J_{\rm s/sc/hc,\star\star}(s,x)$ appearing on the right-hand side of the above identifications are collected in Appendices \ref{app:softMI} and \ref{app:collinearMI}.


\section{NLO massless subtraction formula}
\label{sec:NLOsubtFormula}

We are now in the position of verifying
that the integrated counterterm correctly reproduces all virtual $\eps$ poles, thus providing a valid local subtraction formula for generic NLO processes without massive colourful particles. We separately consider the cases of 0, 1, 2 initial-state QCD partons, relevant to lepton-lepton, lepton-hadron, and hadron-hadron collisions, respectively. For these three process categories, we dub the counterterm $K$ as $K^{\rm F}$, $K^{\rm IF}$, and $K^{\rm IIF}$, respectively.

\subsection{No initial-state QCD partons}
The counterterm for leptonic processes is 
\beq
\label{eq:KFF}
K^{\rm F}
& = &
K_{\rm s}  
+
\final{r} \, 
K_{\hc, \rm F}
\, ,
\eeq
where $K_{\rm s}$ and $K_{\hc, \rm F}$ are defined in Eqs.~(\ref{eq:NLOKcompactS}, \ref{eq:NLOKcompactHCF}), and the notation makes explicit the fact that the emitting dipole $jr$ appearing in the hard-collinear kernels is bound to belong to the final state.

The integration over the radiative phase space, up to $\mc O(\eps)$, yields
\beq
\label{eq:intFF}
I^{\rm F}
\, = \,
I_{\text{poles}} 
+ 
I^{\rm F}_{\text{fin}}
\, ,
\eeq
where\footnote{
The expressions in \eq{eq:intFF_poles} feature sums running on final-state labels only, $\sum_{k\in\rm F}$, as well as on final- and initial-state labels, such as $\sum_{j}$ and $\sum_{c,d\neq c}$. While in the case of leptonic collisions the distinction is immaterial, as $C_{f_a} = \gamma_a=0$ for initial-state particles, such a notation allows us to use \eq{eq:intFF_poles} unmodified for hadronic collisions as well.
}
\beq
\label{eq:intFF_poles}
I_{\text{poles}}
& = &
\frac{\as}{2\pi}
\bigg[
\frac{1}{\eps^2} \, \sum_j C_{f_j} \, B
+
\frac{1}{\eps} 
\Big(
\sum_j \gamma_{j} \, B
+
\sum_{c,d \neq c} \! {\rm L}_{cd} \, B_{cd}
\Big)
\bigg]
\, ,
\\
\label{eq:intFF_fin}
I^{\rm F}_{\text{fin}}
& = &
\frac{\as}{2\pi}
\bigg\{
\Big[
\sum_{k \in \rm F} \phi_k
-
\sum_{j} \gamma_{j}^{\hc} \, {\rm L}_{jr}
\Big] \, B
+
\sum_{c,d \neq c} {\rm L}_{cd} \,
\Big(
2
-
\frac{1}{2} \, {\rm L}_{cd} 
\Big) \, B_{cd}
\nnb\\
&&
\hspace{5mm}
+ \,
2 \, A_2 (\alpha) \,
\Big[
\sum_{j} \, C_{f_j} \, {\rm L}_{jr} \, B
+
\sum_{c,d \neq c} {\rm L}_{cd} \, B_{cd}
\Big]
+ 
\sum_{k \in \rm F} \gamma^{\hc}_k  \,
A_2 (\beta) \,
B
\nnb\\
&&
\hspace{5mm}
+
\Big[
A_2 (\alpha)
\Big(
A_2 (\alpha)
-
2 \, A_2 (\beta)
\Big)
-
A_3 (\alpha)
\Big]
\, \sum_j C_{f_j} \, B
\bigg\}
\, .
\nnb
\eeq
We have introduced some short-hand notation for logarithms, ${\rm L}_{ab} = \ln (s_{ab}/\mu^2)$, and for anomalous dimensions,
\beq
&&
\gamma_a 
\, = \,
\frac{3}{2} \, C_F \, \delta_{f_a \{q, \bar q\}}
+
\frac{1}{2} \, \beta_0 \, \delta_{f_a g}
\, ,
\qquad
\gamma_a^{\hc}
\, = \,
\gamma_a
-
2 \, C_{f_a}
\, ,
\nnb\\
&&
\phi_a 
\, = \,
\frac{13}{3} \, C_F \, \delta_{f_a \{q, \bar q\}}
+
\frac{4}{3} \, \beta_0 \, \delta_{f_a g}
+
\left(
\frac{2}{3}
-
\frac{7}{2} \, \zeta_2
\right) \, C_{f_a}
\, ,
\eeq
where $\beta_0 = \big(11 \, C_A - 4 \, T_R \, N_f \big)/3$ is the first coefficient of the QCD beta function, $C_A=N_c$,  $C_F = (N_c^2-1)/(2N_c)$ 
and $T_R=1/2$. 
The functions $A_{n}(x)$ are defined in Appendix~\ref{app:MasterInt}.

The poles in \eq{eq:intFF_poles} are correctly independent of the damping parameters $\alpha$ and $\beta$, and can be checked to exactly match those of virtual origin, see for instance \cite{Catani:1998bh}, thus verifying the cancellation of singularities in the first line of \eq{sigNLOsub}. As for the finite contribution, the second and third lines collect the full dependence upon
the damping parameters, and cancel out as $\alpha = \beta = 0$.

\subsection{One initial-state QCD parton}
\label{sec:1QCDISR}
The local counterterm relevant for a reaction with one incoming QCD parton is
\beq
\label{eq:KIF}
K^{\rm IF}
& = &
K_{\rm s}  
+
\initial{r} \, 
K_{\hc, \rm F}
+
\final{r} \, 
K_{\hc, \rm I}
\, ,
\eeq
where the singular kernels are listed in Eqs.~(\ref{eq:NLOKcompactS} - \ref{eq:NLOKcompactHCI}). In $K_{\hc, \rm I}$ one assigns a final-state recoiler since the only initial-state coloured parton is identified with $j$. As for $K_{\hc, \rm F}$, one could assign a final-state recoiler only if the process featured at least one massless colourful parton in the final state at Born level, on top of the final-state emitter $j$. Identifying the recoiler with the initial-state colourful parton is instead always allowed.

The integration over the radiative phase space up to $\mc O(\eps)$ gives
\beq
\label{eq:intIF}
\int d \Phi_\npo \, K^{\rm IF}
& = &
\int d \Phi_n (k_a) \, 
\Big( 
I^{\rm F} 
+ 
I^{\rm I}_{\text{fin}} 
\Big)
+
\int_0^1 \frac{dx}{x}
\int d \Phi_n (xk_a) \,
J^{\rm I} (x)
\, ,
\eeq
where $I^{\rm F} $ is the same as 
in \eq{eq:intFF}, while $I^{\rm I}_{\text{fin}}$ is a purely finite contribution reading
\beq
I^{\rm I}_{\text{fin}}
& = &
\frac{\as}{2\pi} \,
2 \, C_{f_a}
\bigg[
1
+
\frac{\zeta_2}{4}
-
A_2 (\alpha)
\Big(
A_1 (\gamma)
-
A_2 (\beta)
-
1
\Big)
+
A_3 (\alpha)
\bigg] \, B
\, ,
\eeq
where $a$ is the label of the initial-state coloured parton. 
The $x$-independent integral on the right-hand side of \eq{eq:intIF} again successfully reproduces the general pole structure of the virtual contribution. The remaining integral over $J^{\rm I}(x)$, whose expression is
\beq
J^{\rm I} (x)
& = &
\frac{\as}{2\pi} \, 
\bigg\{\!
- 
\Big(
\frac{1}{\epsilon} 
-
{\rm L}_{ar}
\Big)
\bar P_a (x) 
+ 
P_{a, \text{fin}}^{(1)} (x)
-
\left(\frac{x^{1+\beta}}{1-x}\right)_{\!\! +}
\sum_{k \in \rm F} \,
\Big(
\gamma^{\hc}_k
-
2 \, C_{f_k} \, A_2 (\alpha)
\Big) 
\nnb\\
&&
\hspace{9mm}
+ \,
2 \, C_{f_a} 
\bigg[\;
\left(\frac{x \ln(1-x)}{1-x}\right)_{\!\! +} \!\!
-
\left(\frac{x}{1-x}\right)_{\!\! +}  \! A_1 (\gamma)
\nnb\\
&&
\hspace{22mm}
+
\left(\frac{x^{1+\alpha}}{1-x}\right)_{\!\! +} \!\!
\Big(
A_1 (\gamma)
-
A_2 (\alpha)
-
1
-
{\rm L}_{ar} 
\Big)
\bigg]
\bigg\} \, B
\nnb\\
&&
\hspace{-4mm}
- \,
\frac{\as}{2\pi} \,
\biggl(\frac{x^{1+ \alpha}}{1-x} \biggr)_+
\,  
\sum_{k \in \rm F} \, 2 \, {\rm L}_{ak} \, B_{ak}
\, ,
\eeq
with $P_{a, \text{fin}}^{(1)} (x)$ defined in Appendix \ref{app:AP}, is instrumental to tame the single pole stemming from collinear factorisation, as contained in \eq{eq:PDFct2}: it is straightforward to check that  the sum $C(x) + J^{\rm I}(x)$ is finite in $d=4$, and features a leftover logarithmic dependence upon the factorisation scale $\mu_F$, in the form
\beq
C(x) + J^{\rm I}(x)
& \supset &
-
\frac{\as}{2 \pi} \, 
\ln \mu_F^2 \, 
\bar P_a(x) \, 
\Bn
\, ,
\eeq
which cancels the $\mc O(\as)$ DGLAP $\mu_F$ dependence from the PDF.

\subsection{Two initial-state QCD partons}
\label{sec:2QCDISR}
The local counterterm for a process featuring two incoming colourful partons is
\beq
\label{eq:KII}
K^{\rm IIF}
& = &
K_{\rm s}  
+
\initial{r} \,
\Big( 
K_{\hc, \rm F}
+
K_{\hc, \rm I}
\Big)
\, ,
\eeq
where the choice of initial recoiler $r$ is dictated by the general availability, for this class of processes, of an extra initial-state QCD parton regardless of the position of the emitter $j$.

Counterterm integration up to ${\mc O(\eps)}$ yields
\beq
\label{eq:intII}
\int d \Phi_\npo \, K^{\rm IIF}
\, = \,
\int \! d \Phi_n (k_a, k_b) 
\Big( 
I^{\rm F} 
\!+ 
I^{\rm II}_{\text{fin}} 
\Big)
+
\int_0^1 \! \frac{dx}{x}
\int_0^1 \! \frac{d \hat x}{\hat x}
\int \! d \Phi_n (xk_a, \hat x k_b) \, 
J^{\rm II} (x,\hat x)
\, .
\qquad
\eeq
As above, $I^{\rm F}$ refers to \eq{eq:intFF}, reproducing the general virtual-pole structure. The remaining $x$-independent contribution is collected in
\beq
I^{\rm II}_{\text{fin}}
& = &
\frac{\as}{2\pi} \,
\bigg\{\;
\bigg[
2
+
\frac{\zeta_2}{2}
+
3 \, A_3 (\alpha)
-
A_2 (\alpha)
\Big(
2 \, A_1 (\gamma)
-
2 \, A_2 (\beta)
+
A_2 (\alpha)
\Big)
\bigg]
\Big(C_{f_a} + C_{f_b} \Big) \,
B
\nnb\\
&&
\hspace{7mm}
+ \,
4
\Big(
\zeta_2 - 1 
+
A_3 (\alpha)
\Big) \,
B_{ab}
\bigg\}
\, ,
\eeq
with $a,b$ labelling the two initial-state coloured partons.

The contribution 
$J^{\rm II}(x,\hat x) \, \equiv \, 
J_{a}^{\rm II}(x)\,\delta(1-\hat x) \,+\, J_{b}^{\rm II}(\hat x)\,\delta(1-x)$
accounts separately for the configurations in which the incoming colourful parton $a$ or $b$, respectively, enters the Born-level amplitude with rescaled momentum. As none of our mappings features a simultaneous rescaling of both initial-state momenta,
the simultaneous dependence on both $x$ and $\hat x$ is trivial in 
$J^{\rm II}(x,\hat x)$. 
Explicitly, one has ($i=a,b$)
\beq
J_{i}^{\rm II} (x)
& = &
\frac{\as}{2\pi} \,
\bigg\{\!
-
\Big(
\frac{1}{\epsilon} 
-
{\rm L}_{ir}
\Big)
\bar P_i (x) 
+ 
P_{i, \text{fin}}^{(2)} (x)
-
\left(\frac{x^{1+\beta}}{1-x}\right)_{\!\! +}
\sum_{k \in \rm F} \,
\Big(
\gamma^{\hc}_k
-
2 \, C_{f_k} \, A_2 (\alpha)
\Big) 
\nnb\\
&&
\hspace{9mm}
+ \,
2 \, C_{f_i} \, 
\bigg[\;
2 \, \left(\frac{x \ln(1-x)}{1-x}\right)_{\!\! +}
-
\left(\frac{x^{1 + \alpha} \ln(1-x)}{1-x}\right)_{\!\! +}
-
\left(\frac{x}{1-x}\right)_{\!\! +} \, A_1 (\gamma)
\nnb\\
&&
\hspace{22mm}
+ \,
\left(\frac{x^{1+\alpha}}{1-x}\right)_{\!\! +} \,
\Big(
A_1 (\gamma)
-
A_2 (\alpha)
-
1
-
{\rm L}_{ab} 
\Big)
\bigg]
\bigg\} \, B
\nnb\\
&&
\hspace{-4mm}
- \,
\frac{\as}{2\pi} \,
2 \, 
\bigg[
\left(\frac{x^{1 + \alpha} \ln(1-x)}{1-x}\right)_{\!\! +}
+
\left(\frac{x^{1 + \alpha}}{1-x}\right)_{\!\! +}
\Big(
A_2 (\alpha)
+
1
+
{\rm L}_{ab} 
\Big)
\bigg] \, 
B_{ab}
\nnb\\
&&
\hspace{-4mm}
- \,
\frac{\as}{2\pi} \,
\biggl(\frac{x^{1+ \alpha}}{1-x} \biggr)_+ \,   
\sum_{k \in \rm F } \, 2 \,
 {\rm L}_{ik} \, B_{ik} 
\, . 
\eeq
The same considerations on collinear-pole cancellation and $\mu_F$ dependence hold as in the case of single initial-state QCD parton, which concludes the proof of $\eps$-pole cancellation by means of the local analytic sector subtraction procedure.


\section{Numerical implementation and validation}
\label{sec:Numerics}
In this Section, we present numerical results obtained by applying
local analytic sector subtraction to the computation of NLO cross
sections for realistic scattering processes. We choose to work in the
\madnk\
framework~\cite{Lionetti:2018gko,Hirschi:2019fkz,Becchetti:2020wof,Bonciani:2022jmb},
which provides a flexible high-level platform suitable for deploying
meta-codes that implement generic subtraction schemes for IRC
divergences at higher orders. \madnk\ builds on the \aMC\
environment~\cite{Alwall:2014hca,Frederix:2018nkq}, relying on the
latter for the generation of tree-level and one-loop matrix
elements.\footnote{We remind the reader that one-loop matrix elements
  in \aMC\ are generated by the {\sc MadLoop}
  module~\cite{Hirschi:2011pa}.} In particular, once the user
specifies the scattering process and the perturbative order (e.g.~NLO
or NNLO in QCD, and possibly mixed QCD-EW corrections), \madnk\
identifies all the building blocks necessary for the corresponding
computations, i.e.~the matrix elements and the counterterms needed in
the singular limits. Matrix elements which can be obtained from \aMC\
are also generated. It is a developer's task to implement those
ingredients which are specific to a given subtraction scheme, such as
the expression of the local and integrated counterterms, momentum
mappings, and possibly sector functions, as well as functions
providing a code in a low-level programming language. In the
following, we will show some numerical results both at the local and
at the integrated level. The interested reader can find details on the
implementation of our subtraction scheme in \madnk\ in
Appendix~\ref{app:madnk}.

\subsection{Validation of local IRC-singularity cancellation}
In this Section we showcase how the cancellation of IRC singularities
is achieved numerically for a selection of processes and of singular
configurations.  Specifically, we evaluate the $(\npo)$-body matrix
element in a randomly-chosen phase-space point, then we progressively deform it in order to approach a specific singular configuration
(soft or collinear). The closeness to the singular configuration is
controlled by a parameter, $\lambda$, whose meaning is described in
details in the Appendix of Ref.~\cite{Hirschi:2019fkz}. For the
purpose of this work, the reader should bear in mind that
$\lambda \sim E_i^2$ ($\lambda \sim \theta_{ij}^2 $) in the soft $\Si$
(collinear $\Cj$) limit.

\begin{figure}
    \includegraphics[width=0.49\textwidth]{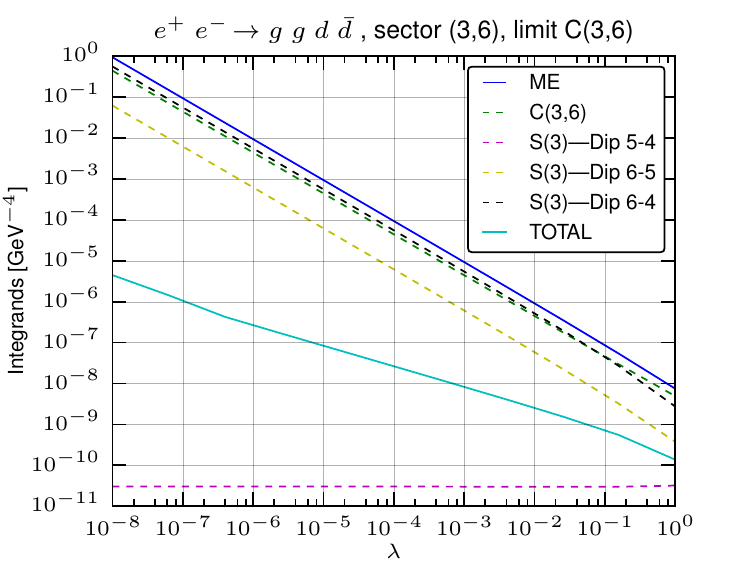}
    \includegraphics[width=0.49\textwidth]{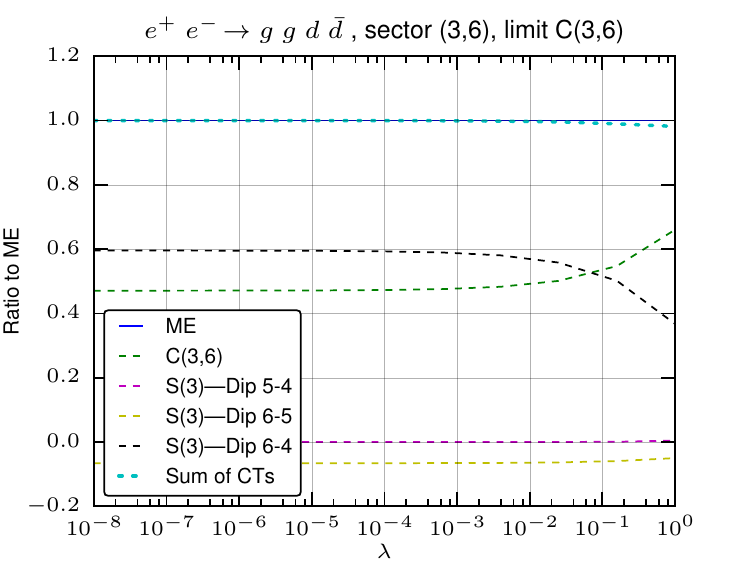}\\
    \includegraphics[width=0.49\textwidth]{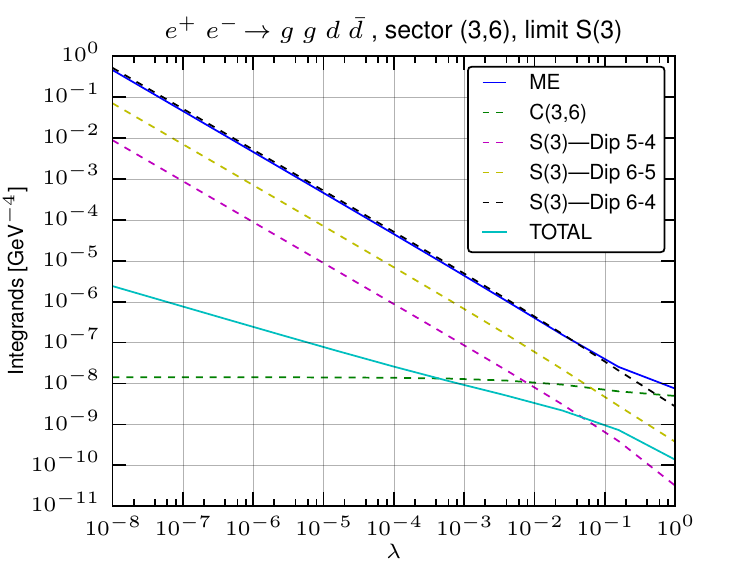}
    \includegraphics[width=0.49\textwidth]{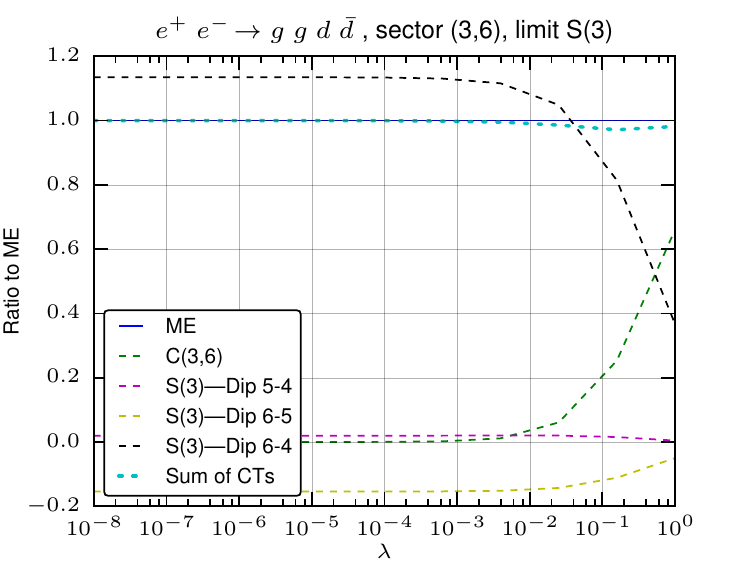}
    \caption{\label{fig:eeddgg}The singular behaviour of the
      real-emission matrix element and counterterms for the process
      $e^+ e^- \to g g d \bar d$, in the sector identified by
      particles 3, 6.  Top row: collinear configuration C(3,6); bottom
      row: soft configuration S(3).}
\end{figure}
We start by showing in Figure \ref{fig:eeddgg} the case for
$e^+ e^- \to g g d \bar d$, and consider the sector identified by the
first gluon and the $\bar d$ quark (labelled as 3, 6 in the particle
list) both in case they become collinear (top row), and in case the
gluon becomes soft (bottom row). Several quantities are displayed: the
solid blue line represents the exact $(\npo)$-body matrix element,
dubbed ME; thin dashed lines of different colours indicate the
collinear counterterms C($x,y$), which include soft-collinear
contributions, and the soft counterterms S($z$), split according to
the different eikonal (or radiating dipole) contributions Dip $a$-$b$;
the subtracted matrix element, labelled with TOTAL, is marked with a
solid teal line, while the sum of all counterterms (Sum of CTs) is
displayed with a thicker dashed line.  Contributions are shown either
in absolute value (left panels) or divided by the matrix element
(right panels).  Both sets of panels help conveying the message that the
local cancellation of singularities has been achieved. In the
left-hand plots, the $\lambda^{-1}$ slope of the real matrix element
and of the counterterms is apparent, reducing to a $\lambda^{-1/2}$
behaviour for the subtracted result, which in turn becomes regular
once combined with the phase-space measure. In the right plots one can
appreciate how the various counterterms combine in such a way that
their sum matches the matrix element in the relevant singular
limit.

\begin{figure}
    \includegraphics[width=0.49\textwidth]{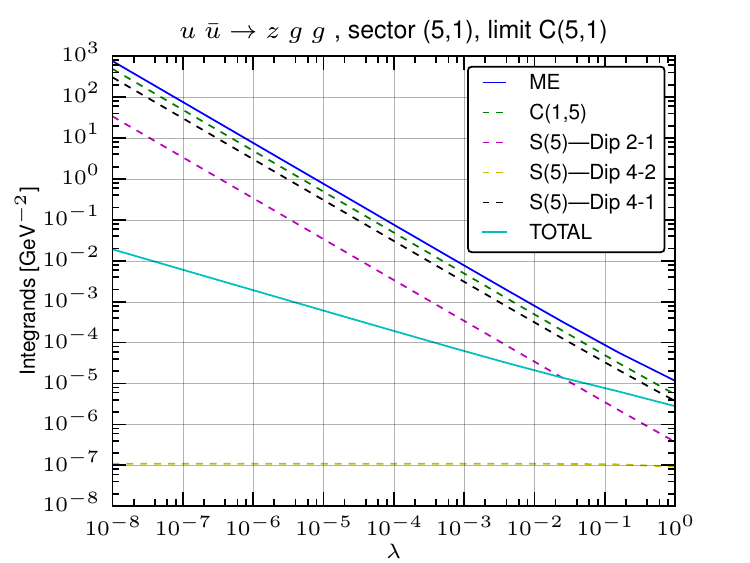}
    \includegraphics[width=0.49\textwidth]{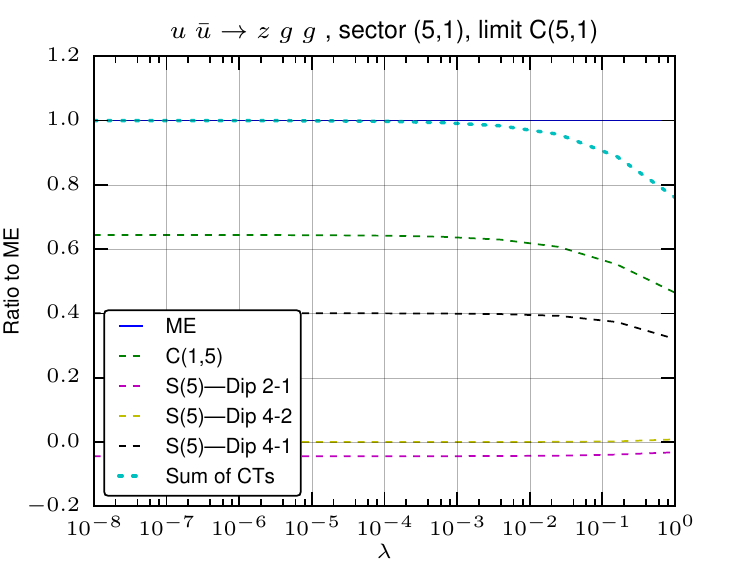}\\
    \includegraphics[width=0.49\textwidth]{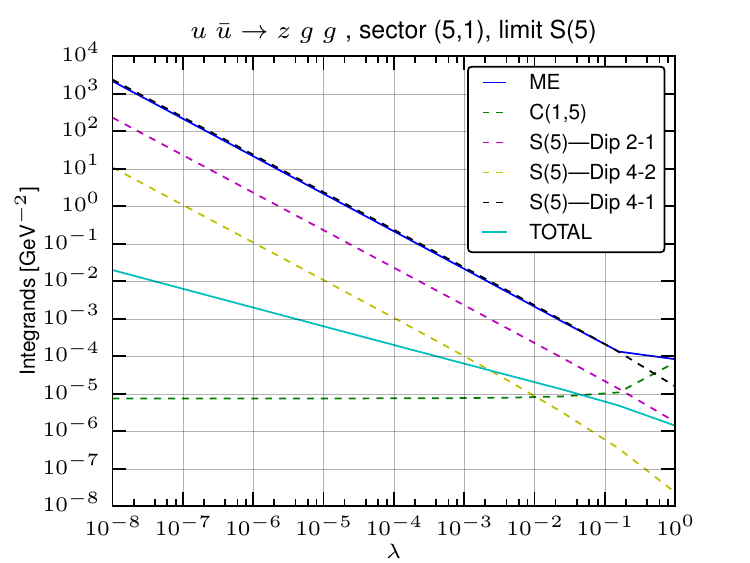}
    \includegraphics[width=0.49\textwidth]{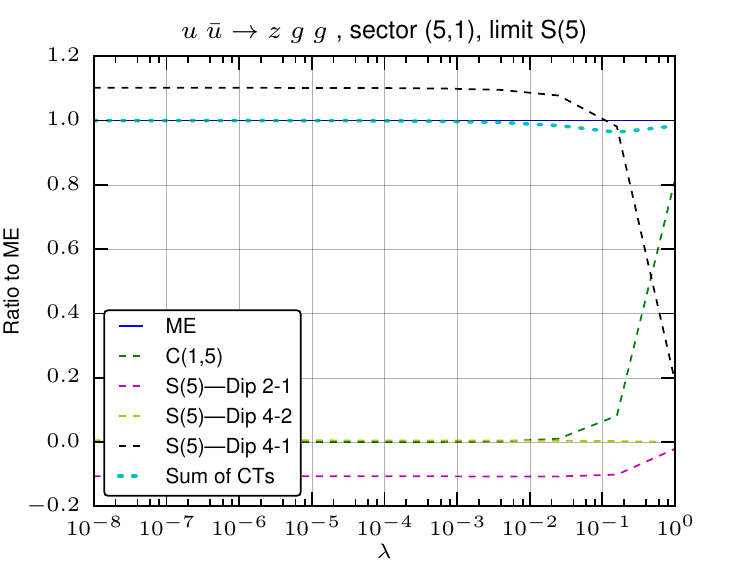}
    \caption{\label{fig:uuzgg}The singular behaviour of the
      real-emission matrix element and counterterms for the process
      $u \bar u \to Z g g$, in the sector identified by particles 1,
      5.  Top row: collinear configuration C(1,5); bottom row: soft
      configuration S(5).}
\end{figure}

Turning to processes initiated by coloured particles, we show in
Figure \ref{fig:uuzgg} the case for $u \bar u \to Z g g$ in the C(1,5)
and S(5) configurations (i.e.~those for which the last gluon (5) is
collinear to the incoming $u$ quark (1), or soft), in the sector
identified by particles 1, 5.
\begin{figure}
    \includegraphics[width=0.49\textwidth]{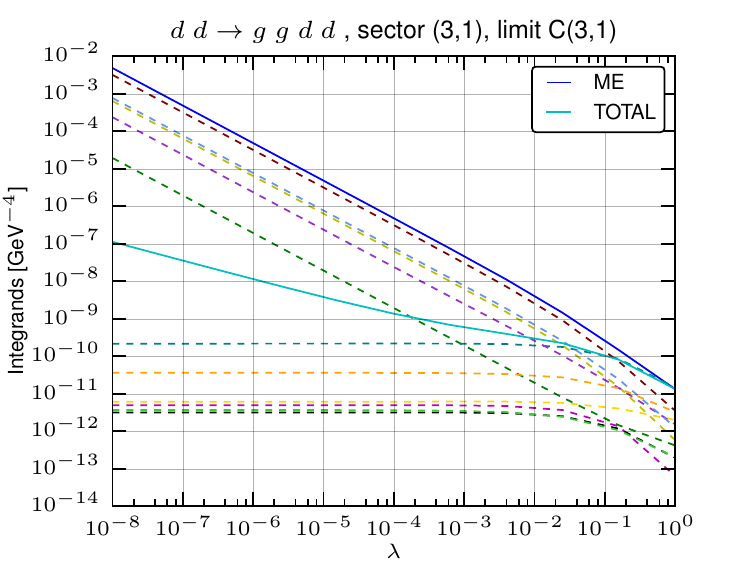}
    \includegraphics[width=0.49\textwidth]{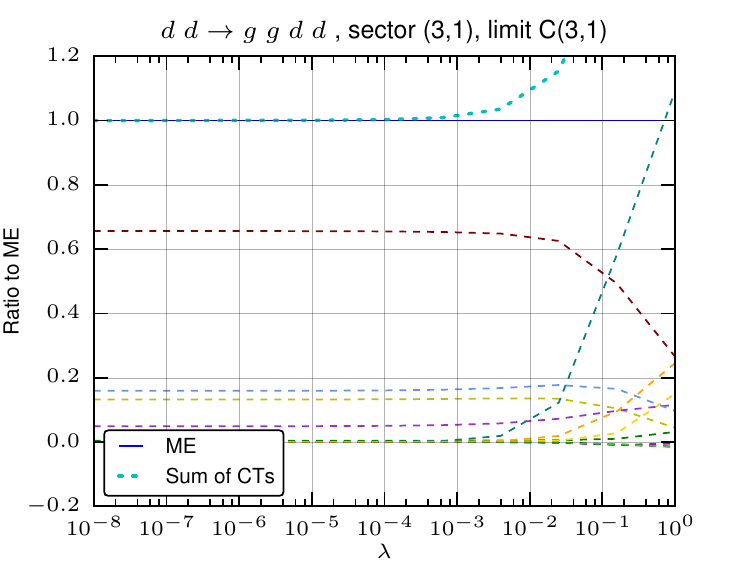}\\
    \includegraphics[width=0.49\textwidth]{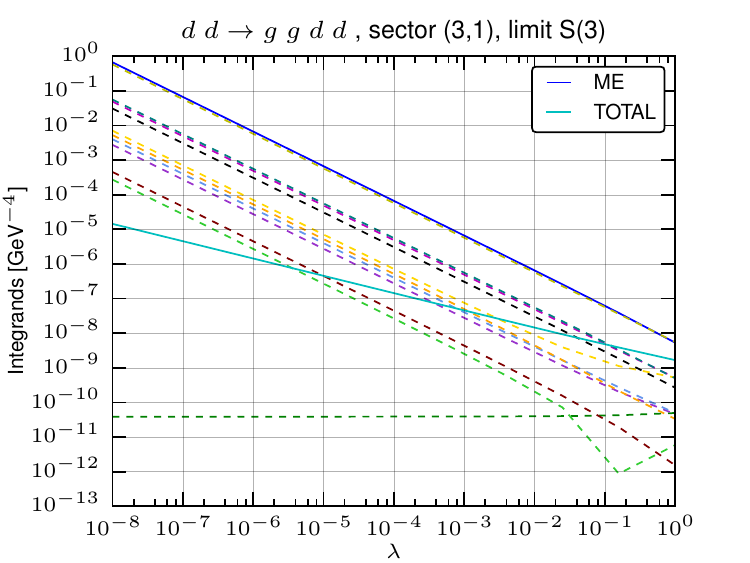}
    \includegraphics[width=0.49\textwidth]{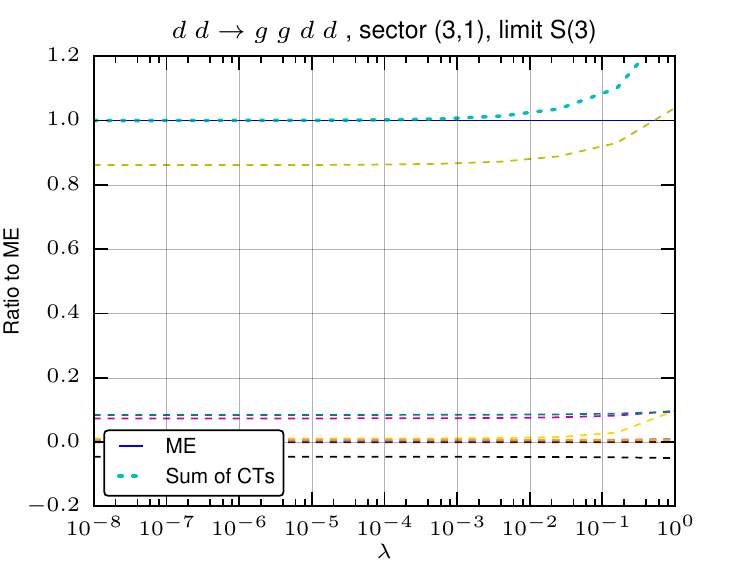}
    \caption{\label{fig:ddggdd}The singular behaviour of the
      real-emission matrix element and counterterms for the process
      $d d \to g g d d$, in the sector identified by particles 1, 3.
      Top row: collinear configuration C(1,3); bottom row: soft
      configuration S(3).}
\end{figure}
Analogously, in Figure \ref{fig:ddggdd}, we consider the case of
$d d \to g g d d$ in the C(1,3) and S(3) configurations, in the sector
identified by particles 1, 3.  Such a process has as many as 11
counterterms in this configuration (1 collinear and 10 soft
dipoles), thus the displayed integrable scaling of the subtracted
matrix element provides a highly non-trivial test of the
correctness of the local subtraction mechanism.

\subsection{Integrated results}
We now turn to the numerical validation of our approach at the level
of integrated cross sections for a selection of processes at NLO,
comparing our results against those obtained with \aMC.  The two main current
limitations of our \madnk-based framework are the absence of a
low-level code implementation, and of optimised phase-space integration
routines. In fact, the integration is steered by a code written in
{\sc Python}, using {\sc Vegas3}~\cite{Lepage:1980dq,Lepage:2020tgj}
as integrator. Such a behaviour somewhat limits the complexity of the
processes that can be run within a reasonable amount of time and
computing resources; still, the processes we consider in the following
cover all radiation topologies and both leptonic and hadronic
collisions, hence we reckon them a sufficient subset for validation
purposes.

The numerical setup we employ is the following: processes at lepton
colliders are run at a centre-of-mass energy of 500 GeV. Hadronic
processes are instead run at the LHC RunII energy of 13 TeV. In the
latter case, the ${\rm PDF4LHC15\_nlo\_30}$ PDFs are
employed~\cite{Butterworth:2015oua}, via the LHAPDF
interface~\cite{Buckley:2014ana}.  The fine-structure and Fermi
constants have the values
\begin{equation}
    \alpha = 1/132.507, \qquad G_f = 1.16639\cdot 10^{-5} \, \textrm{GeV}^{-2},
\end{equation}
while the following values for particle masses are
employed:\footnote{In our model $m_W$ is derived from $\alpha$, $G_f$
  and $m_Z$; also, the presence of a non-zero value for $m_b$ is
  formally inconsistent with the employed PDF set, however this is of
  no relevance as far as validation is concerned.}
\begin{equation}
m_Z = 91.188\, \textrm{GeV},\qquad
m_W = 80.419\, \textrm{GeV},\qquad
m_b = 4.7\, \textrm{GeV},\qquad
m_t = 173\, \textrm{GeV}.
\end{equation}
Renormalisation and factorisation scales are kept fixed to $\mu=\mu_F=m_Z$.\\
Whenever light partons are present in the final state at the Born
level, they are clustered into jets with the anti-$k_t$
algorithm~\cite{Cacciari:2008gp}, as implemented in {\sc
  FastJet}~\cite{Cacciari:2011ma}, with radius parameter $R=0.4$. Jets
are then required to satisfy the following kinematic cuts:
\begin{equation}
    p_T(j) > 20 \, \textrm{GeV}, \qquad |\eta(j)|<5.
\end{equation}
The processes we consider are
\begin{eqnarray}
    e^+ e^- & \to & j j, \\
    e^+ e^- & \to & j j j, \\
    p p & \to & Z, \\
    p p & \to & Z j , \\
    p p & \to & W^+ W^- j .
\end{eqnarray}

For these processes, we have computed the LO cross section and its NLO
correction, which are quoted in Table \ref{tab:valid}. In this case, no
damping factors are applied. Results from \aMC\ (dubbed aMC in the
table) and \madnk\ are in general very well compatible, the largest
deviations being of the order of the combined integration error, which
is at or below the per-mille level.
\begin{table}[ht]
\begin{center}
\begin{tabular}{|c|c|c|c|c|}
\hline
Process &  aMC LO & \madnk\ LO & aMC NLO corr. & \madnk\ NLO corr.\\
\hline
$e^+e^- \to jj$ & 0.53209(6)  & 0.53208(6) & 0.019991(7) & 0.019991(10) \\     
$e^+e^- \to jjj$ & 0.4739(3)  & 0.4740(3) & -0.1461(1) & -0.1463(6) \\     
$pp \to Z$ & 46361(3) & 46362(3) & 6810.9(8) & 6810.8(4) \\     
$pp \to Zj$ & 11270(7)  & 11258(5) & 3770(6) & 3776(17) \\     
$pp \to W^+W^-j$ & 42.42(1)  & 42.39(2) & 10.68(5) & 10.53(13) \\     
\hline
\end{tabular}
\end{center}
\caption{\label{tab:valid} Validation table with predictions for LO
  cross sections and NLO corrections. Numbers are in pb. Integration
  errors, on the last digit(s), are shown in parentheses.}
\end{table}

We also consider the case of non-zero values for the damping
parameters $\alpha,\beta,\gamma$ presented in
Section \ref{sec:localctdamping}. For simplicity, we set the three
parameters to a common value, ranging from 0 to 2. Results for the NLO
corrections are shown in Table \ref{tab:damp}, together with their
breakdown into $n$-body and $(\npo)$-body contributions (the former
including virtual corrections and integrated counterterms, the latter
including subtracted real emissions). While the $n$- and $(\npo)$-body
terms, if consider separately, show a very significant dependence upon
the unphysical damping parameters, their sum remains stable, as
expected.  Results with the three different damping choices are
totally compatible within the respective integration errors, and, in
turn, with the \aMC\ results.

\begin{table}[ht]
\begin{center}
\begin{tabular}{|c|c|c|c|}
\hline
Process &\madnk\ &\madnk\  &\madnk\ \\
 & $\alpha=\beta=\gamma=0$ & $\alpha=\beta=\gamma=1$ &
                                                       $\alpha=\beta=\gamma=2$ \\
\hline
$e^+e^- \to jj$ & & & \\
V+I & 0.02664732(9) & 0.01998531(7) & 0.00666183(2) \\
R-K & -0.00666(1) & 0.000004(6) & 0.013329(6) \\
NLO corr. & 0.019991(10) & 0.019985(6) & 0.019991(6) \\
\hline
$pp \to Z$ & & & \\
V+I+C+J & 3981.5(4) & -3472.7(4) & -9163.2(5) \\
R-K & 2829.3(2) & 10284.3(4) & 15974.1(6) \\
NLO corr. & 6810.8(4) & 6811.6(6) & 6810.9(8) \\     
\hline
$pp \to Zj$ & & & \\
V+I+C+J & 7172(2) & 5246(2) & 3624(2) \\
R-K & -3395(17) & -1469(25) & 156(22) \\
NLO corr. & 3776(17) & 3777(25) & 3780(22)  \\     
\hline
\end{tabular}
\end{center}
\caption{\label{tab:damp} Validation table with predictions for the
  NLO corrections, broken down between $n$ and $\npo$ contributions,
  when different damping factors ($\alpha,\beta,\gamma$) are
  considered. Numbers are in pb. The integration error, on the last
  digit(s), is shown in parentheses.}
\label{tab:limits}
\end{table}

\subsection{Differential validation}
\begin{figure}
    \includegraphics[width=0.49\textwidth]{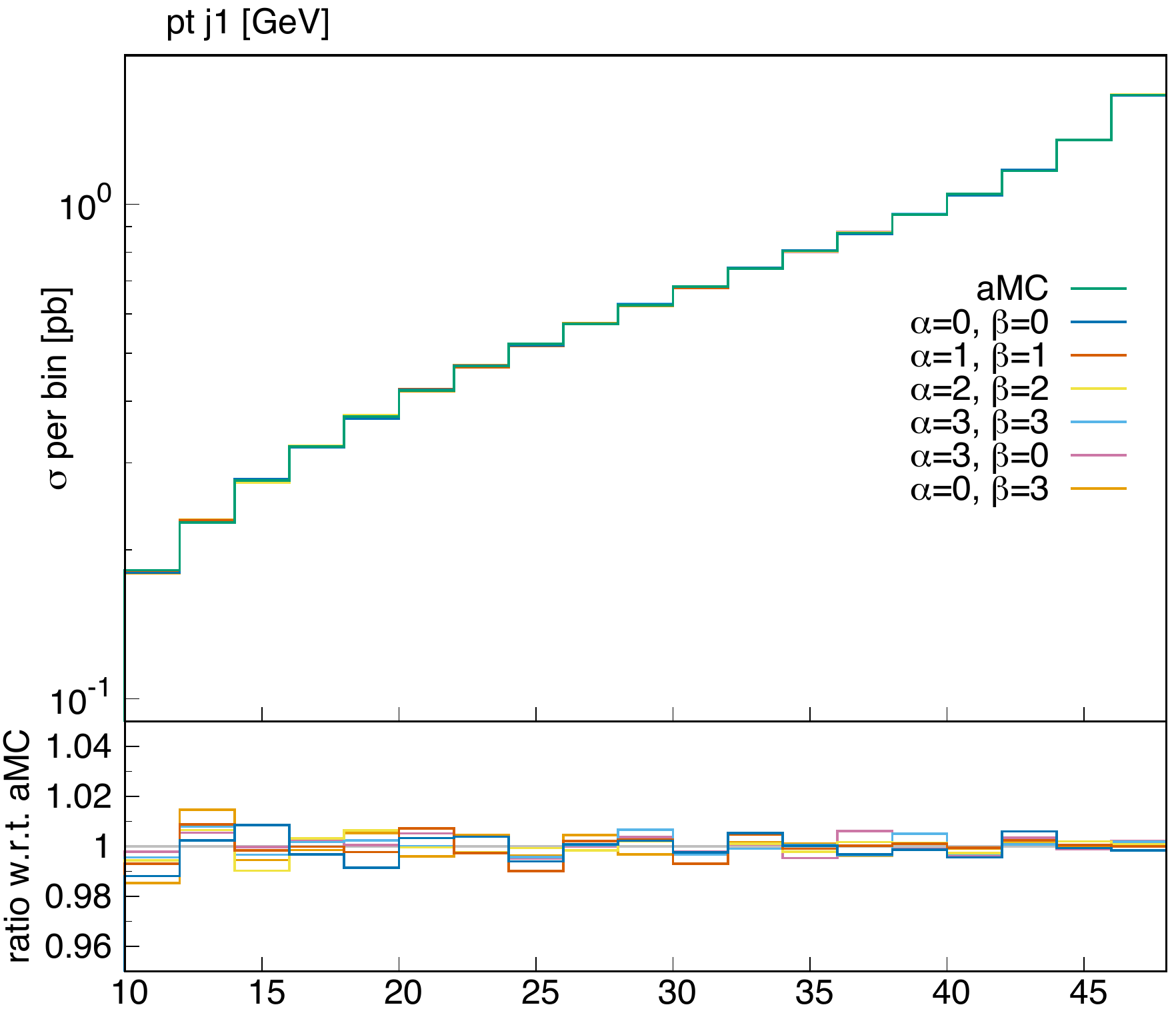}
    \includegraphics[width=0.49\textwidth]{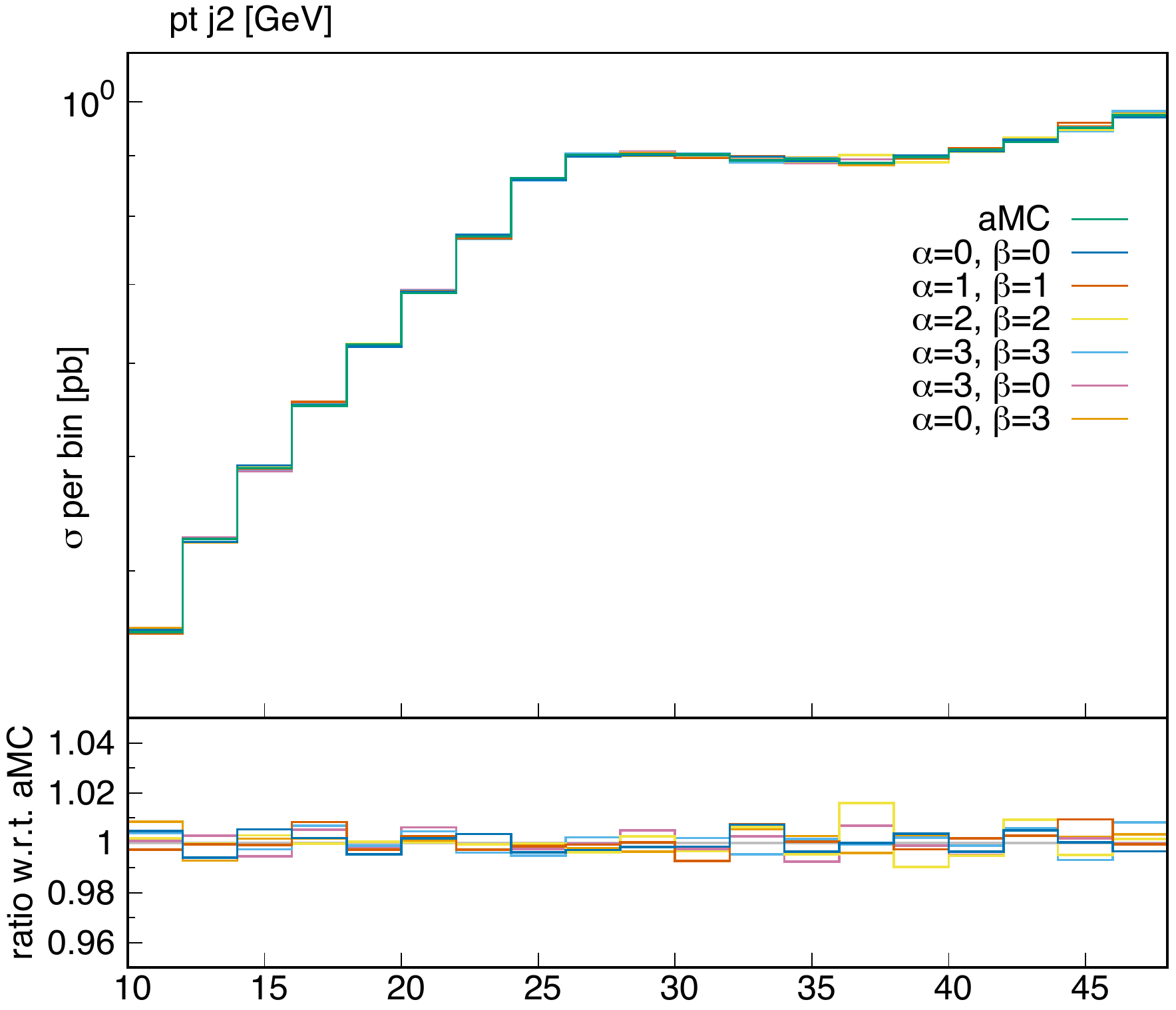}\\\vspace{5mm}\\
    \includegraphics[width=0.49\textwidth]{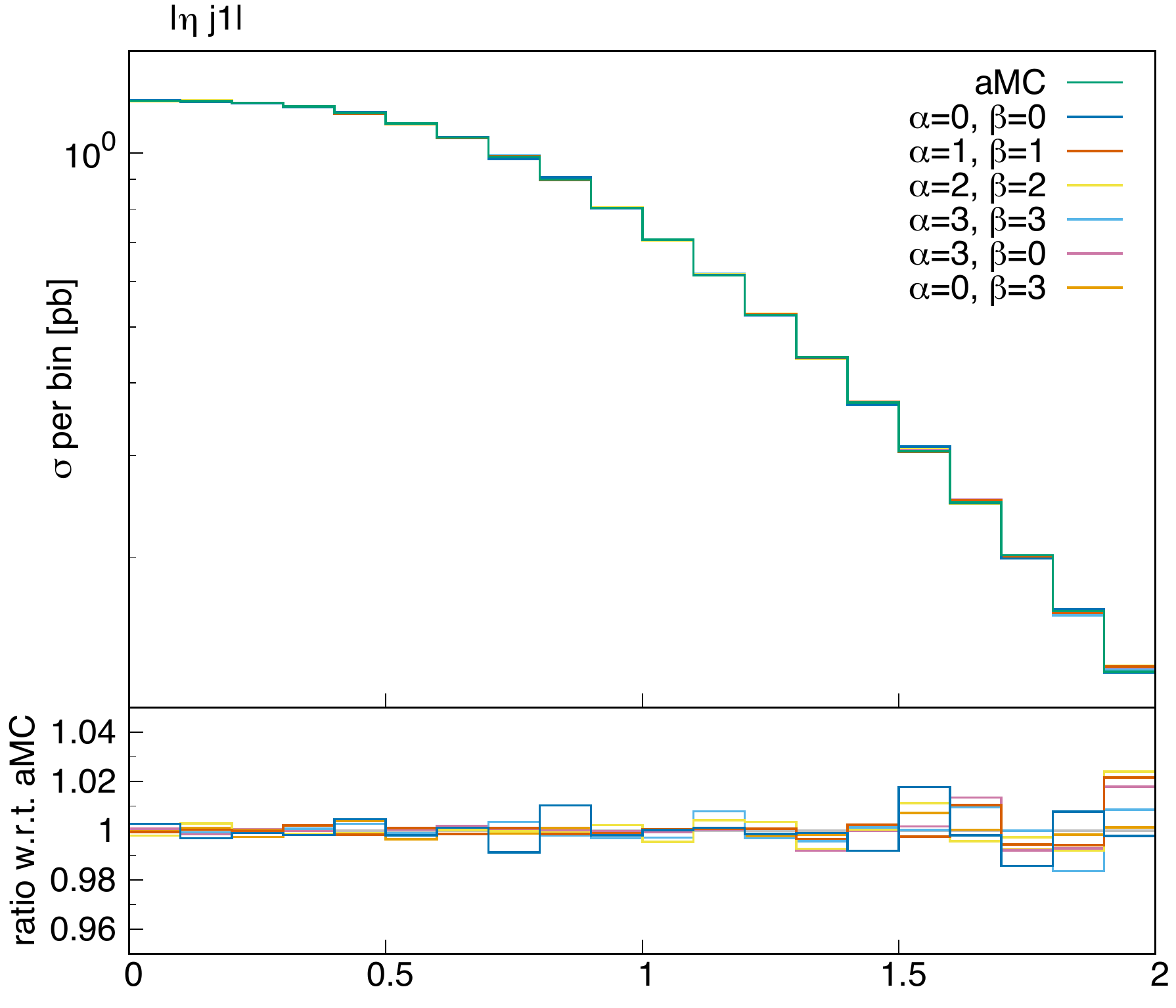}
    \includegraphics[width=0.49\textwidth]{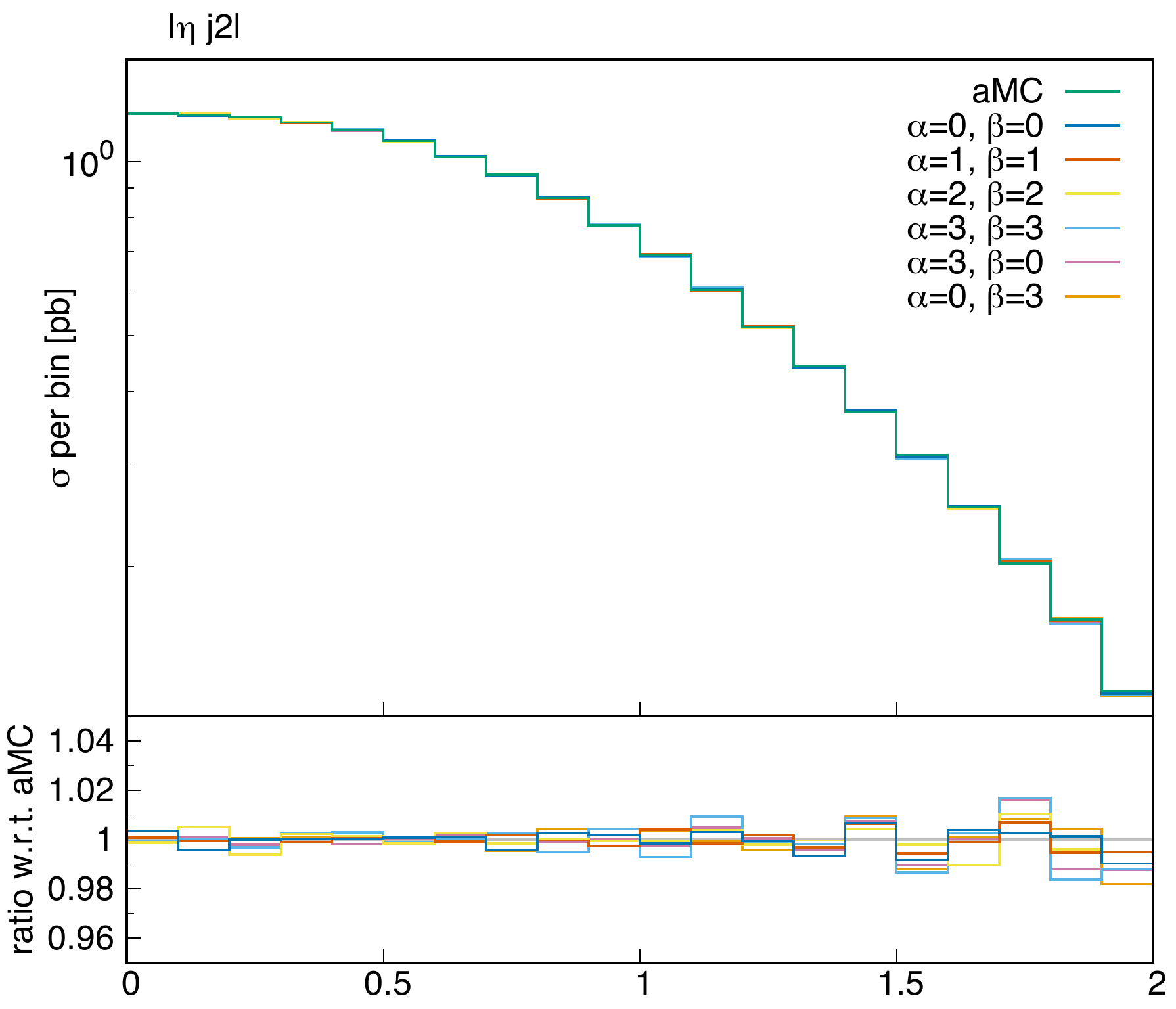}
    \caption{\label{fig:diff_validation}Transverse momenta and
      pseudo-rapidities for the two hardest jets in
      $e^+e^-\to \gamma^*\to j j$ at NLO, comparing aMC and local
      analytic sector subtraction.}
\end{figure}

Finally, we validate the correctness of the damping factors at the
differential level in the simple case of $e^+e^-\to \gamma^*\to j j$,
at centre-of-mass energy $\sqrt s = 100$ GeV, with $\mu=35$ GeV. The
plots in Figure \ref{fig:diff_validation} show differential cross
sections with respect to transverse momentum and (absolute value of)
pseudo-rapidity of the two hardest jets in the events (clustered with the $k_t$ algorithm~\cite{Catani:1993hr,Ellis:1993tq}), which are
NLO-accurate observables receiving contribution from subtraction
counterterms across the whole spectrum. A comparison is provided
between predictions obtained with \aMC\ and an in-house
implementation of local analytic sector subtraction, limited to the
above-mentioned process. Various combinations of parameters $\alpha$
and $\beta$, ranging from 0 to 3, are chosen, in order to cover
different damping possibilities ($\gamma$ is irrelevant for
final-state radiation).

As evident from Figure \ref{fig:diff_validation}, predictions of local
analytic sector subtraction for all chosen damping profiles are in
excellent agreement with those obtained with aMC within the numerical
accuracy used for the runs. A systematic study of the performance of
the various damping choices at the differential level in more complex
processes and setups is however beyond the scope of this paper, and
postponed to future work.


\section{Conclusion}
\label{sec:conclusion}

We have presented the extension of the local analytic sector
subtraction method to the case of initial-state QCD radiation at
NLO. We have shown that the enhanced structural simplicity of the
method, already evident in the case of final-state radiation, carries
over to initial-state radiation, which represents a promising feature
with a view to NNLO subtraction for hadron-collider processes.

Aiming at an improved numerical stability, we have introduced an
optimisation procedure to tune the impact of subtraction terms in the
non-singular regions of phase space, in a way that preserves the
properties of the method, in particular the simplicity of the involved
analytic integrations.

Finally, we have presented the first implementation of the method in
an automated framework, \madnk, and validated its correctness at the
level of singular infrared and collinear limits and, mainly, of
physical cross sections, for a set of simple collider processes
involving initial- and final-state radiation.

The natural directions to be followed after this work are on one hand
a systematic optimisation of the numerical software,
necessary to reduce the time and CPU resources necessary to produce
phenomenological results; on the other hand, the inclusion of a
subtraction scheme for massive final-state particles at NLO, relevant
for top- and bottom-quark physics, and, most importantly, the
definition of local analytic sector subtraction for initial-state
radiation at NNLO.


\section*{Acknowledgements}

We thank Valentin Hirschi for his help with the \madnk\ framework, and for implementing in it some of the functionalities needed by our subtraction method. We also thank Ezio Maina for collaboration in the early stages of this work, and Lorenzo Magnea for carefully reading the manuscript. The work of PT has been partially supported by the Italian Ministry of University and Research (MUR) through grant PRIN 20172LNEEZ and by Compagnia di
San Paolo through grant TORP\_S1921\_EX-POST\_21\_01.
MZ is supported by `Programma per Giovani Ricercatori Rita Levi Montalcini', granted by MUR.

\appendix

\section{Altarelli-Parisi splitting kernels} 
\label{app:AP}
We collect here the expression for the regularised Altarelli-Parisi collinear kernels appearing in the lowest-order DGLAP \cite{Gribov:1972ri,Altarelli:1977zs,Dokshitzer:1977sg} evolution equations.
\beq
\label{eq:AP}
\bar P_a (x) \, B
& \equiv &
\delta_{f_a g} 
\Big[
\bar P_{gg} (x) \, B^{(g)}
+
\bar P_{q\bar q} (x) \,
\big( B^{(q)} + B^{(\bar q)} \big)
\Big]
\nnb\\
&&
+ \,
\delta_{f_a \{q, \bar q\}}
\Big[
\bar P_{gq} (x)  \, B^{(g)}
+
\bar P_{qg} (x)  \, B^{(f_a)}
\Big]
\, ,
\eeq
where
\beq
\bar P_{gg} (x)
& = &
2 \, C_A 
\left[
\frac{x}{(1-x)_+}  \!
+ \frac{1-x}{x} 
+ x(1-x)  
\right] 
+ 
\delta(1-x) \, \frac{\beta_0}{2}
\, ,
\nnb\\
\bar P_{q\bar{q}} (x)
& = &
T_R 
\left[
x^2 
+ 
(1-x)^2
\right]
\, ,
\nnb\\
\bar P_{qg} (x)
& = &
C_F 
\left(
\frac{1+x^2}{1-x} 
\right)_{\!\! +}  \! 
\, ,
\nnb\\
\bar P_{gq} (x)
& = &
C_F \, 
\frac{1 + (1-x)^2}{x}
\, , 
\eeq
$C_A = N_c$, $C_F = (N_c^2-1)/(2N_c)$, $T_R = 1/2$, 
$\beta_0 = (11 \, C_A - 4 \, T_R \, N_f)/3$, and $B^{(f_i)}$ denotes the Born contribution initiated by a parton of flavour $f_i$, stemming from the splitting of parton $a$.

We also collect here finite terms arising from the integration of initial-state collinear counterterms, see Sections \ref{sec:1QCDISR}, \ref{sec:2QCDISR}, which are related to the Altarelli-Parisi kernels:
\beq
\label{eq:APfin}
P_{a, \text{fin}}^{(\lam)} (x) \, B
& \equiv &
\delta_{f_{a}g} 
\Big[
p^{(\lam)}_{gg} (x) \, B^{(g)}
+
p^{(\lam)}_{q\bar q} (x) \,
\big( B^{(q)} + B^{(\bar q)} \big)
\Big]
\nnb\\
&&
+ \,
\delta_{f_{a} \{q, \bar q\}}
\Big[
p^{(\lam)}_{gq} (x) \, B^{(g)}
+
p^{(\lam)}_{qg} (x) \, B^{(f_a)}
\Big]
\, ,
\qquad
\eeq
where $\lam = 1,2$,
\beq
p^{(\lam)}_{gg} (x)
& = &
2 \, C_A 
\left(
\frac{1-x}{x}  
+ x(1-x)  
\right)
\Big[
\lam \, \ln(1-x) 
- 
A_1(\gamma)
\Big]
\, ,
\nnb\\
p^{(\lam)}_{q\bar{q}} (x)
& = &
T_R 
\left(
x^2 
+ 
(1-x)^2
\right)
\Big[
\lam \, \ln(1-x) 
- 
A_1(\gamma)
\Big]
+
T_R \,
2x \, (1-x)
\, ,
\nnb\\
p^{(\lam)}_{qg} (x)
& = &
C_F \, (1-x)
\Big[
\lam \, \ln(1-x) 
+
1
- 
A_1(\gamma)
\Big]
\, ,
\nnb\\
p^{(\lam)}_{gq} (x)
& = &
C_F \, \frac{1 + (1-x)^2}{x}
\Big[
\lam \, \ln(1-x) 
- 
A_1(\gamma)
\Big]
+
C_F \, x
\, ,
\eeq
and $A_1(\gam)$ is defined in Appendix \ref{app:MasterInt}.

\section{Phase-space mappings}
\label{app:mappings}
In this appendix we report the phase-space mappings and parametrisations, used throughout the main text, for initial- and final-state radiation. These mappings are taken from Catani-Seymour dipole subtraction \cite{Catani:1996vz}.

\subsection{Final $a,b,c$}
\label{app:mapping1}
The first configuration we study is shown in leftmost panel of Figure \ref{remapping_setting}. Remapped momenta are defined as
\beq
\label{eq:momentumremap1}
\hspace{-5mm}
\bar{k}^{(abc)}_{b} 
\, = \, 
k_a + k_b - \frac{y}{1-y} \, k_c 
\, ,  
\quad \quad
\bar{k}^{(abc)}_{c} 
\, = \, 
\frac{1}{1-y} \, k_c 
\, ,  
\quad \quad
\bar{k}^{(abc)}_i 
\, = \, 
k_i \, , \, 
\forall i \, \neq \, (a,b,c) 
\, , 
\eeq
as functions of the Catani-Seymour variables $0 \leq y, z \leq 1$
\beq
\label{eq:CSparam1}
y \, = \, \frac{s_{ab}}{s_{abc}} 
\ , 
\quad \quad \quad 
z \, = \, \frac{s_{ac}}{s_{ac} + s_{bc}} 
\ .
\eeq
Given the dipole centre-of-mass squared energy $\bar{s}_{bc}^{(abc)} \equiv 2 \bar{k}_{b}^{(abc)} \cdot \bar{k}_c^{(abc)}$, the Mandelstam invariants satisfy the following relations:
\beq
s_{ab} 
\, = \, 
y \, \bar{s}_{bc}^{(abc)} 
\, , 
\qquad
s_{ac} 
\, = \, 
z (1 - y) \, \bar{s}_{bc}^{(abc)} 
\, , 
\qquad
s_{bc} 
\, = \, 
(1 - z)(1 - y) \, \bar{s}_{bc}^{(abc)} 
\, .
\eeq
The remapping allows to factorise the radiative phase space from the $n$-body phase space  as
\beq
\int d \Phi_\npo  
\, = \, 
\frac{\varsi_\npo}{\varsi_n}
\int d\Phi_n^{(abc)} 
\int d\Phi_{\rm rad}^{(abc)}
\, ,
\label{npo1phsp}
\eeq
where $\varsi_m$ is the multiplicity factor for the $m$-body phase space and 
\beq
d\Phi_n^{(abc)}
\, \equiv \,
d\Phi_n \left( \{\bar k\}^{(abc)} \right)
\, ,
\qquad
d\Phi_{\rm rad}^{(abc)}
\, \equiv \,
d\Phi_{\rm rad} \Big( \bar{s}_{bc}^{(abc)};  y, z, \phi \Big) 
\, ,
\eeq
\beq
\hspace{-4mm}
\int d\Phi_{\rm rad}^{(abc)}
\, = \, 
N (\eps) 
\left( \sk{bc}{abc} \right)^{1 - \eps} \!
\int_0^\pi \!\! d \phi \, \sin^{- 2 \eps} \! \phi \int_0^1 \!\! dy 
\int_0^1 \!\! dz
\Big[ y (1 - y)^2 \, z (1 - z) \Big]^{- \eps} \! (1 - y) \, ,
\eeq
with 
\beq
N(\eps) \, \equiv \, \frac{(4\pi)^{\eps - 2}}{\sqrt \pi \, \Gamma (1/2 - \eps)} \, .
\eeq

\subsection{Final $a,b$, initial $c$}
\label{app:mapping2}
Considering the central panel of Figure(\ref{remapping_setting}), we choose to boost the incoming momentum, i.e.
\beq 
\label{eq:momentumremap2}
\bar{k}^{(abc)}_{b} 
\, = \, 
k_a + k_b - (1-x) \, k_c 
\, ,
\quad \quad  
\bar{k}^{(abc)}_{c} 
\, = \,  
x \, k_c 
\, , 
\quad \quad 
\bar{k}^{(abc)}_i 
\, = \, 
k_i \, , \, \forall i \, \neq \, (a,b,c) 
\, ,  
\eeq
where we introduced kinematic variables $0 \leq x , z \leq 1$ defined as
\beq
x \, = \, \frac{s_{ac} + s_{bc} - s_{ab}}{s_{ac} + s_{bc}} \, , 
\qquad 
z \, = \, \frac{s_{ac}}{s_{ac} + s_{bc}} \, .
\label{eq:CSparam2}
\eeq
As a function of the reference invariant $\bar{s}_{bc}^{(abc)}  =   2 \bar{k}_{b}^{(abc)} \cdot k_c = 2 \bar{k}_{b}^{(abc)} \cdot  \bar{k}_c^{(abc)}/x$, the dot products are
\beq
  s_{ab} \, = \, (1-x)  \ \bar{s}_{bc}^{(abc)} \, , \qquad
  s_{ac} \, = \, z  \, \bar{s}_{bc}^{(abc)} \, , \qquad
  s_{bc} \, = \, (1-z) \, \bar{s}_{bc}^{(abc)} \, .
\eeq
In this case the radiative phase space cannot be exactly factorised as in \eq{npo1phsp} due to the dependence of the reference scale $\bar{s}_{bc}^{(abc)}$ upon the variable $x$ associated to the rescaled momentum of the initial-state parton, over which we are not integrating. Thus we define the following \textit{convolution},
\beq
\int d \Phi_\npo ( k_c)  
\, = \, 
\frac{\varsi_\npo}{\varsi_n}
\int \int d \Phi_n^{(abc)} (x k_c ) \,
d \Phi_{\rm rad}^{(abc)}
\, ,
\eeq
with
\beq
d\Phi_n^{(abc)} \! (x k_c )
\, \equiv \,
d\Phi_n \left( \{\bar k\}^{(abc)} \right) 
\, ,
\qquad
d\Phi_{\rm rad}^{(abc)}
\, \equiv \,
d\Phi_{\rm rad} \Big( \bar{s}_{bc}^{(abc)};  x, z, \phi \Big) 
\, .
\eeq
The single unresolved phase space in terms of the kinematic variables reads
\beq
\int d \Phi_{\rm rad}^{(abc)} 
\, = \, 
N(\eps) 
\left(\sk{bc}{abc}\right)^{1-\eps} 
\int_0^\pi \!\! d \phi \, \sin^{- 2 \eps} \! \phi
\int _0 ^1 \!\! dx  
\int_0^1 \!\! dz \, [(1-x) \, z (1-z)]^{-\eps} 
\,  .  
\eeq

\subsection{Final $a$, initial $b,c$}
\label{app:mapping3}

In the counterterms featuring two incoming partons, we set the incoming momentum $\bar{k}_{b}$ to be parallel to $k_b$ while leaving unchanged the other incoming momentum, $\bar{k}^{\mu}_c = k^{\mu}_c$. Then we shift all other final-state momenta collected in $k_{f} = \{k_j\}_{j \in \rm F}$, as
\beq 
\label{eq:momentumremap3}
\bar{k}^{(abc)}_{b} 
& = & 
x \, k_b  
\, , 
\quad \quad \quad
\bar{k}^{(abc)}_{c} 
\, = \, 
k_c  
\, ,
\nnb\\
\bar{k}^{(abc)}_f 
& = &
k_f  
-  
\frac{2 k_f \cdot (K + \widebar{K})}{(K + \widebar{K})^2} \, (K + \widebar{K}) 
+ 
\frac{2k_f \cdot K}{K^2} \, \widebar{K} 
\, , \qquad
\forall f \, \neq \, (a,b,c)
\, ,
\eeq
with
\beq
K 
\, = \, 
k_b + k_c -  k_a 
\, , 
\quad \quad \quad
\widebar{K} 
\, = \, 
\bar{k}_{b}^{(abc)} + \bar{k}_c^{(abc)} \, .
\eeq
The kinematic variables adopted in this case, satisfying $0 \leq x, v \leq 1$, are 
\beq
x \, = \, \frac{s_{bc} - s_{ab} - s_{ac}}{s_{bc}} \, , 
\qquad  
v \, = \,  \frac{s_{ab}}{s_{ab} + s_{ac}} \, ,
\label{eq:CSparam3}
\eeq
and with respect to the invariant $\bar{s}_{bc}^{(abc)} = 2 k_b \cdot k_c = 2 \bar k_b^{(abc)} \cdot \bar k_c^{(abc)}/x$ we rewrite the dipole Mandelstam invariants as
\beq
s_{ab} \, = \, (1-x)  \, v  \, \bar{s}_{bc}^{(abc)} \, , \qquad 
s_{ac} \, = \, (1-x)  \, (1-v) \, \bar{s}_{bc}^{(abc)} \, , \qquad
s_{bc} \, = \, \bar{s}_{bc}^{(abc)} \, .
\eeq
Then we parametrise  the $(\npo)$-body phase space as a convolution over $x$ of  $d \Phi_n$ and $d \Phi_{\rm rad}$ as 
\beq
\int d \Phi_\npo ( k_b, k_c )  
\, = \, 
\frac{\varsi_\npo}{\varsi_n}
\int \int d \Phi_n^{(abc)} (x k_b, k_c ) \, 
d \Phi_{\rm rad}^{(abc)} 
\, ,
\eeq
with
\beq
d\Phi_n^{(abc)} (x k_b, k_c )
\, \equiv \,
d\Phi_n \left( \{\bar k\}^{(abc)} \right) 
\, ,
\qquad
d\Phi_{\rm rad}^{(abc)} 
\, \equiv \,
d\Phi_{\rm rad} \Big( \bar{s}_{bc}^{(abc)};  x, v, \phi \Big) 
\, ,
\eeq
leading to the explicit expression
\beq
\hspace{-4mm}
\int d\Phi_{\rm rad}^{(abc)} 
\, = \, 
N(\eps) 
\left(\sk{bc}{abc}\right)^{1-\eps} 
\int_0^\pi \!\! d \phi \, \sin^{- 2 \eps} \! \phi
\int _0 ^1 \!\! dx  
\int_0^1 \!\! dv \, [(1-x)^2 \, v (1-v)]^{-\eps} (1- x)
\,  . 
\eeq

\section{Consistency relations}
\label{app:consistency}

In this Section we explicitly verify
the relations in Eqs.~(\ref{limits}) for initial- and final-state radiation, ensuring the locality of the subtraction procedure.

As far as relation $\Si \, \bSi \, \Rl \, = \, \Si \, \Rl $ is concerned, this is trivially verified since $\bS{i} \, \{\bar k\}^{(ikl)} = \bS{i} \, \{\bar k\}^{(ilk)} = \{k \}_{\slashed i}$ for soft emission from both the initial and the final state, hence
\beq
\Si \, \bSi \, \Rl 
& = & 
- \Norm \,
\sum_{k \neq i} \sum_{l \neq i} \,
 \mc I^{(i)}_{kl} \ B_{kl} ( \{k\}_{\slashed{i}} )
\, = \,
\Si \, \Rl 
\, .
\eeq
Analogously, for the collinear relation $\Cj \, \bCj \, \Rl \, = \, \Cj \, \Rl$, the key ingredients are the limits
\beq
\bC{ij} \, x \,
\final{i} \, \initial{j}
\, = \,
x_{[ij]} \,
\final{i} \, \initial{j}
\, ,
\eeq
as well as
\beq
\bC{ij} \,
\{\bar k\}^{(ijr)} \,
\final{i} \, \final{j}
& = &
\big(
\{ k \}_{\slashed i \slashed j},
k_{[ij]}
\big) \,
\final{i} \, \final{j}
\, ,
\nnb\\
\bC{ij} \,
\{\bar k\}^{(ijr)} \,
\final{i} \, \initial{j} \, \initial{r}
& = &
\bC{ij} \,
\{\bar k\}^{(irj)} \,
\final{i} \, \initial{j} \, \final{r}
\, = \,
\big(
\{ k \}_{\slashed i \slashed j},
x_{[ij]} \, k_j
\big) \,
\final{i} \, \initial{j}
\, ,
\eeq
from which one immediately deduces
\beq
\bC{ij} \,
\bbC{ij} \, \Rl  
& = & 
\frac{\Norm}{s_{ij}} \, 
\Bigg[
\final{i} \, \final{j} \,
P_{ij, \rm F}^{\mu \nu}(z_i) \, \Bn_{\mu \nu} \! \left( \{ k \}_{\slashed i \slashed j}, k_{[ij]} \right)
\nnb\\
&&
\qquad
+ \,
\final{i} \, \initial{j} \,
\frac{P_{[ij]i, \rm I}^{\mu \nu}(x_{[ij]})}
{x_{[ij]}}
\, \Bn_{\mu \nu} \! \left( \{ k \}_{\slashed i \slashed j}, x_{[ij]} k_{j} \right) 
\nnb\\
&&
\qquad
+ \,
\final{j} \, \initial{i} \,
\frac{P_{[ji]j, \rm I}^{\mu \nu}(x_{[ji]})}
{x_{[ji]}}
\, \Bn_{\mu \nu} \! \left( \{ k \}_{\slashed i \slashed j}, x_{[ji]} k_{i} \right) 
\Bigg]
\, = \,
\bC{ij} \, R
\, .
\eeq
Moving on to relation $\Si \, \bSi \, \bCj \, \Rl \, = \, \Si \, \bCj \, \Rl$, this is a consequence of the fact that
\beq
\bS{i} \, \bbS{i} \, \bbC{ij} \, R
\, = \,
\, \Norm \, \delta_{f_i g} \,
2 \, C_{f_j} \,
\frac{s_{jr}}{s_{ij} \,
\big (s_{ir}+\initial{j} \, \initial{r} \, s_{ij} \big)} \,
B \Big(\{k\}_{\slashed i} \Big)
\, = \,
\bS{i} \, \bbC{ij} \, R
\, ,
\eeq
having explicitly employed the soft behaviour of Altarelli-Parisi kernels.

The final relation 
$\Cj \, \bSi \, \bCj \, \Rl \, = \,  \Cj \, \bSi \, \Rl$ is instead slightly subtler. 
The explicit collinear action on the soft counterterm is
\beq
\label{eq:CbSR}
\bC{ij} \, \bSi \, \Rl 
& = &
- \, 2 \, \Norm \,
\mc I_{jr}^{(i)} \,
\bC{ij}
\bigg\{
\sum_{\substack{k \neq i \\ k < j }}
\Big[
\big(
\initial{j} \, \initial{k}
\!+\!
\final{j} \, \initial{k}
\!+\!
\final{j} \, \final{k}
\big)
\bar{\Bn}^{(ijk)}_{jk} \!
+
\initial{j} \, 
\final{k} \,
\bar{\Bn}^{(ikj)}_{jk} 
\Big] 
\nnb\\[-1mm]
& &
\hspace{23mm}
+
\sum_{\substack{k \neq i \\ k > j }}
\Big[
\big(
\initial{k} \, \initial{j}
\!+\!
\final{k} \, \initial{j}
\!+\!
\final{k} \, \final{j}
\big)
\bar{\Bn}^{(ikj)}_{jk} \!
+
\initial{k} \, 
\final{j} \,
\bar{\Bn}^{(ijk)}_{jk} 
\Big] 
\bigg\}
\nnb\\
& = &
- \, 2 \, \Norm \,
\mc I_{jr}^{(i)} \,
\bC{ij} \,
\bigg\{
\sum_{\substack{k \neq i \\ k < j }}
\Big[
\big(
\initial{j} \, \initial{k}
+
\final{j} \, \final{k}
\big) \,
\bar{\Bn}^{(ijk)}_{jk} 
\Big] 
\nnb\\
& &
\hspace{24mm}
+
\sum_{\substack{k \neq i \\ k > j }}
\Big[
\big(
\initial{j} \, \initial{k}
+
\final{j} \, \final{k}
\big) \,
\bar{\Bn}^{(ikj)}_{jk} 
\Big] 
\nnb\\[-1mm]
& &
\hspace{24mm}
+
\sum_{\substack{k \neq i, j }}
\Big[
\final{j} \,
\initial{k} \, 
\bar{\Bn}^{(ijk)}_{jk} 
+
\initial{j} \, 
\final{k} \,
\bar{\Bn}^{(ikj)}_{jk} 
\Big]
\bigg\}
\, ,
\qquad
\eeq
where we have used that 
$\bC{ij} \, \mc I^{(i)}_{jk} = \mc I^{(i)}_{jr}$ is independent of $k$, and can be taken out of the sum.
The action of $\bC{ij}$ on the mapped Born kinematics reads
\beq
\label{eq:CijBjk(ijk)}
&&
\bC{ij}\,\final{j} \, \final{k}\,\bar{\Bn}^{(ijk)}_{jk}
=
\bC{ij}\,\final{j} \, \final{k}\,\bar{\Bn}^{(ikj)}_{jk}
=
\final{j} \, \final{k}\,
\Bn_{jk}
\Big( \{ k\}_{ \slashed{i} \slashed{j}},  k_{[ij]} \Big)
\, ,
\nnb\\
&&
\bC{ij}\,\final{j} \, \initial{k}\,\bar{\Bn}^{(ijk)}_{jk}
=
\final{j} \, \initial{k}\,
\Bn_{jk}
\Big( \{ k\}_{ \slashed{i} \slashed{j}},  k_{[ij]} \Big)
\, ,
\nnb\\
&&
\bC{ij}\,\initial{j} \, \final{k}\,\bar{\Bn}^{(ikj)}_{jk}
=
\initial{j} \, \final{k} \,
\Bn_{jk}
\Big( \{ k\}_{ \slashed{i} \slashed{j}},  x_{[ij]} k_j \Big)
\, ,
\nnb\\
&&
\bC{ij}\,\initial{j} \, \initial{k}\,\bar{\Bn}^{(ijk)}_{jk}
=
\bC{ij}\,
\initial{j} \, \initial{k}\,\bar{\Bn}^{(ikj)}_{jk}
=
\initial{j} \, \initial{k}\,
\Bn_{jk}
\Big(
\{ k\}_{ \slashed{i} \slashed{j}}, x_{[ij]} k_{j}
\Big)
\, ,
\eeq
where the latter equality is proven in Appendix \ref{Ci1Remap3}.
At this point, one can recast \eq{eq:CbSR} as
\beq
\bC{ij} \, \bSi \, \Rl 
& = &
- \, 2 \, \Norm \,
\mc I_{jr}^{(i)} \,
\sum_{\substack{k \neq i,j }}
\bigg[
\initial{j} \, 
\Bn_{jk}
\Big(
\{ k\}_{ \slashed{i} \slashed{j}}, x_{[ij]} k_{j}
\Big)
+
\final{j} \, 
\Bn_{jk}
\Big( \{ k\}_{ \slashed{i} \slashed{j}},  k_{[ij]} \Big)
\bigg] 
\, .
\eeq
Upon enforcing colour conservation, $\sum_{k \neq j} {\bf T}_k = - {\bf T}_j $, this becomes
\beq
\label{eq:CbSR2}
\Cj \, \bSi \, \Rl 
& = &
2 \, \Norm \,  C_{f_j} \, \mc I^{(i)}_{jr} \, 
\bigg[
\initial{j} \,
B \Big(\{ k\}_{ \slashed{i} \slashed{j}}, x_{[ij]} k_{j} \Big)
+
\final{j} \,
B \Big(\{ k\}_{ \slashed{i} \slashed{j}},  k_{[ij]} \Big)
\bigg]
\, .
\eeq
Recalling that $\bC{ij} \, z^{(irj)} \, \initial{j} \, \final{r} = \bC{ij} \, v^{(ijr)} \, \initial{j} \, \initial{r} = 0$, it is straightforward at this point to verify that the espression in \eq{eq:CbSR} matches the result of $\Cj \, \bSi \, \bCj \, \Rl$ for all choices of remapping, showing the consistency relation.

\subsection{Collinear limits on mappings with two initial-state partons}
\label{Ci1Remap3} 
In this Appendix we show the last of Eqs.~(\ref{eq:CijBjk(ijk)}), namely that, under the collinear $\bC{ij}$ limit, both $\initial{j} \, \initial{k}\,\bar{\Bn}^{(ijk)}_{jk}$ and $\initial{j} \, \initial{k}\,\bar{\Bn}^{(ikj)}_{jk}$ tend to $ \initial{j} \, \initial{k}\, \Bn_{jk} \Big( \{ k\}_{ \slashed{i} \slashed{j}}, x_{[ij]} k_{j}
\Big)$.
The proof is based on the fact that, although the two sets of momenta do not match in the limit, the colour- (as opposed to spin-) connected Born squared amplitudes depend on kinematics only through Mandelstam invariants, which do coincide in the $\bC{ij}$ limit, as shown below.

Considering particles $j$ and $k$ in the initial state, while $i$ and $f$ in the final state, we analyse the $\bC{ij}$ limit for the mappings $(ijk)$ and $(ikj)$.

\begin{itemize}

\item Mapping $(ijk)$

\beq
\bar{k}_j 
& = &
x \, k_j
\, ,
\\ \nonumber
\bar{k}_k
& = &
k_k
\, ,
\\ \nonumber
\bar{k}_f
& = &
k_f - \frac{2 k_f \cdot (K + \widebar{K}_{(1)})}{(K + \widebar{K}_{(1)})^2} \, (K + \widebar{K}_{(1)}) + \frac{2 k_f \cdot K }{(K )^2} \, \widebar{K}_{(1)}
\nnb \, ,
\eeq
with
\beq
x = \frac{s_{jk} - s_{ij} - s_{ik}}{s_{jk}}
\, ,
\qquad
K = k_j + k_k - k_i
\, ,
\qquad
\widebar{K}_{(1)} = \bar k_j + \bar k_k = xk_j + k_k
\, .
\qquad
\eeq
Under $\bC{ij}$ collinear limit, denoting with $E_a$ the energy of parton $a$ in arbitrary frame, and with $r$ the ratio $E_i/E_j$, one has
\beq
&&
s_{ij} \, \xrightarrow{{\bf C}_{ij}} \, 0
\, ,
\qquad
s_{ik} \, \xrightarrow{{\bf C}_{ij}} \, s_{jk} \, r
\, ,
\qquad
s_{if} \, \xrightarrow{{\bf C}_{ij}} \, s_{jf} \, r
\, ,
\\
&&
x \, \xrightarrow{{\bf C}_{ij}} \, 1 - r
\, ,
\qquad
K \, \xrightarrow{{\bf C}_{ij}} \, k_j (1 - r) + k_k 
\, ,
\qquad
\widebar{K}_{(1)} \, \xrightarrow{{\bf C}_{ij}} \,
k_j (1 - r) + k_k 
\, ,
\nnb\\
&&
2 \, \bar k_j \cdot \bar k_f
\, \xrightarrow{{\bf C}_{ij}} \,
s_{jf} \, (1 - r) 
\, ,
\qquad
2 \, \bar k_k \cdot \bar k_f
\, \xrightarrow{{\bf C}_{ij}} \,
s_{kf} 
\, ,
\qquad
2 \, \bar k_j \cdot \bar k_k
\, \xrightarrow{{\bf C}_{ij}} \,
s_{jk}\, (1-r) 
\, .
\nnb
\eeq

\item Mapping $(ikj)$

\beq
\bar{k}_j 
& = &
k_j
\, ,
\\ \nnb
\bar{k}_k
& = &
x \, k_k
\, ,
\\ \nnb
\bar{k}_f
& = &
k_f - \frac{2 k_f \cdot (K + \widebar{K}_{(2)})}{(K + \widebar{K}_{(2)})^2} \, (K + \widebar{K}_{(2)}) + \frac{2 k_f \cdot K }{(K )^2} \, \widebar{K}_{(2)}
\nnb \, ,
\eeq
with
\beq
x = \frac{s_{jk} - s_{ij} - s_{ik}}{s_{jk}}
\, ,
\qquad
K = k_j + k_k - k_i
\, ,
\qquad
\widebar{K}_{(2)} = \bar k_j + \bar k_k = k_j + x k_k
\, .
\qquad
\eeq
Under $\bC{ij}$ collinear limit, denoting with $E_a$ the energy of parton $a$ in arbitrary frame, and with $r$ the ratio $E_i/E_j$, one has
\beq
&&
s_{ij} \, \xrightarrow{{\bf C}_{ij}} \, 0
\, ,
\qquad
s_{ik} \, \xrightarrow{{\bf C}_{ij}} \, s_{jk} \, r
\, ,
\qquad
s_{if} \, \xrightarrow{{\bf C}_{ij}} \, s_{jf} \, r
\, ,
\\
&&
x \, \xrightarrow{{\bf C}_{ij}} \, 1 - r
\, ,
\qquad
K \, \xrightarrow{{\bf C}_{ij}} \, k_j (1 - r) + k_k 
\, ,
\qquad
\widebar{K}_{(2)} \, \xrightarrow{{\bf C}_{ij}} \,
k_j + k_k (1 - r)
\, ,
\nnb\\
&&
2 \, \bar k_j \cdot \bar k_f
\, \xrightarrow{{\bf C}_{ij}} \,
s_{jf} \, (1 - r) 
\, ,
\qquad
2 \, \bar k_k \cdot \bar k_f
\, \xrightarrow{{\bf C}_{ij}} \,
s_{kf} 
\, ,
\qquad
2 \, \bar k_j \cdot \bar k_k
\, \xrightarrow{{\bf C}_{ij}} \,
s_{jk}\, (1-r) 
\, .
\nnb
\eeq

\end{itemize}
Invariants built with the two different remappings are identical in the collinear $\bC{ij}$ limit. The proof of the last of \eq{eq:CijBjk(ijk)} is completed by the fact that $\bC{ij} \, x \, = \, \bC{ij} \, x_{[ij]} \, = \, 1-r$.

\section{Library of integrals}
\label{app:MasterInt}

The analytical results collected in this Section depend on the following functions
\beq
A_1 (\xi)
& \equiv &
\gamma_E
+
\Psi^{(0)} (\xi +1)
\, ,
\nnb\\
A_2 (\xi)
& \equiv &
\gamma_E
-
1
+
\Psi^{(0)} (\xi +2)
\, = \,
A_1 (\xi+1) - 1
\, ,
\nnb\\
A_3 (\xi)
& \equiv &
1
-
\zeta_2
+
\Psi^{(1)} (\xi +2)
\, ,
\eeq
where $\xi \geq 0$, $\gamma_E = 0.5772156649...$ is the Euler-Mascheroni constant, $\Psi^{(n)} (z)$ is the $n$-th Polygamma function, namely
\beq
\Psi^{(n)} (z)
= 
\frac{d^{\npo}}{d z^{\npo}} \, \ln[\Gamma(z)]
\, ,
\eeq
and all functions $A_{i}(\xi)$ satisfy $A_{i}(0)=0$.

\subsection{Soft counterterms}
\label{app:softMI}

For $\star \star$ taking value in $\rm F \rm F, \rm F \rm I, \rm I \rm I$, we define
\beq
I_{\rm s, \star \star}^{abc}
& \equiv &
\delta_{f_a g} \,
I_{\rm s, \star \star}
\big(\sk{bc}{abc} \big)
\, ,
\\
J_{\rm s, \star \star}^{abc}(x) 
& \equiv &
\delta_{f_a g} \,
J_{\rm s, \star \star}
\big(\sk{bc}{abc}, x \big)
\, ,
\eeq
where the relevant integrals obtained integrating the soft counterterm in \eq{eq:NLOKcompactS} read
\beq
I_{\rm s, \rm F \rm F}(s)
& = &
\frac{\as}{2\pi}
\left( \frac{s}{e^{\euler\!}\mu^2} \right)^{\!\!-\eps} 
\frac{\Gam(1 - \eps) \Gam(2 + \alpha - \eps)}{\eps^2 \, \Gam(2 + \alpha - 3 \eps)}
\\
& = &
\frac{\as}{2\pi}
\left( \frac{s}{\mu^2} \right)^{\!\!-\eps} 
\bigg[
\frac{1}{\eps^2} 
+ 
\frac{2}{\eps} 
- 
\frac{7\pi^2}{12}
+
6 
+  
2 \, A_2(\alpha) \bigg( \frac{1}{\eps} + 2 + A_2(\alpha) \bigg)
-
4 \, A_3(\alpha)
+
\mc O(\eps)
\bigg]
\, ,
\nnb\\[5pt]
I_{\rm s, \rm F \rm I}(s)
& = &
\frac{\as}{2\pi}
\left( \frac{s}{e^{\euler\!}\mu^2} \right)^{\!\!-\eps} 
\frac{\Gam(1-\eps) \, \Gam(2 + \alpha )}{\eps^2 \, \Gam(2 + \alpha - 2 \eps)}
\\
& = &
\frac{\as}{2\pi}
\left( \frac{s}{\mu^2} \right)^{\!\!-\eps} 
\bigg[
\frac{1}{\eps^2} 
+ 
\frac{2}{\eps} 
- 
\frac{\pi^2}{4} 
+
4
+  
2 \, A_2(\alpha) \bigg( \frac{1}{\eps} + 2 + A_2(\alpha) \bigg)
-
2 \, A_3(\alpha)
+
\mc O(\eps)
\bigg]
\, ,
\nnb\\[5pt]
I_{\rm s, \rm I \rm I}(s)
& = &
\frac{\as}{2\pi}
\left( \frac{s}{e^{\euler\!}\mu^2} \right)^{\!\!-\eps} 
\frac{\Gam(1-\eps) \, \Gam(2 + \alpha )}{\eps^2 \, \Gam(2 + \alpha - 2 \eps)}
\, = \,
I_{\rm s, \rm F \rm I}(s)
\, ;
\eeq
\beq
J_{\rm s, \rm F \rm I}(s, x)
& = &
\frac{\as}{2\pi}
\left( \frac{s}{e^{\euler\!}\mu^2} \right)^{\!\!-\eps} 
\frac{\Gam(2 + \alpha - \eps)}{( - \eps) \, \Gam(2 + \alpha - 2 \eps)}
\biggl( \frac{x^{1 + \alpha}}{(1-x)^{1+\eps}} \biggr)_+
\\
& = &
\frac{\as}{2\pi}
\left( \frac{s}{\mu^2} \right)^{\!\!-\eps} \!
\left[
-
\left(\frac{x^{1+ \alpha}}{1-x} \right)_{\!\! +} \!\!
\left(  
\frac{1}{\eps} 
+ 
1 
+ 
A_2(\alpha) 
\right)  
+ 
\left(\frac{x^{1+\alpha}  \ln(1-x)}{1-x} \right)_{\!\! +}
+
\mc O(\eps)
\right]
\, ,
\nnb\\[5pt]
J_{\rm s, \rm I \rm I}(s,x)
& = &
\frac{\as}{2\pi}
\left( \frac{s}{e^{\euler\!}\mu^2} \right)^{\!\!-\eps} 
\frac{\Gam(1  - \eps)}{\eps^2 \, \Gam( - 2 \eps)}
\biggl( \frac{x^{1 + \alpha}}{(1-x)^{1+2\eps}} \biggr)_+
\\
& = &
\frac{\as}{2\pi}
\left( \frac{s}{\mu^2} \right)^{\!\!-\eps} 
\left[
-
\left( \frac{x^{1+\alpha}}{1-x}  \right)_{\!\! +}
\frac{2}{\eps}
+
4
\left(\frac{x^{1+\alpha} \ln(1-x)}{1-x}\right)_{\!\! +}
+
\mc O(\eps)
\right]
\, .
\nnb
\eeq

\subsection{Collinear counterterms}
\label{app:collinearMI}

When a collinear splitting occurs in the final state, one has
\beq
I_{\rm hc, \rm F \star}^{abc}
& \; \equiv \; &
\delta_{ \{f_a f_b\} \{q \bar q\} } \,
I_{\rm hc, \rm F \star}^{\zg}
\big(\sk{bc}{abc} \big)
+ 
\Big( 
\delta_{f_a g} \delta_{f_b \{q, \bar q\}} 
+ 
\delta_{f_b g} \delta_{f_a \{q, \bar q\}} 
\Big) \,
I_{\rm hc, \rm F \star}^{\og}
\big(\sk{bc}{abc}\big)
\nnb\\
&&
+ \,
\delta_{f_a g} \delta_{f_b g} \,
I_{\rm hc, \rm F \star}^{\tg}
\big(\sk{bc}{abc}\big)
\, ,
\nnb\\
J_{\rm hc, \rm F \star}^{abc}(x) 
& \; \equiv \; &
\delta_{ \{f_a f_b\} \{q \bar q\} } \,
J_{\rm hc, \rm F \star}^{\zg}
\big(\sk{bc}{abc}, x \big)
+ 
\Big( 
\delta_{f_a g} \delta_{f_b \{q, \bar q\}} 
+ 
\delta_{f_b g} \delta_{f_a \{q, \bar q\}} 
\Big) \,
J_{\rm hc, \rm F \star}^{\og}
\big(\sk{bc}{abc}, x\big)
\nnb\\
&&
+ \,
\delta_{f_a g} \delta_{f_b g} \,
J_{\rm hc, \rm F \star}^{\tg}
\big(\sk{bc}{abc}, x\big)
\, ,
\nnb\\
I_{\rm sc, \rm F \star}^{abc}
& \; \equiv \; &
\delta_{f_a g} \, 2 \,  C_{f_b} \,
I_{\rm sc, \rm F \star}
\big(\sk{bc}{abc} \big)
\, ,
\nnb\\
J_{\rm sc, \rm F \star}^{abc}(x) 
& \; \equiv \; &
\delta_{f_a g}  \, 2 \, C_{f_b} \,
J_{\rm sc, \rm F \star}
\big(\sk{bc}{abc}, x \big)
\, ,
\eeq
while, if the splitting originates from an initial partonic state, one has
\beq
J_{\rm hc, \rm I \star}^{abc}(x) 
& \; \equiv \; &
\delta_{ \{f_a f_{[ab]}\} \{q \bar q\} } \,
J_{\rm hc, \rm I \star}^{\zg}
\big(\sk{bc}{abc}, x \big)
+ 
\delta_{f_a g} \delta_{f_{[ab]} \{q, \bar q\}} \,
J_{\rm hc, \rm I \star}^{\og, qg}
\big(\sk{bc}{abc}, x \big)
\nnb\\
&&
+ \,
\delta_{f_{[ab]} g} \delta_{f_a \{q, \bar q\}} \,
J_{\rm hc, \rm I \star}^{\og, gq}
\big(\sk{bc}{abc}, x \big)
+ 
\delta_{f_a g} \delta_{f_{[ab]} g} \, 
J_{\rm hc, \rm I \star}^{\tg}
\big(\sk{bc}{abc}, x \big)
\, ,
\nnb\\
I_{\rm sc, \rm I \star}^{abc} 
& = & 
\delta_{f_a g} \, 2 \, C_{f_b} \,
I_{\rm sc, \rm I \star}
\big(\sk{bc}{abc} \big)
\, ,
\nnb\\
J_{\rm sc, \rm I \star}^{abc}(x) 
& = &
\delta_{f_a g}  \, 2 \,  C_{f_b} \,
J_{\rm sc, \rm I \star}
\big(\sk{bc}{abc}, x \big)
\, ,
\eeq
where $\star = \rm F, \rm I$. Explicitly, the integrals obtained through integration of the collinear counterterms in Eqs.~(\ref{eq:NLOKcompactHCF}, \ref{eq:NLOKcompactHCI}) read as follows.
\begin{itemize}
\item Final $j$,  final $r$:
\beq
I_{\rm hc, \rm F \rm F}^{\zg} \left(s\right)
& = &
\frac{\as}{2\pi}
\left( \frac{s}{e^{\euler\!}\mu^2} \right)^{\!\!-\eps} \!\!
4 \, T_R
\frac{\Gam(2-\eps)^2  \, \Gam(2 + \beta - 2 \eps )}{(-\eps) \, \Gam(4-2\eps) \, \Gam(2 + \beta - 3 \eps) }
\nnb\\
& = &
\frac{\as}{2\pi}
\left( \frac{s}{\mu^2} \right)^{\!\!-\eps} \!\!
T_R
\left[
- 
\frac{2}{3} \, \frac1\eps
- 
\frac{16}{9}  
- 
\frac{2}{3} \, A_2(\beta)
+
\mc O(\eps)
\right]
\, ,
\\[5pt]
I_{\rm hc, \rm F \rm F}^{\og} (s)
& = & 
\frac{\as}{2\pi}
\left( \frac{s}{e^{\euler\!}\mu^2} \right)^{\!\!-\eps} \!\!
(3-2\eps) \, C_F
\frac{\Gam(2-\eps)^2  \, \Gam(2 + \beta - 2 \eps )}{(-\eps) \, \Gam(4-2\eps) \, \Gam(2 + \beta - 3 \eps) }
\nnb\\
& = &
\frac{\as}{2\pi}
\left( \frac{s}{\mu^2} \right)^{\!\!-\eps} \!\!
C_F 
\left[
- 
\frac{1}{2} \, \frac1\eps
- 
1
- 
\frac{1}{2} \, A_2(\beta)
+
\mc O(\eps)
\right]
\, ,
\\[5pt]
I_{\rm hc, \rm F \rm F}^{\tg} (s)
& = &
\frac{\as}{2\pi}
\left( \frac{s}{e^{\euler\!}\mu^2} \right)^{\!\!-\eps} \!\!
2 \, C_A
\frac{\Gam(2-\eps)^2 \, \Gam(2 + \beta - 2 \eps )}{(-\eps) \, \Gam(4-2\eps) \, \Gam(2 + \beta - 3 \eps) }
\nnb\\
& = &
\frac{\as}{2\pi}
\left( \frac{s}{\mu^2} \right)^{\!\!-\eps} \!\!
C_A
\left[
- 
\frac{1}{3} \, \frac1\eps
- 
\frac{8}{9}  
- 
\frac{1}{3} \, A_2(\beta)
+
\mc O(\eps)  
\right]
\, ,
\\[5pt]
\nnb\\
I_{\rm sc, \rm F \rm F} (s)
& = &
\frac{\as}{2\pi}
\left( \frac{s}{e^{\euler\!}\mu^2} \right)^{\!\!-\eps} 
\frac{\Gam(1-\eps) \, \Gam(2 +\beta - 2 \eps)}{\eps^2 \, \Gam(2 +\beta - 3 \eps)}
\left[
\frac{\Gam(2-\eps)}{\Gam(2 -2 \eps)}
-
\frac{\Gam(2 +\alpha - 2 \eps)}{\Gam(2 +\alpha - 3 \eps)}
\right]
\nnb\\
& = &
\frac{\as}{2\pi}
\left( \frac{s}{\mu^2} \right)^{\!\!-\eps} 
\bigg[
- 
1 
+ 
\frac{\pi^2}{6} 
- 
A_2(\alpha) \,
\bigg( \frac{1}{\eps} + 2 + \frac12 \, A_2(\alpha) + A_2(\beta) \bigg) 
\nnb\\
&&
\hspace{37mm}
+ \,
\frac{5}{2} \,  A_3(\alpha) 
+
\mc O(\eps)
\bigg] 
\, .
\eeq
\end{itemize}
\begin{itemize}
\item Final $j$,  initial $r$:
\beq
I_{\rm hc, \rm F \rm I}^{\zg} (s)
& = &
\frac{\as}{2\pi}
\left( \frac{s}{e^{\euler\!}\mu^2} \right)^{\!\!-\eps} \!\!
4 \, T_R
\frac{\Gam(2-\eps)^2 \, \Gam(2 + \beta )}{(-\eps) \, \Gam(4-2\eps) \, \Gam(2 + \beta - \eps)}
\nnb\\
& = &
\frac{\as}{2\pi}
\left( \frac{s}{\mu^2} \right)^{\!\!-\eps} \!\!
T_R
\left[
- 
\frac{2}{3} \, \frac1\eps
- 
\frac{16}{9}  
- 
\frac{2}{3} \, A_2(\beta)
+
\mc O(\eps)
\right]
\, ,
\\[5pt]
I_{\rm hc, \rm F \rm I}^{\og} (s)
& = &
\frac{\as}{2\pi}
\left( \frac{s}{e^{\euler\!}\mu^2} \right)^{\!\!-\eps} \!\!
(3-2\eps) \, C_F
\frac{\Gam(2-\eps)^2 \, \Gam(2 + \beta )}{(-\eps) \, \Gam(4-2\eps) \, \Gam(2 + \beta - \eps)}
\nnb\\
& = &
\frac{\as}{2\pi}
\left( \frac{s}{\mu^2} \right)^{\!\!-\eps} \!\!
C_F 
\left[
- 
\frac{1}{2} \, \frac1\eps
- 
1
- 
\frac{1}{2} \, A_2(\beta)
+
\mc O(\eps)
\right]
\, ,
\\[5pt]
I_{\rm hc, \rm F \rm I}^{\tg} (s)
& = &
\frac{\as}{2\pi}
\left( \frac{s}{e^{\euler\!}\mu^2} \right)^{\!\!-\eps} \!\!
2 \, C_A
\frac{\Gam(2-\eps)^2 \, \Gam(2 + \beta )}{(-\eps) \, \Gam(4-2\eps) \, \Gam(2 + \beta - \eps)}
\nnb\\
& = &
\frac{\as}{2\pi}
\left( \frac{s}{\mu^2} \right)^{\!\!-\eps} \!\!
C_A
\left[
- 
\frac{1}{3} \, \frac1\eps
- 
\frac{8}{9}  
- 
\frac{1}{3} \, A_2(\beta)
+
\mc O(\eps)  
\right]
\, ,
\eeq
\beq
J_{\rm hc, \rm F \rm I}^{\zg} \left(s, x \right)
& = &
\frac{\as}{2\pi}
\left( \frac{s}{e^{\euler\!}\mu^2} \right)^{\!\!-\eps} \!
4 \, T_R
\frac{(1  - \eps) \Gam(2  - \eps)}{\Gam(4 - 2 \eps)}
\biggl( \frac{x^{1 + \beta}}{(1-x)^{1+\eps}} \biggr)_+
\nnb\\
& = &
\frac{\as}{2\pi}
\left( \frac{s}{\mu^2} \right)^{\!\!-\eps} \!\!
T_R 
\left( \frac{x^{1+\beta}}{1-x}  \right)_{\!\! +}
\left[
\frac{2}{3}  
+
\mc O(\eps)
\right]
\, ,
\\[5pt]
J_{\rm hc, \rm F \rm I}^{\og} \left(s, x \right)
& = &
\frac{\as}{2\pi}
\left( \frac{s}{e^{\euler\!}\mu^2} \right)^{\!\!-\eps} \!
(3-2\eps) \, C_F
\frac{(1  - \eps) \Gam(2  - \eps)}{\Gam(4 - 2 \eps)}
\biggl( \frac{x^{1 + \beta}}{(1-x)^{1+\eps}} \biggr)_+
\nnb\\
& = & 
\frac{\as}{2\pi}
\left( \frac{s}{\mu^2} \right)^{\!\!-\eps} \!\!
C_F 
\left( \frac{x^{1+\beta}}{1-x}  \right)_{\!\! +}
\left[
\frac{1}{2}  
+
\mc O(\eps)
\right]
\, ,
\\[5pt]
J_{\rm hc, \rm F \rm I}^{\tg} \left(s, x \right)
& = &
\frac{\as}{2\pi}
\left( \frac{s}{e^{\euler\!}\mu^2} \right)^{\!\!-\eps} \!
2 \, C_A
\frac{(1  - \eps) \Gam(2  - \eps)}{\Gam(4 - 2 \eps)}
\biggl( \frac{x^{1 + \beta}}{(1-x)^{1+\eps}} \biggr)_+
\nnb\\
& = &
\frac{\as}{2\pi}
\left( \frac{s}{\mu^2} \right)^{\!\!-\eps} \!\!
C_A
\left( \frac{x^{1+\beta}}{1-x}  \right)_{\!\! +}
\left[
\frac{1}{3}  
+
\mc O(\eps)
\right]
\, ,
\\[5pt]
\nnb\\
I_{\rm sc, \rm F \rm I} (s)
& = &
\frac{\as}{2\pi}
\left( \frac{s}{e^{\euler\!}\mu^2} \right)^{\!\!-\eps} 
\frac{\Gam(1-\eps) \, \Gam(2 +\beta )}{\eps^2 \, \Gam(2 +\beta -  \eps)}
\left[
\frac{\Gam(2-\eps)}{\Gam(2 -2 \eps)}
-
\frac{\Gam(2 +\alpha - \eps)}{\Gam(2 +\alpha - 2 \eps)}
\right]
\qquad
\nnb\\
& = &
\frac{\as}{2\pi}
\left( \frac{s}{\mu^2} \right)^{\!\!-\eps} 
\bigg[
- 
A_2(\alpha) 
\bigg( \frac{1}{\eps} + 2 + \frac12 \, A_2(\alpha) + A_2(\beta) \bigg) 
\nnb\\
&&
\hspace{22mm}
+ \,
\frac{3}{2} \, A_3(\alpha)
+
\mc O(\eps)
\bigg] 
\, ,
\\[5pt]
J_{\rm sc, \rm F \rm I} (s,x)
& = &
\frac{\as}{2\pi}
\left( \frac{s}{e^{\euler\!}\mu^2} \right)^{\!\!-\eps} \!\!
\left(-\frac{1}{\eps} \right) \!
\left[
\frac{\Gam(2-\eps)}{\Gam(2 -2 \eps)}
-
\frac{\Gam(2 +\alpha - \eps)}{\Gam(2 +\alpha - 2 \eps)}
\right]
\biggl( \frac{x^{1 + \beta}}{(1-x)^{1+\eps}} \biggr)_+
\nnb\\
& = &
\frac{\as}{2\pi}
\left( \frac{s}{\mu^2} \right)^{\!\!-\eps} 
\left( \frac{x^{1+\beta}}{1-x}  \right)_{\!\! +}
\Big[
A_2(\alpha) 
+
\mc O(\eps)
\Big]
\, .
\eeq
\end{itemize}
\begin{itemize}
\item Initial $j$, final $r$:
\beq
J_{\rm hc, \rm I \rm F}^{\zg} (s,x)
& = &
\frac{\as}{2\pi}
\left( \frac{s}{e^{\euler\!}\mu^2} \right)^{\!\!-\eps} \!
T_R 
\left( 
1
- 
\frac{2 x (1-x)}{1-\eps} 
\right) 
\frac{ (1-x)^{-\eps} \, \Gam(1 +\gamma - \eps) }{ (- \eps) \, \Gam(1 +\gamma - 2 \eps)}
\\
& = &
\frac{\as}{2\pi}
\left( \frac{s}{\mu^2} \right)^{\!\!-\eps} \!
T_R 
\biggl[
\Big( x^2 + (1-x)^2 \Big)
\left(
- 
\frac{1}{\eps}
+ 
\ln(1-x)
- 
A_1(\gamma)
\right)
\nnb\\
&&
\hspace{72mm}
+ \,
2 \, x(1-x) 
+
\mc O(\eps)
\biggr]
\, , \nnb
\\[5pt]
J_{\rm hc, \rm I \rm F}^{\og, qg} (s,x)
& = &
\frac{\as}{2\pi}
\left( \frac{s}{e^{\euler\!}\mu^2} \right)^{\!\!-\eps} \!\!
C_F \,
(1-x) \, (1-\eps) 
\frac{ (1-x)^{-\eps} \, \Gam(1 +\gamma - \eps) }{ (- \eps) \, \Gam(1 +\gamma - 2 \eps)}
\\
& = & 
\frac{\as}{2\pi}
\left( \frac{s}{\mu^2} \right)^{\!\!-\eps} \!\!
C_F \,
(1-x)
\left[
- 
\frac{1}{\eps}
+ 
\ln(1-x)
+
1  
- 
A_1(\gamma)
+
\mc O(\eps)
\right]
\, , \nnb
\\[5pt]
J_{\rm hc, \rm I \rm F}^{\og, gq} (s,x)
& = &
\frac{\as}{2\pi}
\left( \frac{s}{e^{\euler\!}\mu^2} \right)^{\!\!-\eps} \!\!
C_F 
\left(
\frac{1+(1-x)^2}{x}
-
\eps x
\right) 
\frac{ (1-x)^{-\eps} \, \Gam(1 +\gamma - \eps) }{ (- \eps) \, \Gam(1 +\gamma - 2 \eps)}
\\
& = & 
\frac{\as}{2\pi}
\left( \frac{s}{\mu^2} \right)^{\!\!-\eps} \!
C_F 
\biggl[
\frac{1 + (1-x)^2}{x} 
\left(
- 
\frac{1}{\eps}
+ 
\ln(1-x)
- 
A_1(\gamma)
\right) 
+ 
x   
+
\mc O(\eps)
\biggr]
\, , \nnb
\\[5pt]
J_{\rm hc, \rm I \rm F}^{\og} (s,x)
& \equiv &
J_{\rm hc, \rm I \rm F}^{\og,qg} (s,x)
+
J_{\rm hc, \rm I \rm F}^{\og,gq} (s,x)
\\
& = &
\frac{\as}{2\pi}
\left( \frac{s}{e^{\euler\!}\mu^2} \right)^{\!\!-\eps} \!\!
C_F 
\left(
\frac2x-1-\eps
\right) 
\frac{ (1-x)^{-\eps} \, \Gam(1 +\gamma - \eps) }{ (- \eps) \, \Gam(1 +\gamma - 2 \eps)}
\nnb\\
& = & 
\frac{\as}{2\pi}
\left( \frac{s}{\mu^2} \right)^{\!\!-\eps} \!
C_F 
\biggl[
\left(\frac2x-1\right)
\left(
- 
\frac{1}{\eps}
+ 
\ln(1-x)
- 
A_1(\gamma)
\right) 
+ 
1   
+
\mc O(\eps)
\biggr]
\, , \nnb
\\[5pt]
J_{\rm hc, \rm I \rm F}^{\tg} (s,x)
& = &
\frac{\as}{2\pi}
\left( \frac{s}{e^{\euler\!}\mu^2} \right)^{\!\!-\eps} \!\!
2 \, C_A
\left( 
\frac{1-x}{x} 
+ 
x(1-x) 
\right) 
\frac{ (1-x)^{-\eps} \, \Gam(1 +\gamma - \eps) }{ (- \eps) \, \Gam(1 +\gamma - 2 \eps)}
\\
& = &
\frac{\as}{2\pi}
\left( \frac{s}{\mu^2} \right)^{\!\!-\eps} \!\!
2 \, C_A 
\left( 
\frac{1-x}{x} 
+ 
x(1-x) 
\right)  
\left[
- 
\frac{1}{\eps}
+ 
\ln(1-x)
- 
A_1(\gamma) 
+
\mc O(\eps)
\right]
\, ,
\nnb\\
\nnb\\[5pt]
I_{\rm sc, \rm I \rm F} (s)
& = &
\frac{\as}{2\pi}
\left( \frac{s}{e^{\euler\!}\mu^2} \right)^{\!\!-\eps} 
\frac{\Gam(1 -\eps) \, \Gam(1+\gamma-\eps) }{\eps^2 \, \Gam(1 + \gamma -  2 \eps)}
\left[
\frac{1}{\Gam(2 - \eps)}
-
\frac{\Gam(2 +\alpha)}{\Gam(2 +\alpha - \eps)}
\right]
\\
& = &
\frac{\as}{2\pi}
\left( \frac{s}{\mu^2} \right)^{\!\!-\eps} 
\bigg[
- 
A_2(\alpha)
\bigg( \frac{1}{\eps} + 1 + \frac12 \, A_2(\alpha) + A_1(\gamma) \bigg)
+
\frac{1}{2} \, A_3(\alpha) 
+
\mc O(\eps)
\bigg] 
\, ,
\nnb\\[5pt]
J_{\rm sc, \rm I \rm F} (s,x)
& = &
\frac{\as}{2\pi}
\left( \frac{s}{e^{\euler\!}\mu^2} \right)^{\!\!-\eps} \!\!
\frac{\Gam(1 +\gamma-\eps)}{(-\eps) \, \Gam(1 +\gamma-2\eps)}
\biggl( \frac{x(1-x^\alpha)}{(1-x)^{1+\eps}} \biggr)_+
\\
& = &
\frac{\as}{2\pi}
\left( \frac{s}{\mu^2} \right)^{\!\!-\eps} 
\bigg[
\left(\frac{x(x^\alpha-1)}{1-x}\right)_{\!\! +} 
\bigg( 
\frac{1}{\eps} 
+
A_1(\gamma) 
\bigg)
\nnb\\
&&
\hspace{30mm}
+  \,
\left(\frac{x(1-x^\alpha) \ln (1-x)}{1-x}\right)_{\!\! +} 
+
\mc O(\eps)
\bigg]
\, .
\nnb
\eeq
\end{itemize}
\begin{itemize}
\item Initial $j$, initial $r$:
\beq
J_{\rm hc, \rm I \rm I}^{\zg} (s,x)
& = &
\frac{\as}{2\pi}
\left( \frac{s}{e^{\euler\!}\mu^2} \right)^{\!\!-\eps} \!
T_R 
\left( 
1
- 
\frac{2 x (1-x)}{1-\eps} 
\right) 
\frac{ (1-x)^{-2\eps} \, \Gam(1 +\gamma - \eps) }{ (- \eps) \, \Gam(1 +\gamma - 2 \eps)}
\\
& = &
\frac{\as}{2\pi}
\left( \frac{s}{\mu^2} \right)^{\!\!-\eps} \,
T_R 
\biggl[
\Big( x^2 + (1-x)^2 \Big)
\left(
- 
\frac{1}{\eps}
+ 
2 \ln(1-x)
- 
A_1(\gamma)
\right)
\nnb\\
&&
\hspace{76mm}
+ \,
2 \, x(1-x) 
+
\mc O(\eps)
\biggr]
\, , \nnb
\\[5pt]
J_{\rm hc, \rm I \rm I}^{\og, qg} (s,x)
& = &
\frac{\as}{2\pi}
\left( \frac{s}{e^{\euler\!}\mu^2} \right)^{\!\!-\eps} \!
C_F \,
(1-x) \, (1-\eps) 
\frac{ (1-x)^{-2\eps} \, \Gam(1 +\gamma - \eps) }{ (- \eps) \, \Gam(1 +\gamma - 2 \eps)}
\\
& = & 
\frac{\as}{2\pi}
\left( \frac{s}{\mu^2} \right)^{\!\!-\eps} 
C_F \,
(1-x)
\left[
- 
\frac{1}{\eps}
+ 
2 \ln(1-x)
+
1
- 
A_1(\gamma)
+
\mc O(\eps)
\right]
\, ,
\nnb\\[5pt]
J_{\rm hc, \rm I \rm I}^{\og, gq} (s,x)
& = &
\frac{\as}{2\pi}
\left( \frac{s}{e^{\euler\!}\mu^2} \right)^{\!\!-\eps} \!
C_F 
\left(
\frac{1+(1-x)^2}{x}
-
\eps x
\right) 
\frac{ (1-x)^{-2\eps} \, \Gam(1 +\gamma - \eps) }{ (- \eps) \, \Gam(1 +\gamma - 2 \eps)}
\\
& = & 
\frac{\as}{2\pi}
\left( \frac{s}{\mu^2} \right)^{\!\!-\eps} 
C_F 
\biggl[
\frac{1 + (1-x)^2}{x} 
\left(
- 
\frac{1}{\eps}
+ 
2 \ln(1-x)
- 
A_1(\gamma)
\right) 
+
x   
+
\mc O(\eps)
\biggr]
\, , \nnb
\\[5pt]
J_{\rm hc, \rm I \rm I}^{\og} (s,x)
& \equiv &
J_{\rm hc, \rm I \rm I}^{\og,qg} (s,x)
+
J_{\rm hc, \rm I \rm I}^{\og,gq} (s,x)
\\
& = &
\frac{\as}{2\pi}
\left( \frac{s}{e^{\euler\!}\mu^2} \right)^{\!\!-\eps} \!
C_F 
\left(
\frac2x-1-\eps
\right) 
\frac{ (1-x)^{-2\eps} \, \Gam(1 +\gamma - \eps) }{ (- \eps) \, \Gam(1 +\gamma - 2 \eps)}
\nnb\\
& = & 
\frac{\as}{2\pi}
\left( \frac{s}{\mu^2} \right)^{\!\!-\eps} 
C_F 
\biggl[
\left(\frac2x-1\right)
\left(
- 
\frac{1}{\eps}
+ 
2 \ln(1-x)
- 
A_1(\gamma)
\right) 
+
1   
+
\mc O(\eps)
\biggr]
\, ,
\nnb\\[5pt]
J_{\rm hc, \rm I \rm I}^{\tg} (s,x)
& = &
\frac{\as}{2\pi}
\left( \frac{s}{e^{\euler\!}\mu^2} \right)^{\!\!-\eps} \!
2 \, C_A
\left( 
\frac{1-x}{x} 
+ 
x(1-x) 
\right) 
\frac{ (1-x)^{-2\eps} \, \Gam(1 +\gamma - \eps) }{ (- \eps) \, \Gam(1 +\gamma - 2 \eps)}
\\
& = &
\frac{\as}{2\pi}
\left( \frac{s}{\mu^2} \right)^{\!\!-\eps} 
2 \, C_A 
\left( 
\frac{1-x}{x} 
+ 
x(1-x) 
\right)  
\left[
- 
\frac{1}{\eps}
+ 
2 \ln(1-x)
- 
A_1(\gamma) 
+
\mc O(\eps)
\right]
\, ,
\nnb\\
\nnb\\[5pt]
I_{\rm sc, \rm I \rm I} (s)
& = &
\frac{\as}{2\pi}
\left( \frac{s}{e^{\euler\!}\mu^2} \right)^{\!\!-\eps} 
\frac{\Gam(1 - 2\eps) \, \Gam(1+\gamma-\eps) }{2\eps^2 \, \Gam(1 + \gamma -  2 \eps)}
\left[
\frac{1}{\Gam(2 - 2\eps)}
-
\frac{\Gam(2 +\alpha)}{\Gam(2 +\alpha - 2\eps)}
\right]
\\
& = &
\frac{\as}{2\pi}
\left( \frac{s}{\mu^2} \right)^{\!\!-\eps} 
\bigg[
-
A_2(\alpha) \bigg( \frac{1}{\eps} + 2 + A_2(\alpha) + A_1(\gamma) \bigg)
+
A_3(\alpha)
+
\mc O(\eps)
\bigg] 
\, ,
\nnb\\[5pt]
J_{\rm sc, \rm I \rm I} (s,x)
& = &
\frac{\as}{2\pi}
\left( \frac{s}{e^{\euler\!}\mu^2} \right)^{\!\!-\eps} \!\!
\frac{\Gam(1 +\gamma-\eps)}{(-\eps) \, \Gam(1 +\gamma-2\eps)}
\biggl( \frac{x(1-x^\alpha)}{(1-x)^{1+2\eps}} \biggr)_+
\\
& = &
\frac{\as}{2\pi}
\left( \frac{s}{\mu^2} \right)^{\!\!-\eps} 
\bigg[
\left(\frac{x(x^\alpha-1)}{1-x}\right)_{\!\! +} 
\bigg( 
\frac{1}{\eps} 
+
A_1(\gamma) 
\bigg)  
\nnb\\
&&
\hspace{30mm}
+  \,  
2 \left(\frac{x(1-x^\alpha) \ln (1-x)}{1-x}\right)_{\!\! +} 
+
\mc O(\eps)
\bigg]
\, .
\nnb
\eeq
\end{itemize}

\section{Implementation of the subtraction scheme in \madnk}
\label{app:madnk}
In this Appendix, we provide some more details on the implementation of local analytic sector subtraction at NLO within \madnk.
On top of the general operations that \madnk\ automatically handles, the user must provide and code the building blocks defining the subtraction algorithm to be 
used. In our case, the scheme-specific ingredients to be introduced are the \textit{sector partition}, the \textit{kinematic mappings}, 
the \textit{local counterterms}, and their \textit{integrations} over the radiative phase space.\\

\noindent
{\bf Sector partition}\\
First, we introduce the \Verb|SectorGenerator| class, which implements the functions responsible for building the correct list of sectors to be considered for a given process. This procedure consists of two steps, namely the generation of all sectors, which basically follows what was done in {\sc MadFKS}~\cite{Frederix:2009yq}, and the identification of the counterterms, or rather \textit{singular currents}, belonging to each of them. The assignment of currents to sectors, depending on the current type, is exemplified in the following code snippet.
\begin{python}
...
for s in all_sectors:
    ...
    if counterterms is not None:
        s['counterterms'] = []
        for i_ct, ct in enumerate(counterterms):
            current = ct.nodes[0].current
            singular_structure = current.get('singular_structure').substructures[0]
            all_legs = singular_structure.get_all_legs()
            if singular_structure.name()=='S':
                ...
                    s['counterterms'].append(i_ct)      # soft CT
            if singular_structure.name()=='C':
                if not singular_structure.substructures: 
                    ...
                        s['counterterms'].append(i_ct)  # pure-collinear CT
                else:   
                    ... 
                        s['counterterms'].append(i_ct)  # soft-collinear CT
\end{python}
Being aware of the problem of \textit{spurious} singularities that plague collinear kernels, two possibilities for introducing a unitary phase-space partition have been proposed in Section~\ref{sec:localctdef}. Of these, we choose to code the sector symmetrisation, which consists in performing the sum of the mirror sectors $\W{ij} + \W{ji}$. To this aim, we incorporate the different weight functions for the standard/soft/collinear/soft-collinear sector functions by using the definitions in Eqs.~(\ref{eq:sfunNLO} - \ref{eq:Wij_lim2}), where in particular, for the standard case, one has
\begin{python}
def get_sector_wgt(q, p_sector): # sigma_ij without the normalisation
    """
     - q is the total momentum of the incoming particles
     - p_sector : list of the momenta of the particles defining the sector
    """
    s = q.square()
    s_ij = 2 * p_sector[0].dot(p_sector[1])
    s_qi = 2 * p_sector[0].dot(q)
    s_qj = 2 * p_sector[1].dot(q)
    e_i = s_qi / s
    w_ij = s * s_ij / s_qi / s_qj

    return 1 / e_i / w_ij
\end{python}
Then, we implement the \Verb|Sector| class, whose purpose is to return the full weight of the sector function by calling the correct numerator function based on the sector types listed above, and associating the correct normalisation (i.e.~dividing by the sum of all the relevant sector functions), given a kinematic configuration and the flavours of the external states.
\\

\noindent
{\bf Kinematic mappings} \\
All of the kinematic mappings introduced in this work and listed in Appendix \ref{app:mappings} are encoded in a single class, dubbed \Verb|TRNMapping|, which provides the \Verb|map_to_lower_multiplicity| attribute: given an $(\npo)$-body kinematics and a specific counterterm, this function exploits the information about the structure of the singular current and the position of the particles defining the singularity, and applies the suitable transformation of momenta, eventually returning a valid on-shell and momentum-conserving Born-level kinematics. \\

\noindent
{\bf Local counterterms} \\
The implementation of the counterterms at the local level relies on the definition of \textit{singular currents}, whose structure depends on the specific type of divergence being treated. 
Let us consider first the collinear and soft-collinear contributions in Eqs.~(\ref{eq:dampedC}, \ref{eq:dampedSC}): labelling with $ab$  the flavours of a pair of massless QCD particles, we recognise the sets $(gg, gq, q \bar q)$ and $(gg, gq, qg, q \bar q)$~\footnote{Following the notation used in Altarelli-Parisi kernels in Eq.~(\ref{eq:APkernels1}), the first index identifies the parton entering the Born-level amplitude after the collinear emission.} to be the possible singular structures defining  a final or initial collinear splitting event, respectively. 
As an exemplary case, we focus on the final-state splitting with $gq$-flavour resulting particles, namely the $
\final{i} \, \final{j} ( \delta_{f_i g} \delta_{f_j \{q , \bar q\}}  + \delta_{f_j g} \delta_{f_i \{q , \bar q\}})$ contribution in Eqs.~(\ref{eq:dampedC}, \ref{eq:dampedSC}). In
\madnk, singular currents need to specify the so-called singular structure, i.e.~the particle types, and the kinematic limit, for which a specific
current needs to be employed. In the case at hand, we have two final-state legs, a (anti-)quark and a gluon, becoming collinear. Thus we define 
the \Verb|QCD_TRN_C_FgFq| class with this specific singular structure: 
\begin{python}
class QCD_TRN_C_FgFq(general_current.GeneralCurrent):
    ...
    coll_structure_q = sub.CollStructure(
        substructures=tuple([]),
        legs=(
            sub.SubtractionLeg(10, 21, sub.SubtractionLeg.FINAL),
            sub.SubtractionLeg(11, +1, sub.SubtractionLeg.FINAL),
        )
    )
    coll_structure_qx = ...
    ...
\end{python} 
This structure is then introduced in the \Verb|currents| definition, so as it can be found and picked by the code when needed.
\begin{python}
    ...
    currents = [
        sub.Current({
            'resolve_mother_spin_and_color'     : True,
            'n_loops'                           : 0,
            'squared_orders'                    : {'QCD': 2},
            'singular_structure'                : sub.SingularStructure(substructures=(coll_structure_q,)),
        }),
            ...
    ]
    ...
\end{python}
Next, we add the \Verb|mapping_rules| block: here the counterterm is linked to the functions that, respectively, apply the suitable momentum mapping (the \Verb|'mapping'| key), provide the kinematic variables involved in the singular kernel (\Verb|'variables'|) and return the identity of the recoiler particle (\Verb|'reduced_recoiler'|), which is selected according to the rules made explicit by the $\theta_{r \in \star}$ prescriptions in Eqs.~(\ref{eq:KFF}, \ref{eq:KIF}, \ref{eq:KII}).
\begin{python}
    ...
    mapping_rules = [
        {
            'singular_structure'    : ... ,
            'mapping'               : torino_config.final_coll_mapping,
            'variables'             : general_current.CompoundVariables(
                                      kernel_variables.TRN_FFn_variables
                                      ),
            'is_cut'                : torino_config.generalised_cuts,
            'reduced_recoilers'     : torino_config.get_recoiler,
            ... ,
        },
    ]
    ...
\end{python}
Then, the \Verb|kernel| function is specified: given a phase-space point and the (soft-)collinear prefactor weighted by the corresponding sector function and modulated by damping factors, it evaluates the singular kernel and finally stores the results, including the possible spin correlation, in the \Verb|evaluation| vector.
\begin{python}
    def kernel(self, evaluation, all_steps_info, global_variables):
        """ Evaluate this counterterm given the variables provided. """   
        ...
        prefactor = 1./s_rs 
        prefactor *= compensate_sector_wgt( ... , 'C' )
        CS_prefactor = 1./s_rs
        CS_prefactor *= compensate_sector_wgt( ... , 'SC' )

        # Modulate prefactor with damping factors
        recoiler = all_steps_info[0]['mapping_vars']['ids']['c']
        if recoiler > 2:    # final-state
            prefactor *= (1. - y)**beta_FF
            CS_prefactor *= (1. - y)**beta_FF * (1. - z)**alpha
        else:               # initial-state
            prefactor *= x**beta_FI
            CS_prefactor *= x**beta_FI * (1. - z)**alpha
            
        # collinear CT
        spin_correlation, weight = AltarelliParisiKernels.P_qg(self, z, kT)
        # soft_collinear CT 
        soft_col = EpsilonExpansion({0: self.CF * 2. * (1.- z)/z , 1: 0.})
        
        ...
        return evaluation
\end{python}
The \Verb|compensate_sector_wgt| function replaces the resolved sector function, applied by the code, to the corresponding collinear or soft-collinear one.

Let us focus now on the remaining soft contribution. Formally the implementation of this singular structure follows the steps of the  previous collinear example, for which we included the definition of the current, here reading
\begin{python}
class QCD_TRN_S_g(dipole_current.DipoleCurrent):
    ...
    soft_structure = sub.SoftStructure(
        substructures=tuple([]),
        legs=(
            sub.SubtractionLeg(11, 21, sub.SubtractionLeg.FINAL),
        )
    )
    ...
\end{python}
In this case, the singular structure consists in just one final-state gluon being soft. 
In addition, it inherits from the \Verb|DipoleCurrent| class, which handles the different remapped kinematics associated to the $(ikl)$ momentum sets needed by the structure
of the soft kernel, see Eq.~(\ref{eq:dampedS}). Thus, given a $(\npo)$-body phase-space point and a single identified soft particle $i$, the \Verb|kernel| function defines the prefactor weighted by the corresponding sector function, for each $(ikl)$ dipole it evaluates the eikonal kernels with damping factors, and finally stores the results and the colour correlation due to the involved particles in the \Verb|evaluation| vector. \\

\noindent
{\bf Integrated counterterms} \\
The implementation of the integrated counterterms closely follows that of the local counterpart, for both soft and collinear contributions. Again, a
singular structure has to be specified, but this time mapping rules are not needed, as all counterterms share a common Born-level kinematics, as well as sector functions, which are summed away before performing the actual integration. Consider again as a case study the collinear counterterm for the $gq$ final-state splitting, namely the $
\final{i} \, \final{j} ( \delta_{f_i g} \delta_{f_j \{q , \bar q\}}  + \delta_{f_j g} \delta_{f_i \{q , \bar q\}})$ part of Eq.~(\ref{eq:inthcF}) (in the integrated currents we code the collinear and soft-collinear contributions separately, at variance with the local case). Here, in the case of initial-state recoiler, the \Verb|current| definition requires the further specification of the \textit{endpoint} and \textit{bulk+counterterm} receptacles, which collect the $x$-independent and the $x$-dependent contributions, along with the emerging plus-distributions, respectively. In the \Verb|kernel| function we report the results of the counterterm analytic integrations over the radiative phase spaces, which are different depending on the position of the chosen recoiler, as briefly sketched below. 
\begin{python}
class QCD_integrated_TRN_C_FqFg(general_current.GeneralCurrent):
    ...
        def kernel(self, evaluation, all_steps_info, global_variables):
        ...
        recoiler = global_variables['recoiler']
        color_factor = self.CF
        overall = 1./2.

        if recoiler > 2:    # final-state recoiler
            kernel = {
                'bulk': 0. ,
                'counterterm': 0. ,
                'endpoint': color_factor * overall * (EpsilonExpansion({
                    -2: 0.,
                    -1: - 1.,
                    0: - (2. + A2(beta_FF))
                }))
            }
            ...
        elif recoiler <= 2: # initial-state recoiler
            kernel = {
                'bulk': color_factor * overall * (EpsilonExpansion({
                    -2: 0. ,
                    -1: 0. ,
                    0: x**(1. + beta_FI) / (1. - x) 
                })) ,
                'counterterm': color_factor * overall * (EpsilonExpansion({
                    -2: 0. ,
                    -1: 0. ,
                    0: x**(1. + beta_FI) / (1. - x)
                })) ,
                'endpoint': color_factor * overall * (EpsilonExpansion({
                    -2: 0. ,
                    -1: - 1. ,
                    0: - (2. + A2(beta_FI))
                }))
            }
            ...
\end{python}
Next, we translate the evaluation of the coefficients of $\eps$ poles and finite parts to the parameter convention used in \madnk\ through the \Verb|torino_to_madnk_epsexp| function and finally store the resulting computation.


\bibliographystyle{JHEP}
\bibliography{ISR}

\end{document}